\documentclass[10pt,times,a4paper]{iopart}

\expandafter\let\csname equation*\endcsname\relax
\expandafter\let\csname endequation*\endcsname\relax

\usepackage{times}
\usepackage[usenames]{color}
\usepackage{amsfonts}
\usepackage{amsmath}
\usepackage{amssymb}
\usepackage{bm}
\usepackage{dcolumn}
\usepackage{enumerate}
\usepackage{epsfig}
\usepackage{graphicx}
\usepackage{graphics}
\usepackage{multirow}
\usepackage[utf8]{inputenc}
\usepackage{latexsym}
\usepackage{rotating}
\usepackage{hyperref}
\usepackage[caption=false]{subfig}
\usepackage{soul}
\usepackage{cite}
\usepackage{float}
\newcommand{\<}{\begin{equation}}
\newcommand{\?}{\end{equation}}

\newcommand{\R}{\mathbb{R}}
\newcommand{\ri}{\mathrm{i}}
\newcommand{\cA}{\mathcal{A}}
\newcommand{\cO}{\mathcal{O}}

\newcommand{\bO}{\bar{\mathcal{O}}}
\newcommand{\polidx}{\mathfrak{p}}

\DeclareMathOperator{\Real}{Re}

\begin{document}

\title[Fast FD expression for time-dependent GW detector response to CBCs]{A fast frequency-domain expression for the time-dependent detector response of ground-based gravitational-wave detectors to compact binary signals}

\author{Anson Chen}
%\address{Department of Physics, Chinese University of Hong Kong}
%\address{Department of Physics and Astronomy, University College London, Gower Street, London WC1E 6BT, UK}
\address{School of Physics and Astronomy, Queen Mary University of London, Mile End Road, Bethnal Green, London E1 4NS, UK}
\author{Nathan~K~Johnson-McDaniel}
\address{Department of Applied Mathematics and Theoretical Physics, Centre for Mathematical Sciences, University of Cambridge, Wilberforce Road, Cambridge, CB3 0WA, UK}
\address{Department of Physics and Astronomy, University of Mississippi, University, Mississippi 38677, USA}

\date{\today}

\begin{abstract}
For proposed third-generation gravitational-wave detectors such as the Einstein Telescope and Cosmic Explorer, whose sensitive bands are proposed to extend down to $5$~Hz or below, the signals of low-mass compact binaries such as binary neutron stars remain in the detector's sensitive band long enough (up to a few days for the smallest proposed low-frequency cutoff of $1$~Hz) that one cannot neglect the effects of the Earth's rotation on the detector's response and the changing Doppler shift of the signal. In the latter case, one also needs to consider the effects of the Earth's orbital motion, which is currently only included in analyses of compact binary signals using continuous wave techniques. These effects are also relevant for current detectors and signals from putative subsolar-mass binaries. Here we present simple Fourier series methods for computing these effects in the frequency domain, giving explicit expressions for the Earth's orbital motion in terms of low-order Fourier series, which will be sufficiently accurate for all compact binary signals except for those from very low-mass subsolar-mass binaries. The expression for the effects of the Earth's rotation on the antenna pattern functions does not use the stationary phase approximation (SPA), so we are able to show that the SPA is indeed quite accurate in these situations and present a Fourier series expression equivalent to it which is an order of magnitude faster. We also provide illustrations of these effects on detector sensitivity and the accumulation of information about various binary parameters with frequency.
\end{abstract}

%%%%%%%%%%%%%%%%%
\section{Introduction}
%%%%%%%%%%%%%%%%%

Now that gravitational wave observations are becoming a standard astronomical tool, there is much excitement about the promise of the third generation of ground-based gravitational wave detectors, notably the proposed Einstein Telescope (ET)~\cite{Hild:2010id,ET_design} and Cosmic Explorer (CE)~\cite{Reitze:2019iox,Evans:2021gyd} detectors. These detectors will lead to new challenges for the data analysis of compact binary coalescences, such as overlapping signals (e.g.\ \cite{Meacher:2015rex,Samajdar:2021egv,Pizzati:2021apa,Himemoto:2021ukb,Relton:2021cax,Antonelli:2021vwg,Relton:2022whr,Hu:2022bji,Langendorff:2022fzq,Janquart:2023hew,Wang:2023ldq,Alvey:2023naa,Dang:2023xkj,Johnson:2024foj}), correlated noise~\cite{Cireddu:2023ssf,Wong:2024hes} and very loud signals, where the likelihood is very sharply peaked and thus difficult to sample accurately (see, e.g.\ the discussion in~\cite{Smith:2021bqc}). Detectors like Cosmic Explorer also have a frequency-dependent response at high frequencies, where their long armlengths become comparable to or greater than the wavelength of gravitational radiation (see, e.g.~\cite{Essick:2017wyl}). Here we are concerned with the effects of Earth's rotation and orbital motion on the analysis of very long compact binary signals, particularly from binary neutron stars (BNSs), which can last for up to days in band, due to the increased low-frequency sensitivity of the detectors: CE is expected to be sensitive down to $5$~Hz~\cite{Hall:2020dps}, while ET is expected to be sensitive down to $3$~Hz~\cite{Harms:2022jth} and perhaps even as low as $\sim 1$~Hz (\cite{McClelland:2021wqy} mentions $1.8$~Hz).

The dominant effect of the Earth's rotation is through the time-varying antenna pattern functions, though the time-varying Doppler shift due to Earth's rotation can have a significant effect on sky localization for loud BNS signals~\cite{Baral:2023xst}. As far as we are aware, there are no studies taking into account the time-varying Doppler shift due to the Earth's orbital motion, except for ones using continuous-wave analysis techniques (discussed below), even though this is the larger effect for signals lasting around a day or longer: The Earth's orbital motion differs from a straight line by $\sim 3$ Earth radii over a day. The effects of the Earth's ration (sometimes without including the time-varying Doppler shift) have been included in previous studies, either directly in the time domain~\cite{Chan:2018csa,Grimm:2020ivq}, by dividing up the analysis into smaller frequency intervals over which one can neglect the effect's of the Earth's rotation~\cite{Singh:2021bwn,Li:2021mbo,Singh:2021zah,Hu:2023hos}, or using the stationary phase approximation (SPA)~\cite{Smith:2021bqc,Baral:2023xst,Zhao_and_Wen,Nishizawa:2019rra,Takeda:2019gwk,Tsutsui:2020bem,Borhanian:2020ypi,Nitz:2021pbr,Tsutsui:2021izf,Liu:2021dcr,Borhanian:2022czq,Dupletsa:2022scg,Iacovelli:2022bbs,Zhou:2022nmt,Reali:2022aps,Singh:2023cxn,Gardner:2023znk,Yang:2023zxk,Johnson:2024foj}. %Of these, all but~\cite{Chan:2018csa,Takeda:2019gwk,Grimm:2020ivq,Tsutsui:2020bem,Zhao_and_Wen} restrict to the dominant effects from the time-varying antenna pattern functions.

The time-dependent response of the detectors allows one to better localize sources on the sky, particularly when not all detectors in the planned network are operational (see, e.g.\ \cite{Baral:2023xst,Chan:2018csa,Li:2021mbo,Zhao_and_Wen,Nitz:2021pbr,Borhanian:2022czq}), and also when providing pre-merger localizations (see, e.g.\ \cite{Hu:2023hos,Tsutsui:2020bem,Nitz:2021pbr,Liu:2021dcr,Borhanian:2022czq}). These effects are also relevant for searches for (or analyses of) subsolar-mass binaries with the Advanced LIGO~\cite{LIGOScientific:2014pky} and Advanced Virgo~\cite{VIRGO:2014yos} detectors when using the full band of the detector down to $\sim 20$~Hz or below---current analyses~\cite{Abbott:2019qbw,Nitz:2020bdb,Nitz:2021mzz,Phukon:2021cus,Nitz:2021vqh,LIGOScientific:2021job,Nitz:2022ltl,LIGOScientific:2022hai,Prunier:2023cyv} restrict to higher frequencies to reduce the computational burden (see the discussion in~\cite{Magee:2018opb}), or ignore these effects when making forecasts~\cite{Bandopadhyay:2022tbi,Wolfe:2023yuu}. (In~\cite{Morras:2023jvb}, the total mass is high enough that these effects are not relevant.)
However, for sufficiently low masses, continuous wave search techniques can be employed, which already take the effects of the detectors' motion into account~\cite{Miller:2020kmv,Miller:2021knj,KAGRA:2022dwb,Miller:2023rnn} (see also~\cite{Miller:2024fpo}, though this restricts to cases where the effect of the detectors' motion is negligible). The continuous wave techniques also allow one to perform pre-merger detection and sky localization on BNSs in third-generation detectors using the effects of Earth's orbital motion~\cite{Miller:2023rnn}.
%at a average loss in signal-to-noise ratio (SNR) of $8\%$, for the LVC paper % How much SNR is lost with design sensitivity?

For continuous wave signals, one can even distinguish true signals from instrumental artefacts using the time-varying Doppler shift (see, e.g.\ \cite{Zhu:2017ujz}) and time-dependent detector response (see~\cite{DAntonio:2021cfv}). There is also a recent analytic computation of a continuous gravitational wave signal including the Doppler shift in the frequency domain in~\cite{Valluri:2020cqe}.

Low-frequency gravitational wave detectors in space also have time-varying responses. These are more complicated than the ones for ground-based detectors we consider here, though they can also be calculated in the frequency domain using different techniques~\cite{Marsat:2018oam}.

The detectors' motion may even be relevant for very loud binary black hole signals in third-generation detectors, where~\cite{Borhanian:2022czq} finds using a Fisher matrix calculation that a few events have an uncertainty in the sky localization with a $90\%$ credible region of area $< 10^{-3} \text{ deg}^2$.
Thus, even though these signals do not last as long in band as the signals from BNSs (see, e.g.\ figure~8 in~\cite{Borhanian:2022czq}), the effects of the time-dependent detector response may still be relevant, due to the exquisite accuracy of the measurement. We leave a quantitative study of this for future work. %Moreover, such high accuracy in the sky localization could potentially allow one to infer the binary's host galaxy or if it is not located in a galaxy (see, e.g.\ \cite{Chen:2016tys,Howell:2017wvf}).

In~\cite{Chen:2020fzm}, we introduced a simple Fourier series method to account for the frequency-domain effects of the Earth's rotation through the time-varying antenna pattern functions. This method is exact, up to minuscule (fractional errors of $\lesssim 10^{-5}$) corrections due to the time dependence of the Earth's rotational sidereal angular velocity. Here we further explore the method, comparing its speed and accuracy with the SPA, and extend it to include the Doppler shift due to the Earth's rotation and orbital motion, which involves further approximations. While we find that the SPA is quite accurate for BNS signals and the signal from a $0.2+0.2M_\odot$ subsolar-mass binary we consider, it is less accurate than a na{\"\i}ve estimate of the error. Moreover, the SPA calculation (at least in the obvious simple implementation) is much slower than the Fourier series method. %Moreover, if intermediate mass-ratio binaries exist, their signals can last long enough in band that one needs to take the time-dependent response into account, while they may already be relativistic enough that the SPA is not a good approximation. \nkjm{Give more details.}
As applications, we show how the time-dependent response affects the sky dependence of the signal-to-noise ratio (SNR) accumulated at low frequencies and provide the analogues of the plot in~\cite{Harry:2018hke} showing how the information about different BNS parameters is accumulated as a function of frequency, illustrating the effects of the time-dependent response for ET and CE. We expect that the speed of this method will allow parameter estimation studies with third-generation detectors and/or subsolar-mass binaries to use the time-dependent response as a matter of course. We have released a simple Python implementation of the method~\cite{time_dep_response_Git} and plan to implement it in the Bayesian inference package Bilby~\cite{Ashton:2018jfp}. The Bilby implementation will also interface with other work on including the effects of the detector's size at high frequencies. These effects are necessary for analysis of BNS or other low-mass signals in CE (see, e.g.~\cite{Baral:2023xst}, which gives a Bilby implementation of these effects), but are not included in the present study, where we are concerned with the low-frequency effects due to the detector's motion. Nevertheless, since the detector size effects depend on the detector's orientation with respect to the source, it is possible that some of the techniques considered here can speed up the computation of these effects in stochastic sampling calculations.

The paper is structured as follows: We consider the effects of the Earth's rotation on the detector's response, including the speed of our method and accuracy of the SPA, in section~\ref{sec:rot} and then consider the Doppler shift in section~\ref{sec:Doppler}. We show the effects of the time-dependent response on the dependence on sky location of the SNR accumulated at low frequencies in section~\ref{sec:SNR} and on the accumulation of information about binary parameters with frequency in section~\ref{sec:information}. We conclude in section~\ref{sec:concl}. In \ref{app:resp_coeff_dep}, we give information about the sky location and polarization angle dependence of the Fourier coefficients describing the time dependence of the detector's response, while in \ref{app:Doppler_2nd_order}, we give the second order expressions for the Earth's orbital motion used in the Doppler shift computation. We use $c = 1$ units throughout.

%%%%%%%%%%%%%%%%%
\section{Computing the effects of the Earth's rotation on the detector's response}
\label{sec:rot}
%%%%%%%%%%%%%%%%%

We first recall the expression for the effects of the Earth's rotation on the detector response in the frequency domain given in appendix~B of \cite{Chen:2020fzm}. This expression relies on the time dependence of the antenna pattern functions having a Fourier series that terminates with the $2\Omega_\oplus t$ terms, where $\Omega_\oplus$ is the Earth's rotational sidereal angular velocity, which can be taken to be constant to a very good approximation [fractional errors of $\lesssim 10^{-5}$; see, e.g.\ equations~(2.11-14) of~\cite{Kaplan:2006nv}] \footnote{This property of the effects of the Earth's rotation on the detector response is also used in continuous wave analyses, e.g.\ \cite{Astone:2010zz}.}. %\nkjm{It appears that the PyCBC calculation of the response just uses GMST, and thus doesn't include nutation, so that effect, which is dominant, but still small, isn't included here.}].
This expression also neglects the Doppler shift, which we will account for in the next section. Thus, the antenna pattern functions can be written as
\<\label{eq:Fourier_series}
R_{+,\times} = a^{+,\times}_0 + a^{+,\times}_{1\text{c}}\cos(\Omega_\oplus t) + a^{+,\times}_{1\text{s}}\sin(\Omega_\oplus t) + a^{+,\times}_{2\text{c}}\cos(2\Omega_\oplus t) + a^{+,\times}_{2\text{s}}\sin(2\Omega_\oplus t).
\?
The coefficients can be determined algebraically in terms of the antenna pattern functions evaluated at a set of five times (i.e.\ a collocation method), for which we chose $0$, $T_\oplus/8$, $T_\oplus/4$, $T_\oplus/2$ and $3T_\oplus/4$, where $T_\oplus := 2\pi/\Omega_\oplus$ is the Earth's rotational period\footnote{One could alternatively compute the coefficients directly from Eqs.~(10--13) in~\cite{Jaranowski:1998qm}, but it is easier for our applications to extract them from existing numerical implementations in this manner.}. We refer to $R_{+,\times}$ evaluated at those times as $R_1$, $R_2$, $R_3$, $R_4$ and $R_5$ respectively (dropping the polarization subscript for notational simplicity) %\nkjm{Should we also drop it on the $a_k$s?}
and have
\begin{subequations}
\<
a^{+,\times}_0 = \frac{R_1+R_3+R_4+R_5}{4},
\?\<
a^{+,\times}_{1\text{c}} = \frac{R_1-R_4}{2},
\?\<
a^{+,\times}_{1\text{s}} = \frac{R_3-R_5}{2},
\?\<
a^{+,\times}_{2\text{c}} = \frac{R_1+R_4-R_3-R_5}{4},
\?\<
a^{+,\times}_{2\text{s}} = R_2 + \frac{(\sqrt{2}-1)(R_4+R_5) - (\sqrt{2}+1)(R_1+R_3)}{4}.
\?
\end{subequations}
One can compute the $R_\bullet$ values using standard functions in LALSuite~\cite{LALSuite} or PyCBC~\cite{PyCBC}, e.g.\ the \verb,antenna_pattern, function in the PyCBC Detector module.

\begin{figure}[tb]
\centering
\includegraphics[width=0.6\textwidth]{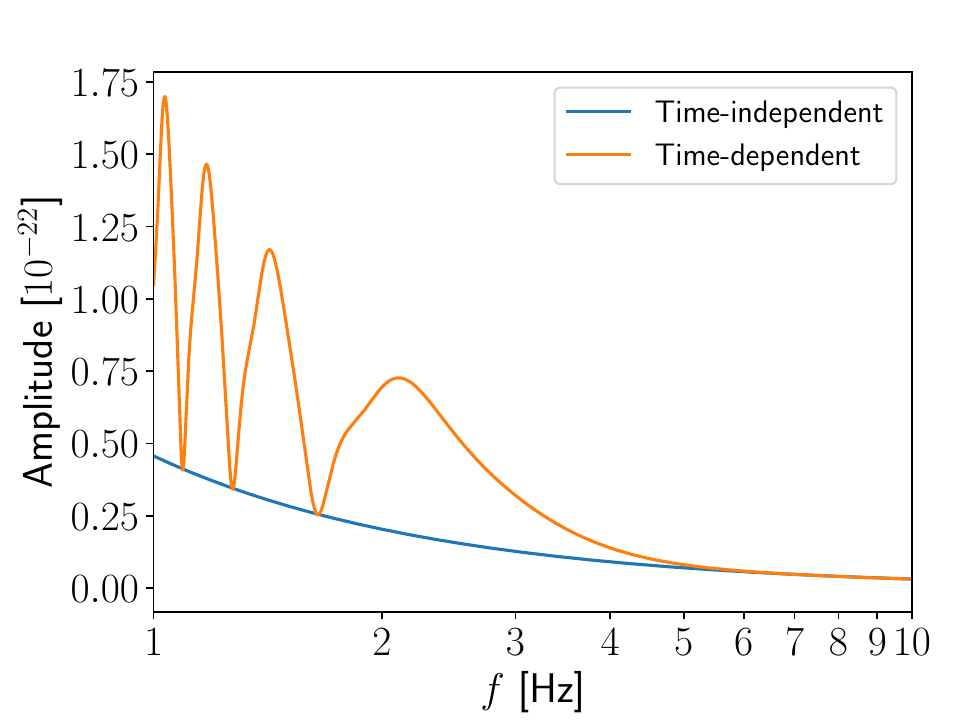}
\caption{The amplitude of the ET response in the frequency domain for the BNS system given in section~\ref{ssec:SPA_comp}, comparing the time-dependent and time-independent responses.}
\label{fig:low_freq_test}
\end{figure}

The detector's response in the time domain to a gravitational wave signal $h_{+,\times}$ is
\<
h(t) = R_+(t)h_+(t) + R_\times(t) h_\times(t),
\?
and we can compute the Fourier transform of the response exactly in terms of the Fourier transform of the gravitational wave signal, $\tilde{h}_{+,\times}$, using equation~\eqref{eq:Fourier_series}. Here, we use the opposite Fourier transform sign convention to \cite{Chen:2020fzm} so that we are consistent with the convention used for the TaylorF2 post-Newtonian approximant in \cite{Buonanno:2009zt}, viz.\
\<\label{eq:fourier}
\tilde{h}(f) := \int_\R h(t)e^{2\pi\ri ft}\mathrm{d}t.
\?
We thus have
\<\label{eq:htilde_response}
\begin{split}
\tilde{h}(f) &= \sum_{\polidx\in\{+,\times\}}\Biggl(a^\polidx_0\tilde{h}_\polidx(f) + \sum_{k\in\{1,2\}}\biggl\{\frac{a^\polidx_{k\text{c}}}{2}[\tilde{h}_\polidx(f + kF_\oplus) + \tilde{h}_\polidx(f - kF_\oplus)]\\
&\quad + \frac{a^\polidx_{k\text{s}}}{2\ri}[\tilde{h}_\polidx(f + kF_\oplus) - \tilde{h}_\polidx(f - kF_\oplus)]\biggr\}\Biggr).
\end{split}
\?
where $F_\oplus := \Omega_\oplus/(2\pi) \simeq 10^{-5}$~Hz. See figure~\ref{fig:low_freq_test} for an illustration of the time-dependent and time-independent response in the frequency domain.

Since $F_\oplus$ is much smaller than the minimum frequencies it is possible to detect with ground-based detectors ($\gtrsim 1$~Hz), one might consider approximating the shifts in frequency in equation~\eqref{eq:htilde_response} using a Taylor series of $\tilde{h}_{+.\times}$. Such an approximation was suggested in \cite{Chen:2020fzm} as being a potentially efficient way to do the computation, computing the derivatives by taking finite differences of a single waveform. It turns out that the specific form of the frequency-domain gravitational-wave signal from a compact binary, where the phase diverges rapidly for small frequencies (see, e.g.\ \cite{Buonanno:2009zt}), makes this approximation quite inaccurate if one applies it to the waveform directly. However, if one applies this Taylor series approximation to the phase and amplitude separately, without expanding the complex exponential, then one obtains a good approximation, giving the SPA result with corrections, as discussed below. %\nkjm{Reference Fig.~\ref{fig:low_freq_SPA_corrections} somewhere in here.}

We now must make a small correction to the results in~\cite{Chen:2020fzm}, in part from the correction to the Fourier transform convention mention above. In~\cite{Chen:2020fzm}, we considered the effect of the time-dependent response on the predicted SNR of some BNS signals as observed by the Einstein Telescope. Specifically, we compared the SNR computed using the time-independent and time-dependent responses, finding that using the time-dependent response reduces the total SNR from $5$~Hz by at most $1$ in all cases. We also checked the difference in SNR from $5$ to $10$~Hz and found that the largest change there was for the lowest-mass case considered, where using the time-dependent response reduces the SNR from $93.5$ to $91.8$. However, there were two mistakes in that computation. First, the Fourier convention we used in \cite{Chen:2020fzm} is opposite to the one used to compute the TaylorF2 waveform we used~\cite{Buonanno:2009zt}, which causes the change of the sign for the sine coefficient $a^A_{k\text{s}}$ in equation~(\ref{eq:htilde_response}). %The comparison with the SPA in our later section has confirmed that the application of Fourier transform in equation~(\ref{eq:fourier}) gives the correct coefficients in equation~(\ref{eq:htilde_response}).
Second, we had computed the SNR in the time-independent case using the IMRPhenomPv2\_NRTidal waveform~\cite{Dietrich:2017aum,Dietrich:2018uni}, but had used the TaylorF2 waveform to compute the SNR with the time-dependent response at low frequencies. However, while the phase of TaylorF2 has negligible differences compared to the IMRPhenomPv2\_NRTidal for such low frequencies, the amplitude does not, since TaylorF2 just uses the Newtonian approximation. We thus used the IMRPhenomD~\cite{Khan:2015jqa} amplitude implemented in the gwent package~\cite{Kaiser:2020tlg}, which is the same as the IMRPhenomPv2\_NRTidal amplitude prior to merger. With these corrections, we find that using the time-dependent response actually \emph{increases} the SNR in the $5$ to $10$~Hz band from $93.5$ to $95.0$ for the case for which this has the largest effect (an equal-mass non-spinning BNS with component masses of 1.35 $M_\odot$ at a luminosity distance of 369 Mpc). The total SNR of this system with the third-generation network from $5$~Hz increases by $\sim0.5$ when using the time-dependent response.

%---------------
\subsection{Comparison with the SPA and a simplified approximate expression}
\label{ssec:SPA_comp}
%---------------

We now compare our exact result with the SPA calculation used in, e.g.\ \cite{Zhao_and_Wen}, where one computes the effect of the time-dependent detector response in the frequency domain by evaluating the detector's response for each frequency at the time associated with a given frequency by the SPA. We start by illustrating how the SPA result can be obtained as an approximation to our result. For simplicity of exposition, we first consider only one of the terms in the Fourier series for the response. Specifically, we take $h_s(t) := \sin(\Omega_\oplus t)h(t)$, so the SPA gives a Fourier transform of $\tilde{h}^\text{SPA}_s(f) = \sin(\Omega_\oplus t_f)\tilde{h}^\text{SPA}(f)$. Here $t_f = \Psi'(f)/2\pi$ is the time associated with the frequency $f$ and $\Psi$ is the waveform's frequency-domain phase, i.e.\ $\tilde{h}^\text{SPA}(f) := A(f)e^{\ri\Psi(f)}$, where $A$ is the waveform's frequency-domain amplitude. Here we are taking the Fourier transform of the waveform to be computed using the SPA. The exact expression for the Fourier transform is
\<
\tilde{h}_s(f) = \frac{\tilde{h}(f + F_\oplus) - \tilde{h}(f - F_\oplus)}{2\ri}.
\?

We now evaluate this using $\tilde{h}^\text{SPA}$ and linearize the phase in $F_\oplus$, since we are interested in the case where $f \gg F_\oplus$. We also drop the correction to the amplitude from evaluating at $f \pm F_\oplus$ instead of $f$, so we have
\<
\begin{split}
\tilde{h}_s(f) &\simeq A(f)\frac{e^{\ri[\Psi(f) + \Psi'(f)F_\oplus]} - e^{\ri[\Psi(f) - \Psi'(f)F_\oplus]}}{2\ri}\\
&= \sin[\Psi'(f)F_\oplus]\tilde{h}^\text{SPA}(f)\\
&= \tilde{h}^\text{SPA}_s(f).
\end{split}
\?
The same sort of calculation applies to the case with the full response, where we recover the SPA expression in this approximation. The explicit expression is given below.

\begin{figure}[tb]
\centering
\includegraphics[width=0.48\textwidth]{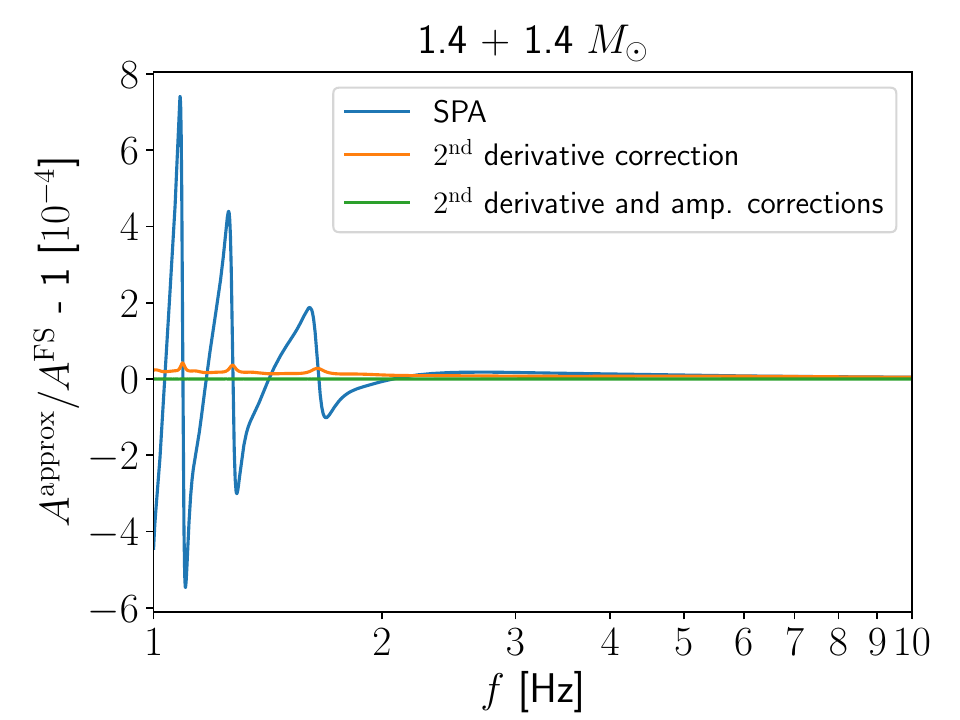}
\includegraphics[width=0.48\textwidth]{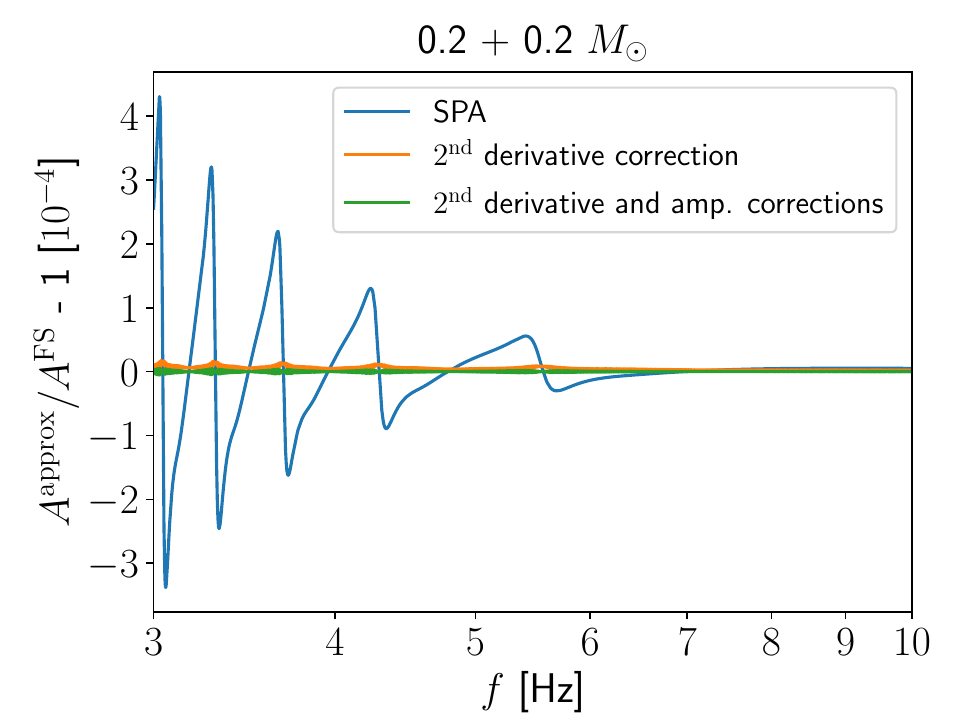}\\
\includegraphics[width=0.48\textwidth]{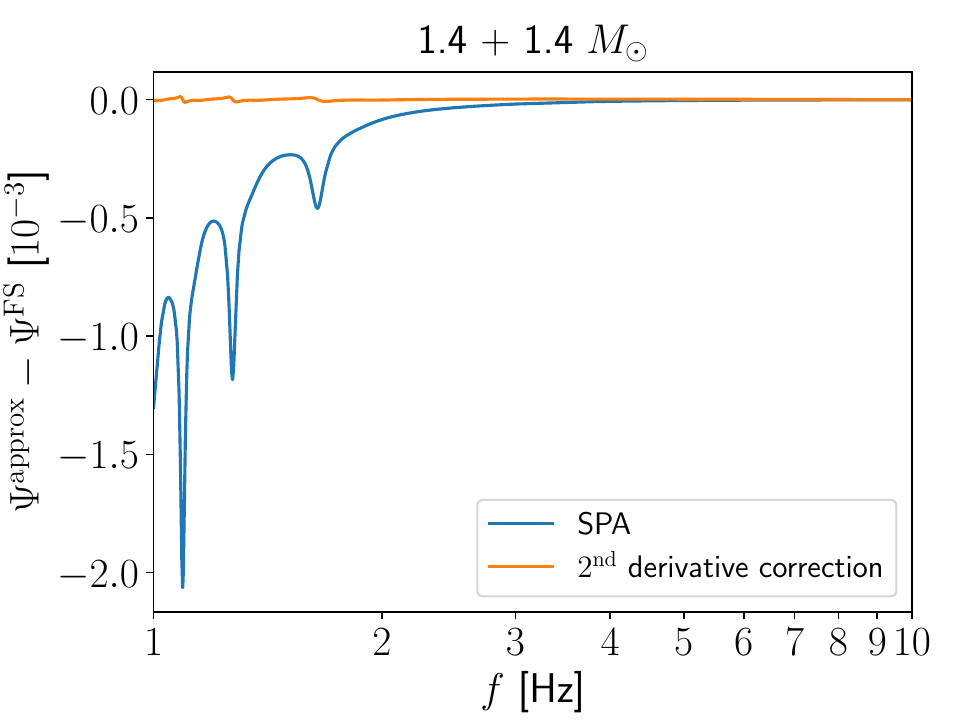}
\includegraphics[width=0.48\textwidth]{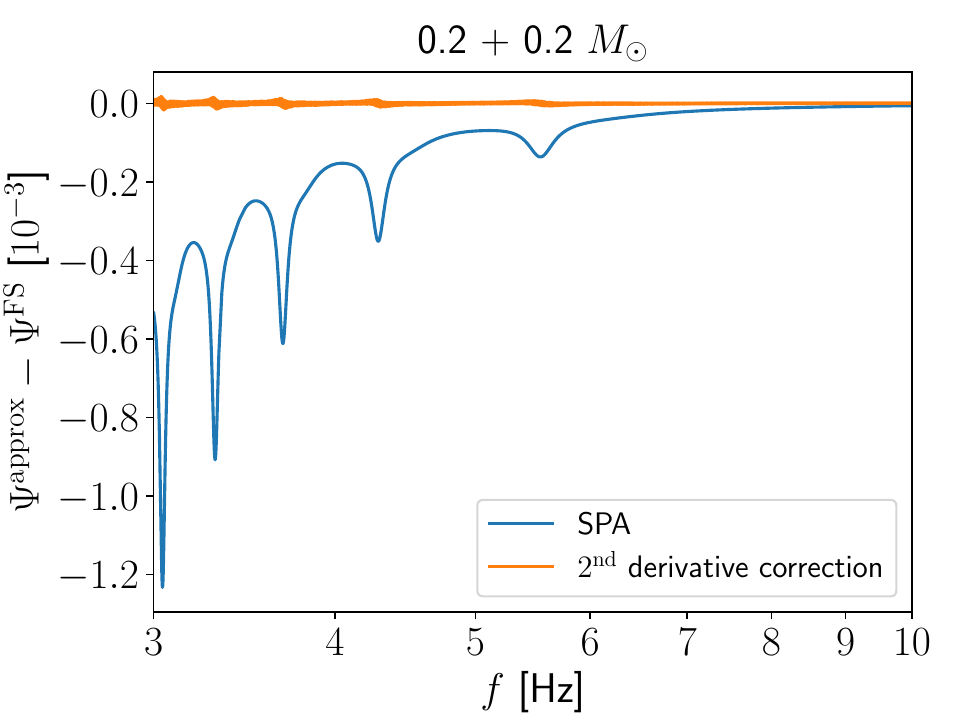}
\caption{The error of various approximations to the time-dependent response given by including some or all of the terms in equation~(\ref{eq:htilde_response_approx}), computed with respect to the exact Fourier series expression. The top panel shows the fractional difference in the amplitude, and the bottom panel shows the difference in the phase. The left panel shows the difference for the BNS case, and the right panel shows the difference for the subsolar-mass case; the parameters are given in the text.}
\label{fig:low_freq_SPA_corrections}
\end{figure}

We find that the error introduced by using the SPA is small, but increases as one lowers the frequency, as one would expect---see figure~\ref{fig:low_freq_SPA_corrections}. The size of the maximum fractional error in the amplitude for the $1.4+1.4 M_\odot$ system down to $1$~Hz is $\sim 8\times 10^{-4}$, which is almost two orders of magnitude larger than one might expect from the na{\"\i}ve scaling of $F_\oplus/f$, which is only $\sim 10^{-5}$ for $f = 1$~Hz. The maximum fractional error in the phase is even larger in this case, $\sim 2\times 10^{-3}$~rad. %\nkjm{This is presumably due to the divergence of the (frequency-domain) amplitude and phase of the signal as $f$ goes to zero, but how exactly does this come about? Is it just from the shifts in minima of the response? To check this, it might be worth plotting the responses together, at least for our own checks---we can include such a plot in the paper if it's illuminating.}
These errors are still likely smaller than the accuracy necessary for observational applications. However, we show how one can improve on the SPA result by using higher derivatives of the phase and also including the amplitude corrections. These derivatives could be computed by finite differencing of the waveform on a sufficiently fine mesh, instead of having to evaluate the waveform at five different frequencies for each frequency for which one wishes to compute the response. Specifically, we expand to second order in the phase and first order in the amplitude, giving the approximation
\<\label{eq:htilde_response_approx}
\tilde{h}(f) \simeq \sum_{\polidx\in\{+,\times\}}\Biggl(a^\polidx_0 + \sum_{k\in\{1,2\}}e^{\ri\Psi''_\polidx(f)k^2F_\oplus^2/2}\biggl\{\cA^\polidx_{k\mathrm{c}}(f)\cos[\Psi'_\polidx(f)kF_\oplus ] + \cA^\polidx_{k\text{s}}(f)\sin[\Psi'_\polidx(f)kF_\oplus]\biggr\}\Biggr)\tilde{h}_\polidx(f)
\?
where $h_\polidx(f) = A_\polidx(f)e^{\ri\Psi_\polidx(f)}$ and
\begin{subequations}
\begin{align}
\cA^\polidx_{k\mathrm{c}}(f) &:= a^\polidx_{k\text{c}} - \ri k F_\oplus a^\polidx_{k\text{s}}A'_\polidx(f)/A_\polidx(f),\\
\cA^\polidx_{k\mathrm{s}}(f) &:= a^\polidx_{k\text{s}} + \ri k F_\oplus a^\polidx_{k\text{c}}A'_\polidx(f)/A_\polidx(f).
\end{align}
\end{subequations}
One obtains the SPA result by only including the first derivative terms and neglecting the corrections to the amplitude [so $\cA^\polidx_{k\mathrm{c,s}}(f) \to a^\polidx_{k\mathrm{c,s}}$]. Explicitly, one has
\<\label{eq:SPA_FS}
\tilde{h}^\text{SPA}(f) = \sum_{\polidx\in\{+,\times\}}\Biggl(a^\polidx_0 + \sum_{k\in\{1,2\}}\biggl\{a^\polidx_{k\mathrm{c}}\cos[\Psi'_\polidx(f)kF_\oplus] + a^\polidx_{k\text{s}}\sin[\Psi'_\polidx(f)kF_\oplus]\biggr\}\Biggr)\tilde{h}_\polidx(f).
\?
Thus, one can obtain the SPA result using this expression with only five calls to the antenna pattern function for each sky location and polarization angle (to compute the Fourier coefficients), as opposed to having to call it for each frequency at which one evaluates the waveform. This gives a significant speedup in our tests (described below), and would likely give a speedup even in cases like the one considered in~\cite{Smith:2021bqc}, where one is applying the SPA expression to a waveform evaluated a relatively small set of frequency nodes ($\sim 500$) using reduced order modelling. In fact, one can even write the dependence of the Fourier coefficients on the sky location and polarization angle explicitly, to avoid needing to call the antenna pattern function when changing these quantities. We give the explicit expressions (based on the expressions in~\cite{Jaranowski:1998qm}) in \ref{app:resp_coeff_dep}.

In figure~\ref{fig:low_freq_SPA_corrections} we show the effect of the corrections to the SPA given by equation~(\ref{eq:htilde_response_approx}) for both a standard BNS and a subsolar-mass binary, both as observed by ET. Here we consider the TaylorF2 waveform for an equal-mass non-spnning system with source-frame component masses of $1.4~M_\odot$ (BNS) or $0.2M_\odot$ (subsolar mass). In both cases, we place the binary at a distance of $400$~Mpc, setting the inclination angle to $30^\circ$, the polarization angle to $45.3^\circ$, and the sky location to $\theta=30.41^{\circ}, \phi=201.85^{\circ}$ in the Earth fixed frame from~\cite{Anderson:2000yy}, corresponding to right ascension of $257.26^\circ$ and declination of $59.59^\circ$. The cosmological redshift of the system is $z \simeq 0.085$, using the Planck~2018 TT,TE,EE+lowE+lensing+BAO parameters from table~2 in~\cite{Aghanim:2018eyx}. We use the Virgo location for ET, since the location for ET is not yet known, and use the orientation of the `E1' detector given in~\cite{LALDetectors}, taking the GPS time to be $1152346754$~s. The lowest frequency considered in the BNS case is $1$~Hz, which is the lowest possible frequency for the ET sensitivity curve. In the subsolar-mass case, we only consider frequencies down to $3$~Hz, since the computation is very expensive for lower frequencies (and ET is not likely to have sensitivity all the way down to $1$~Hz). We find that the contribution from the second derivative of the phase is the most important correction, while it is only necessary to include the amplitude correction for the systems considered if one requires fractional accuracies as small as $\sim 10^{-7}$. %We also find that the third derivative has very little correction effect.

We now consider the accuracy of the computation of the derivatives of the phase in equation~(\ref{eq:htilde_response_approx}) by finite differencing the waveform. While it is easy to compute the derivatives of the TaylorF2 waveform analytically, this is not the fastest method, as we will see later, and it may be simpler to compute derivatives numerically for some waveform models. We first consider direct centred second order finite differencing. To increase the accuracy of the computation for low frequencies, we use multiple frequency bands, where each band has different frequency spacing. The frequency spacing is given by the largest negative power of 2 that is smaller than $1/t(f)$ as in the multibanding implementation in~\cite{Vinciguerra:2017ngf} \footnote{See~\cite{Garcia-Quiros:2020qlt} for an updated implementation and~\cite{Morisaki:2021ngj} for an improved method.}. Here $t(f)$ is the TaylorT2 time before coalescence corresponding to the lower frequency $f$ of each band given in equation~(3.8b) of~\cite{Buonanno:2009zt}. The specifics of the frequency bands we use are shown in Table~\ref{tab:freq_band}. We use a frequency spacing of $10^{-6}$ Hz in the finite differencing.

\begin{figure}[tb]
\centering
\includegraphics[width=0.32\textwidth]{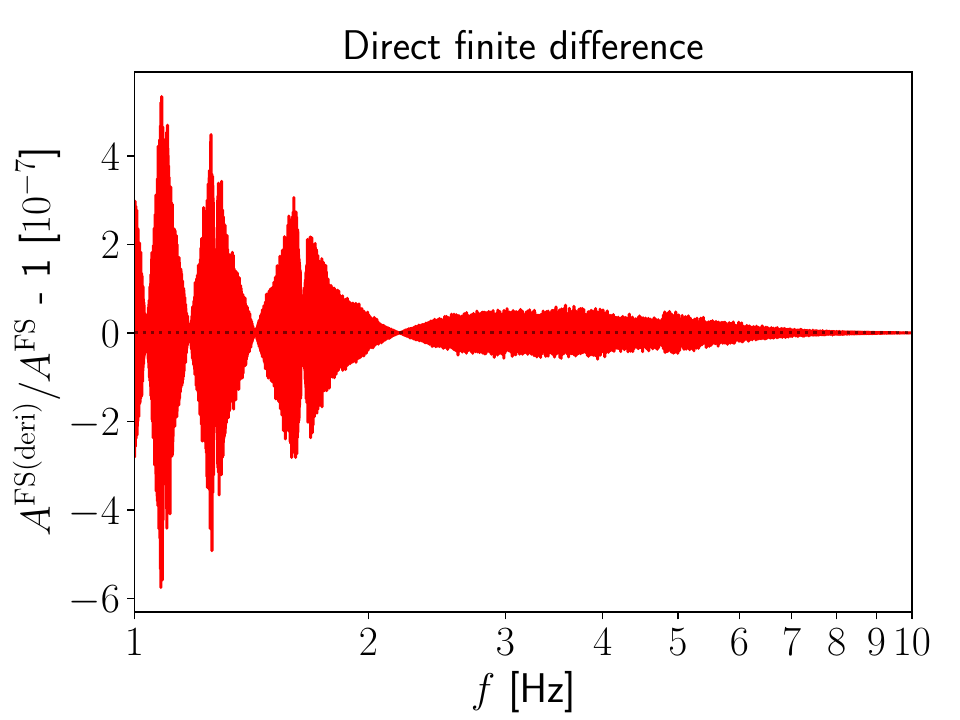}
\includegraphics[width=0.32\textwidth]{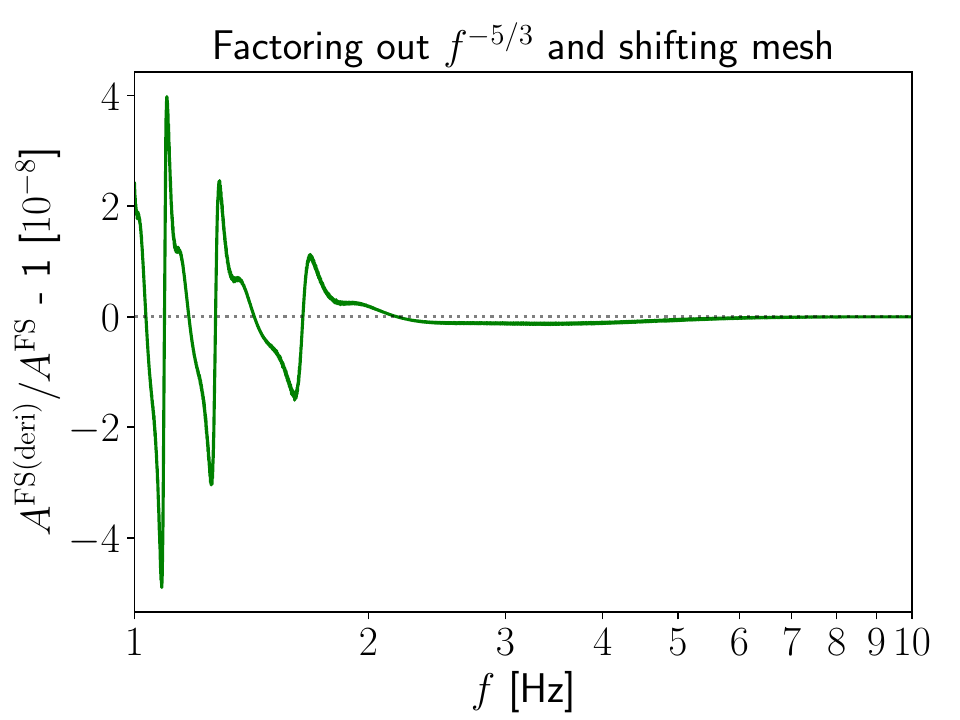}
\includegraphics[width=0.32\textwidth]{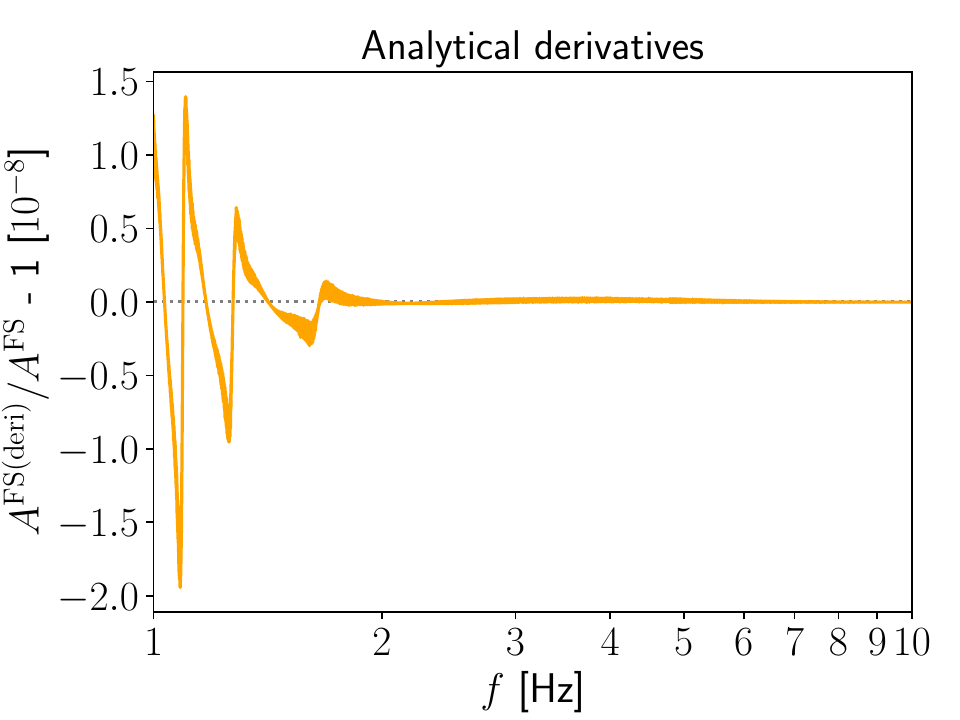}
\caption{The error of the amplitude of the time-dependent response computed using the approximation in equation~(\ref{eq:htilde_response_approx}) with different methods of differentiation of the phase, compared to the exact Fourier series method. Note the differences in the vertical scales of the three plots, in particular the order of magnitude larger scale for the left-hand plot. The waveform used is the same as that for the BNS case in figure~\ref{fig:low_freq_SPA_corrections}. The direct finite differencing and analytical derivative use the multiband frequency mesh in table~\ref{tab:freq_band}, while the method of factoring out $f^{-5/3}$ and shifting mesh uses a uniform frequency mesh with $\delta f=10^{-14}$ Hz.}
\label{fig:low_freq_ratio_deri}
\end{figure}

Due to the use of multibanding, we evaluate $\tilde{h}(f+\delta f)$ separately from $\tilde{h}(f)$; we call this the direct finite difference approach. However, one can speed up the finite difference approach by using the spacing of the frequency mesh as the finite difference $\delta f$, so one can compute the derivatives with only one evaluation of the waveform. Since the value of $\tilde{h}(f+\delta f)$ is just the value of $\tilde{h}(f)$ with a shift of $\delta f$ on the frequency mesh, we call this method ``finite difference with shifting mesh". Here, for simplicity, we apply this to a uniform frequency mesh with a spacing of $2^{-14}$~Hz, though it would be possible to apply in the multibanding case with some additional care taken at the boundaries between the bands. We increase the accuracy of this method by factoring out the leading $f^{-5/3}$ divergence of the phase and computing its derivatives analytically, while finite differencing the remainder with shifting mesh. We show the resulting errors in amplitudes in figure~\ref{fig:low_freq_ratio_deri}. We have also checked that the errors in phases follow the same pattern as those in amplitudes. As expected, the analytical derivatives are the most accurate, and factoring out the divergence improves the accuracy of the finite differencing, but the basic finite differencing method is already quite accurate for the multiband frequency mesh we consider in table~\ref{tab:freq_band}.

\begin{table}[t]
  \centering
  \caption{The lower frequency $f_\text{low}$ and frequency spacing $\delta f$ for each band in the simple multibanding implementation used in our timing test.}
  \begin{tabular}{cccccccccccc}        
    \hline 
   	$f_\text{low}$ [Hz] & $1$ & $2$ & $3$ & $4$ & $5$ & $7$ & $10$ & $20$ & $30$ & $40$ & $60$ \\ 
	\hline
	$\delta f$ [Hz] & $2^{-19}$ & $2^{-17}$ & $2^{-15}$ & $2^{-14}$ & $2^{-13}$ & $2^{-12}$ & $2^{-11}$ & $2^{-8}$ & $2^{-6}$ & $2^{-5}$ & $2^{-4}$ \\
    \hline 
  \end{tabular}
  \label{tab:freq_band}
\end{table}

%\begin{figure}[tb]
%\centering
%\includegraphics[width=0.48\textwidth]{fig/SPA_corrections_subsolar.pdf}
%\includegraphics[width=0.48\textwidth]{fig/SPA_corrections_phase_diff_subsolar.pdf}
%\caption{The differences of higher-order approximations from the Fourier series as in figure~\ref{fig:low_freq_SPA_corrections}, but in subsolar masses. The constructed waveform uses the same parameters as in figure~\ref{fig:low_freq_SPA_corrections}, except that the masses are $0.2+0.2~M_\odot$. }
%\label{fig:low_freq_SPA_corrections_subsolar}
%\end{figure}

%---------------
\subsection{Computational speed}
%---------------

We compare the computational speed of the time-dependent response with the Fourier series method given in this work to the SPA method and the approximations given in equation~\eqref{eq:htilde_response_approx}. We use the same BNS system and detector orientation as in figure~\ref{fig:low_freq_SPA_corrections}. We use our own implementation of the TaylorF2 waveform (to 3.5PN order in the phase and Newtonian order in the amplitude, as in~\cite{Buonanno:2009zt}). We time the computation of the time-dependent response over the frequency range of $1$--$100$~Hz and quote the average speed of 100 runs of the code. We chose the higher frequency cutoff of $100$ Hz because the effect of the time-dependent response is negligible above $100$ Hz, as illustrated in section~\ref{sec:information}.

We run the computation on a 12th Gen Intel Core i7-1260P processor. We find that the computation time is $ 0.7$~s for the Fourier series method given in equation~\eqref{eq:htilde_response}, and $11.1$~s for the SPA method [implemented directly, not using the Fourier series expression in equation~\eqref{eq:SPA_FS}], both on the multibanding mesh. If we perform the same computation over a frequency range of $5$--$100$~Hz with multibanding, the time spent is $0.04$~s for the Fourier series method, and $0.7$~s for the SPA method. This shows that the Fourier series method is faster than the SPA method by about an order of magnitude. However, this implementation of the SPA method is likely overly slow. For instance, the implementation in Baral~\emph{et al}~\cite{Baral:2023xst} only evaluates the antenna pattern functions every $4$~s and interpolates. We compare the computation time between the Fourier series method and the implementation by Baral~\emph{et al} over the multiband frequency array covering $5$--$2048$~Hz (the range used in \cite{Baral:2023xst}). While the implementation by Baral~\emph{et al} spends $\sim 0.07$~s, our method spends $\sim 0.05$~s, a $\sim 30\%$ improvement.

In addition, we study the speed when using different differentiation methods in the approximate expression in equation~(\ref{eq:htilde_response_approx}), obtaining the results given in table~\ref{tab:derivative_speed}. We find that using analytical derivatives is the most costly, due to the lengthy expressions. With the multibanding frequency mesh, using analytic derivatives is more than twice as slow as direct finite differencing. As expected, the direct finite difference approach is slower when using this finer mesh compared to the multibanding mesh. However, computing the derivatives just using the values on the mesh increases the speed by a factor of two, as expected. Factoring out the divergent factor of $f^{-5/3}$ in the phase and computing its derivative analytically while finite differencing the remainder with shifting mesh to improve the accuracy (see figure~\ref{fig:low_freq_ratio_deri}) is only slightly slower. 

\begin{table}[t]
  \centering
  \caption{The evaluation time of the approximate expressions for the time-dependent response in equations~\eqref{eq:SPA_FS} (the first-order expression equivalent to the SPA) and~\eqref{eq:htilde_response_approx} (the second-order expression) for different differentiation methods, evaluated from $1$ to $100$~Hz.}
  \begin{tabular}{c|c|c|c}        
    \hline 
   	Approx. order & Frequency mesh & Method & Time [s] \\ 
	\hline
	\multirow{5}{*}{1$^{\rm st}$ \eqref{eq:SPA_FS}} & \multirow{2}{*}{Multiband} & Analytic & $0.4$ \\
	 & & Direct finite difference & $0.4$ \\
	 \cline{2-4}
	 & \multirow{3}{*}{Uniform} & Direct finite difference & $0.9$ \\
	 & & Shifting mesh & $0.4$ \\
	 & & Factoring out $f^{-5/3}$ and shifting mesh & $0.5$ \\
	 \hline
	\multirow{5}{*}{2$^{\rm nd}$ \eqref{eq:htilde_response_approx}} & \multirow{2}{*}{Multiband} & Analytic & $0.8$ \\
	 & & Direct finite difference & $0.5$\\
	\cline{2-4}
	 & \multirow{3}{*}{Uniform} & Direct finite difference & $1.1$ \\
	 & & Shifting mesh & $0.7$ \\
	 & & Factoring out $f^{-5/3}$ and shifting mesh & $0.8$ \\
    \hline 
  \end{tabular}
  \label{tab:derivative_speed}
\end{table}

%%%%%%%%%%%%%%%
\section{Computing the effects of the time-dependent Doppler shift from the Earth's rotation and orbital motion}
\label{sec:Doppler}
%%%%%%%%%%%%%%%

%---------------
\subsection{Computation without the SPA just including the Earth's rotation}
%---------------

The Doppler shift is due to the change in the retarded time at which the waveform is observed due to the detector's motion. This modulates the waveform $h^\text{SSB}_{+,\times}(t)$ that would be detected at the solar system barycentre (SSB) to $h^\text{det}_{+,\times}(t) = h^\text{SSB}_{+,\times}(t - \hat{n}\cdot\vec{x}(t))$, where $\hat{n}$ denotes the direction of the source and $\vec{x}(t)$ denotes the position of the detector relative to the SSB. If we neglect the Earth's orbital motion here (i.e.\ take the Earth to be located at the SSB), for simplicity of exposition, we can write $\hat{n}\cdot\vec{x}(t)$ as a Fourier series with period $T_\oplus$, which only contains terms with the same period as the Earth's rotation, in the constant rotational angular velocity approximation:
\<
\hat{n}\cdot\vec{x}(t) = \hat{n}\cdot\vec{x}(0) + b_{1\text{c}}[\cos(\Omega_\oplus t) - 1] +  b_{1\text{s}}\sin(\Omega_\oplus t).
\?

In order to obtain an approximation that is well behaved at high frequencies (where the time variation of the Doppler shift is negligible for compact binary signals), we want to expand around the waveform shifted by the retarded time at merger, which we take to occur at $t = 0$. We thus define $\bar{t} := t - \hat{n}\cdot\vec{x}(0)$.
We now compute the Fourier transform by expanding $h^\text{SSB}_{+,\times}(t)$ as a Taylor series about $\bar{t}$, giving (dropping the polarization subscripts for ease of notation here and henceforth)
\<
\label{eq:Doppler_Taylor}
\begin{split}
h^\text{det}(t) &= h^\text{SSB}(\bar{t}) - \{b_{1\text{c}}[\cos(\Omega_\oplus t) - 1] +  b_{1\text{s}}\sin(\Omega_\oplus t)\}\dot{h}^\text{SSB}(\bar{t})\\
&\quad + \frac{1}{2}\{b_{1\text{c}}[\cos(\Omega_\oplus t) - 1] +  b_{1\text{s}}\sin(\Omega_\oplus t)\}^2\ddot{h}^\text{SSB}(\bar{t}) + \cdots.
\end{split}
\?
Thus, defining $\tilde{h}^\text{SSB}_\text{shift}(f) := e^{2\pi\ri f\hat{n}\cdot\vec{x}(0)}\tilde{h}^\text{SSB}(f)$ as the Fourier transform of $h^\text{SSB}(\bar{t})$ with respect to $t$, i.e.\ the waveform with the time shift associated with the location of the detector at coalescence time, we have
\<
\label{eq:Doppler_Taylor_eval}
\begin{split}
\tilde{h}^\text{det}(f) &= [1 - 2\pi\ri b_{1\text{c}} f - \pi^2(3b_{1\text{c}}^2 + b_{1\text{s}}^2)f^2]\tilde{h}^\text{SSB}_\text{shift}(f)\\
&\quad + \pi\{\ri b_{1\text{c}} f + b_{1\text{s}}F_\oplus + 2\pi b_{1\text{c}} [b_{1\text{c}}(f^2 + F_\oplus^2) - 2\ri b_{1\text{s}}F_\oplus f]\}[\tilde{h}^\text{SSB}_\text{shift}(f + F_\oplus) + \tilde{h}^\text{SSB}_\text{shift}(f - F_\oplus)]\\
&\quad + \pi\{b_{1\text{s}} f + \ri b_{1\text{c}}F_\oplus - 2\pi b_{1\text{c}} [\ri b_{1\text{s}}(f^2 + F_\oplus^2) - 2b_{1\text{c}}F_\oplus f]\}[\tilde{h}^\text{SSB}_\text{shift}(f + F_\oplus) - \tilde{h}^\text{SSB}_\text{shift}(f - F_\oplus)]\\
&\quad  - \pi^2\left[(b_{1\text{c}}^2 - b_{1\text{s}}^2)\left(\frac{f^2}{2} + 2F_\oplus^2\right) - 4\ri b_{1\text{c}}b_{1\text{s}} F_\oplus f\right][\tilde{h}^\text{SSB}_\text{shift}(f + 2F_\oplus) + \tilde{h}^\text{SSB}_\text{shift}(f - 2F_\oplus)]\\
&\quad  + \pi^2 [\ri b_{1\text{c}}b_{1\text{s}}(f^2 + 4F_\oplus^2) - 2(b_{1\text{c}}^2 - b_{1\text{s}}^2)F_\oplus f][\tilde{h}^\text{SSB}_\text{shift}(f + 2F_\oplus) - \tilde{h}^\text{SSB}_\text{shift}(f - 2F_\oplus)] + \cdots\\
&\simeq [1 - 2\pi\ri b_{1\text{c}} f - \pi^2(3b_{1\text{c}}^2 + b_{1\text{s}}^2)f^2]\tilde{h}^\text{SSB}_\text{shift}(f)\\
&\quad + (\pi\ri + 2\pi^2 b_{1\text{c}} f)b_{1\text{c}} f[\tilde{h}^\text{SSB}_\text{shift}(f + F_\oplus) + \tilde{h}^\text{SSB}_\text{shift}(f - F_\oplus)]\\
&\quad + (\pi - 2\pi^2\ri b_{1\text{c}} f)b_{1\text{s}} f [\tilde{h}^\text{SSB}_\text{shift}(f + F_\oplus) - \tilde{h}^\text{SSB}_\text{shift}(f - F_\oplus)]\\
&\quad  - \frac{\pi^2}{2}(b_{1\text{c}}^2 - b_{1\text{s}}^2)f^2[\tilde{h}^\text{SSB}_\text{shift}(f + 2F_\oplus) + \tilde{h}^\text{SSB}_\text{shift}(f - 2F_\oplus)]\\
&\quad  + \pi^2\ri b_{1\text{c}}b_{1\text{s}}f^2[\tilde{h}^\text{SSB}_\text{shift}(f + 2F_\oplus) - \tilde{h}^\text{SSB}_\text{shift}(f - 2F_\oplus)].
\end{split}
\?
Here we have noted that we can integrate by parts freely when computing the Fourier transform, since we assume that the GW signal goes to zero at early and late times, as is the case for the compact binary coalescence signals in which we are interested. We have obtained the simpler approximate expression by recalling that $f \gg F_\oplus$, so we can neglect all the $F_\oplus$ terms that occur out of the argument of the GW signal (with its fast oscillations) to a good approximation (as well as neglecting the higher-order Taylor series terms denoted by $\cdots$ in the first expression). This is equivalent to neglecting the derivatives acting on the terms that vary at the Earth's rotational frequency when integrating by parts. %\nkjm{Can one obtain something useful by doing the full series this way?}

We now write $\tilde{h}^\text{SSB}_\text{shift}(f) = A(f)e^{\ri\Psi(f)}$ and expand to leading order in the phase, dropping the amplitude corrections, for a simple approximate expression, obtaining
\<
\label{eq:Doppler_Taylor_exp}
\begin{split}
\tilde{h}^\text{det}(f) &\simeq %\Big\{1 + (2\pi\ri f + 4\pi^2 b_{1\text{c}} f^2)\big(b_{1\text{c}}\{\cos[\psi'(f)F_\oplus] - 1\} + b_{1\text{s}}\sin[\psi'(f)F_\oplus]\big)\\
%&\quad - \pi^2(b_{1\text{c}}^2 - b_{1\text{s}}^2)f^2 \{\cos[2\psi'(f)F_\oplus] - 1\} - 2\pi^2 b_{1\text{c}}b_{1\text{s}}f^2 \sin[2\psi'(f)F_\oplus] \Bigr\}\tilde{h}^\text{SSB}_\text{shift}(f)\\
%&=
\Big\{1 + 2\pi\ri f\big(b_{1\text{c}}\{\cos[\Psi'(f)F_\oplus] - 1\} + b_{1\text{s}}\sin[\Psi'(f)F_\oplus]\big)\\
&\quad - 2\pi^2 f^2\big(b_{1\text{c}}\{\cos[\Psi'(f)F_\oplus] - 1\} + b_{1\text{s}}\sin[\Psi'(f)F_\oplus]\big)^2 \Bigr\}\tilde{h}^\text{SSB}_\text{shift}(f),%\\
%&= \biggl(1 + 2\pi\ri b_{1\text{s}} f \psi'(f)F_\oplus - \pi (\ri b_{1\text{c}}f + 2 \pi b_{1\text{s}}^2 f^2)[\psi'(f)F_\oplus]^2 + O\big\{[\psi'(f)F_\oplus]^3\big\}\biggr)\tilde{h}^\text{SSB}_\text{shift}(f).
\end{split}
\?
where the part in the large curly braces goes to unity for high frequencies (or $F_\oplus\to 0$), as desired. %We also give the expansion for small $\psi'(f)F_\oplus$ to show explicitly how well behaved it is, with various cancellations, so there are, e.g., no terms of the form $f^2\psi'(f)F_\oplus$.
[Recall that $\Psi'(f) \propto f^{-8/3}$ in the Newtonian approximation.]
We also see from the second expression that this approximation reproduces the SPA result from~\cite{Zhao_and_Wen} expanded to second order in $b_{1\text{c,s}}f$ \footnote{This application of the SPA in~\cite{Zhao_and_Wen} evaluates the time delay due to the motion of the detector using the time as a function of frequency from the SPA applied to the signal with no detector motion. It would be interesting to see if one recovers some of the amplitude corrections we have derived if one applies the SPA directly to the time-domain signal modulated by the detector's motion.}. The unexpanded SPA result from~\cite{Zhao_and_Wen} [cf.\ their equation~(7), noting that their $\mathbf{n}$ is the negative of our $\hat{n}$] is
\<
\tilde{h}^\text{det}_\text{SPA}(f) = e^{2\pi\ri f\left(b_{1\text{c}}\{\cos[\Psi'(f)F_\oplus] - 1\} + b_{1\text{s}}\sin[\Psi'(f)F_\oplus]\right)}\tilde{h}^\text{SSB}_\text{shift}(f).
\?
It would be interesting to see if it is possible to write the corrections to the SPA expression in such a simple form (possibly by changing variables to the retarded time and using the Fourier series representation of the solution to the generalized Kepler equation one obtains or using the techniques from~\cite{Marsat:2018oam}), but since these are likely not relevant in practice for the applications considered here, we do not pursue this at present. Indeed, we find quite small differences comparing the full corrections given in equation~\eqref{eq:Doppler_Taylor_eval} with the expanded version in equation~\eqref{eq:Doppler_Taylor_exp}, and considering either the Earth's rotation or the leading (circular) Fourier coefficients for the orbital motion of the Earth-Moon barycentre considered in the next subsection (here $F_\oplus$ is replaced by the inverse of a sidereal year). Specifically, we consider the cases (and frequency ranges) shown in figure~\ref{fig:low_freq_SPA_corrections} for a selection of sky locations, and find a maximum fractional difference of the amplitude of a few times $10^{-4}$ and a maximum phase difference of $\sim 10^{-4}$. For the case of the Earth-Moon barycentre motion, we used $\tilde{h}(f + F) + \tilde{h}(f - F) \simeq 2\tilde{h}(f)\cos[\Psi'(f)F]$ and $\tilde{h}(f + F) - \tilde{h}(f - F) \simeq \tilde{h}(f)\left\{2\ri\sin[\Psi'(f)F] - (7/3)F/f\right\}$ for $F \ll f$ to evaluate the finite differences numerically to sufficient accuracy.

%---------------
\subsection{Including both the Earth's rotation and its orbital motion}
%---------------

We now include the Earth's orbital motion in the computation of the Doppler shift. Since we are only interested in accounting for this effect for signals that last less than a month in the sensitive band of the detector, we can approximate the Earth's orbital motion to good accuracy by using low-order Fourier series expressions for the motion of the Earth-Moon barycentre (EMB) about the SSB and the motion of the Earth's centre of mass about the EMB. Specifically, we write the position of the detector with respect to the SSB as
\<\label{eq:x_decomp}
\vec{x}(t) = \vec{x}_{\text{SSB}\to\text{EMB}}(t) + \vec{x}_{\text{EMB}\to\oplus}(t) + \vec{x}_{\oplus\to\text{det}}(t),
\?
where $\vec{x}_{A\to B}(t)$ gives the position of $B$ with respect to $A$. We then specialize to the projection of the position along the direction to the source and write
\begin{subequations}
\label{eq:xdn_1st}
\begin{align}
\hat{n}\cdot\vec{x}_{\text{SSB}\to\text{EMB}}(t) &\simeq \hat{n}\cdot\vec{x}_{\text{SSB}\to\text{EMB}}(t_\text{mid}) + b_{1\text{c}}^\text{EMB}\{\cos[\Omega_\text{EMB}(t - t_\text{mid})] - 1\} + b_{1\text{s}}^\text{EMB}\sin[\Omega_\text{EMB}(t - t_\text{mid})],\\
\hat{n}\cdot\vec{x}_{\text{EMB}\to\oplus}(t) &\simeq \hat{n}\cdot\vec{x}_{\text{EMB}\to\oplus}(t_\text{mid}) + b_{1\text{c}}^\text{EM}\{\cos[\Omega_\text{EM}(t - t_\text{mid})] - 1\} + b_{1\text{s}}^\text{EM}\sin[\Omega_\text{EM}(t - t_\text{mid})],\\
\hat{n}\cdot\vec{x}_{\oplus\to\text{det}}(t) &\simeq \hat{n}\cdot\vec{x}_{\oplus\to\text{det}}(t_\text{mid}) + b_{1\text{c}}\{\cos[\Omega_\oplus(t - t_\text{mid})] - 1\} + b_{1\text{s}}\sin[\Omega_\oplus(t - t_\text{mid})].
\end{align}
\end{subequations}
Here we take $\Omega_\text{EMB}$ to be the Earth's mean motion, using the value of the sidereal year of $31558149.7635456$~s used by LALSuite (from~\cite{1994A&A...282..663S}), and similarly take $\Omega_\text{EM}$ to be given by the mean motion of the Moon, obtaining a value of $2\pi/(27.32166155\cdot86400)\text{ rad}/\text{s}$ from~\cite{2002A&A...387..700C}.
We obtain the Fourier coefficients from an Astropy~\cite{astropy1,astropy2,astropy3} ephemeris (accessed through \verb,get_body_barycentric, for the orbital motion and \verb,get_gcrs_posvel, for the detector's location) using a collocation method. Specifically, if $t_\text{mid} \pm \Delta t$ gives the start and end of the segment whose data we want to model, we use
\begin{subequations}
\begin{align}
d^{\text{EMB}, \pm} &:= \hat{n}\cdot\vec{x}_{\text{SSB}\to\text{EMB}}(t_\text{mid} \pm \Delta t) - \hat{n}\cdot\vec{x}_{\text{SSB}\to\text{EMB}}(t_\text{mid}),\\
d^{\text{EM}, \pm} &:= \hat{n}\cdot\vec{x}_{\text{EMB}\to\oplus}(t_\text{mid} \pm \Delta t) - \hat{n}\cdot\vec{x}_{\text{EMB}\to\oplus}(t_\text{mid}),\\
d^{\text{det}, k} &:= \hat{n}\cdot\vec{x}_{\oplus\to\text{det}}(t_\text{mid} + kT_\oplus/4) - \hat{n}\cdot\vec{x}_{\oplus\to\text{det}}(t_\text{mid}),
\end{align}
\end{subequations}
where $k\in\{1,2\}$. Our approximation for the orbital motion will thus be exact at the endpoints and midpoint of the time interval considered.
We then have (neglecting the uncontrolled remainders)
\begin{subequations}
\begin{align}
b_{1\text{c}}^\bullet &= -\frac{d^{\bullet, +} + d^{\bullet, -}}{2 - 2\cos(\Omega_\bullet\Delta t)},\\
b_{1\text{s}}^\bullet &= \frac{d^{\bullet, +} - d^{\bullet, -}}{2\sin(\Omega_\bullet\Delta t)},\\
b_{1\text{c}} &= -\frac{d^{\text{det}, 2}}{2},\\
b_{1\text{s}} &= d^{\text{det}, 1 } - \frac{d^{\text{det}, 2}}{2},
\end{align}
\label{eq:b_coef}
\end{subequations}
where $\bullet$ is either EMB or EM.
Since the expressions for the orbital Fourier coefficients have small denominators, we find that in the case of the Earth's orbit around the EMB, we have to impose a minimum value of $\Delta t$ used in this calculation to be a fifth of a sidereal day in order to prevent roundoff error from leading to inaccurate results.
We give expressions to second order in the orbital frequencies for increased accuracy in \ref{app:Doppler_2nd_order}.
The sky location dependence of each of the Fourier coefficients can be expressed in three Cartesian coordinates, since the dependence just comes from $\hat{n}$, e.g.\ $b_{1\text{c}} = b_{1\text{c}}^x\cos\delta\cos\alpha + b_{1\text{c}}^y\cos\delta\sin\alpha + b_{1\text{c}}^z\sin\delta$, where $\alpha$ and $\delta$ are the source's right ascension and declination, respectively.

\begin{figure}[H]
\centering
\includegraphics[width=0.48\textwidth]{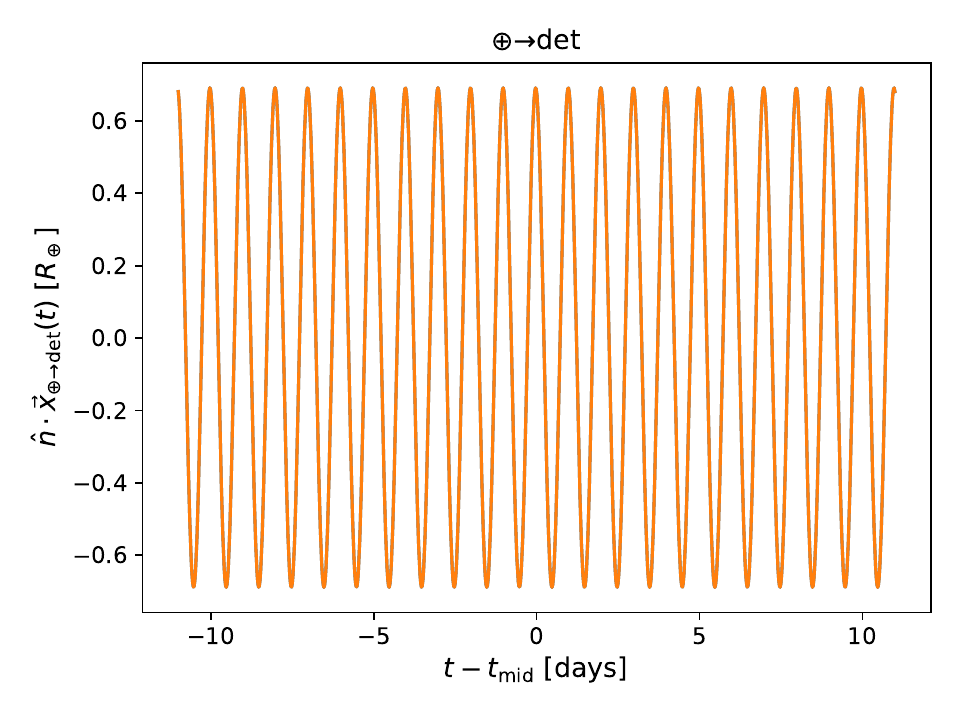}
\includegraphics[width=0.48\textwidth]{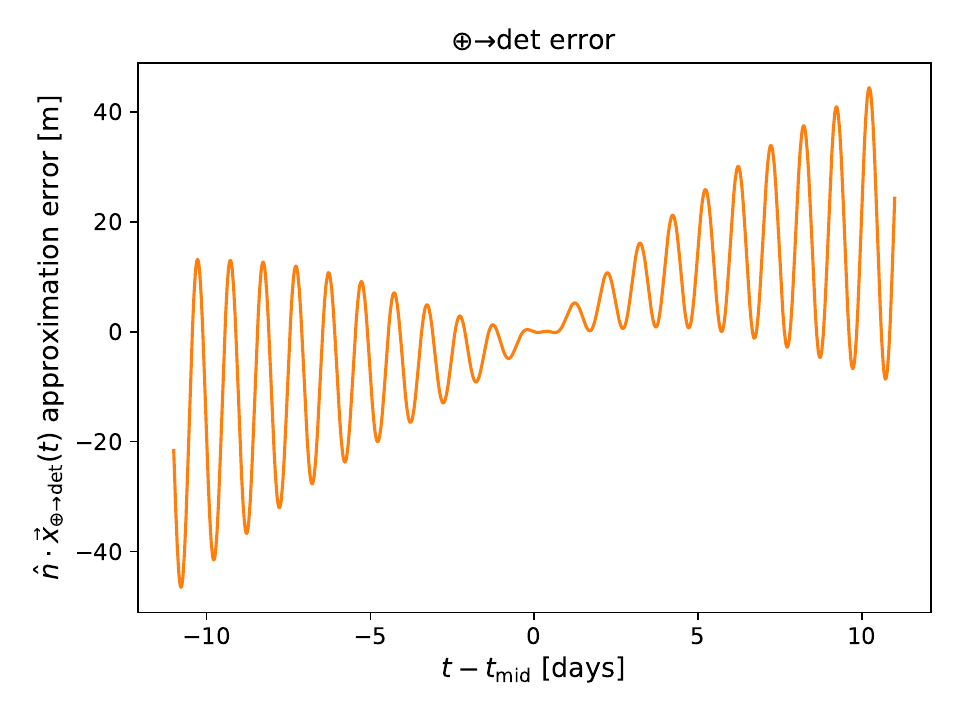}\\
\includegraphics[width=0.48\textwidth]{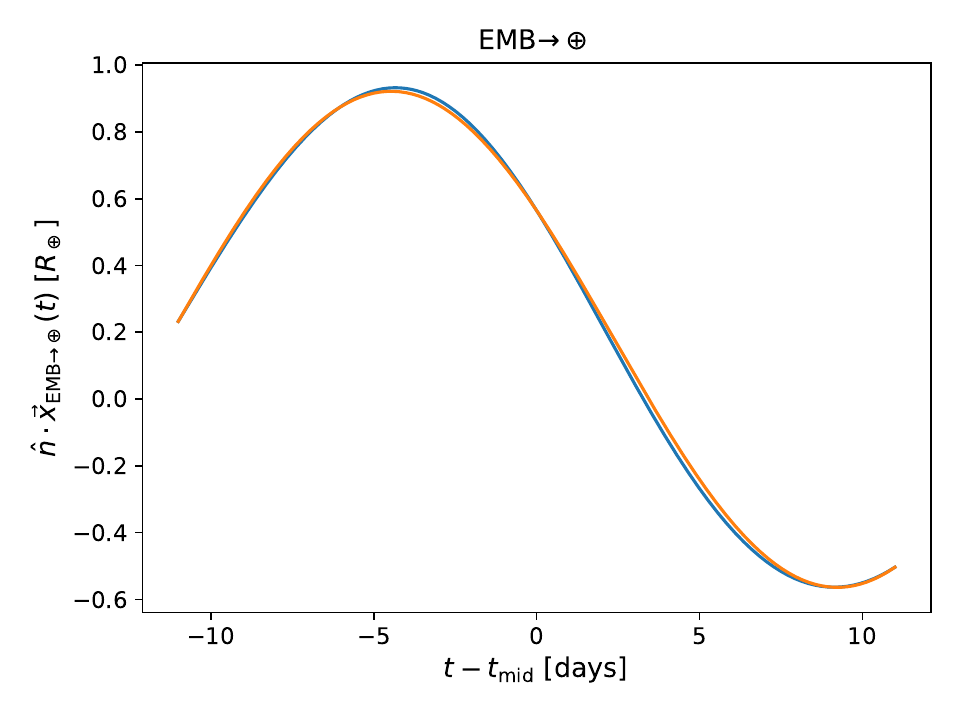}
\includegraphics[width=0.48\textwidth]{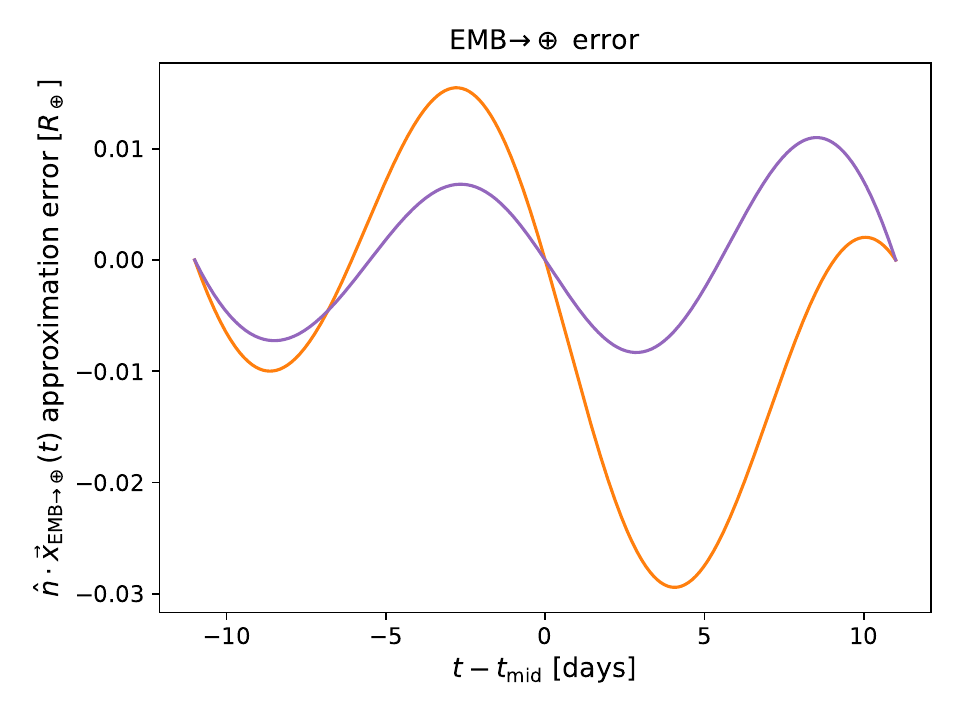}\\
\includegraphics[width=0.48\textwidth]{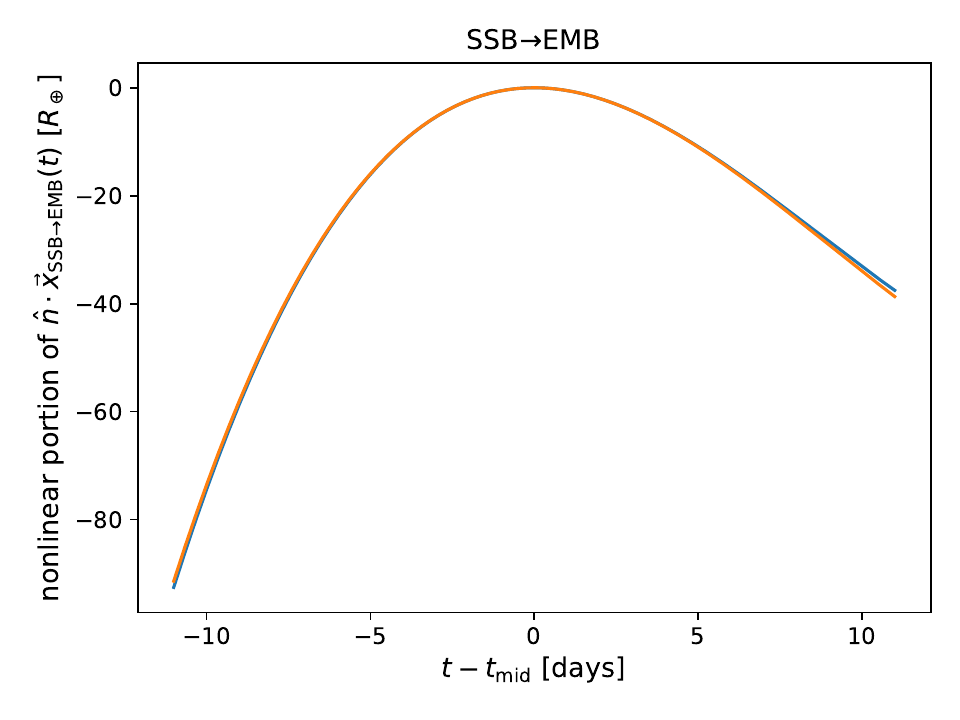}
\includegraphics[width=0.48\textwidth]{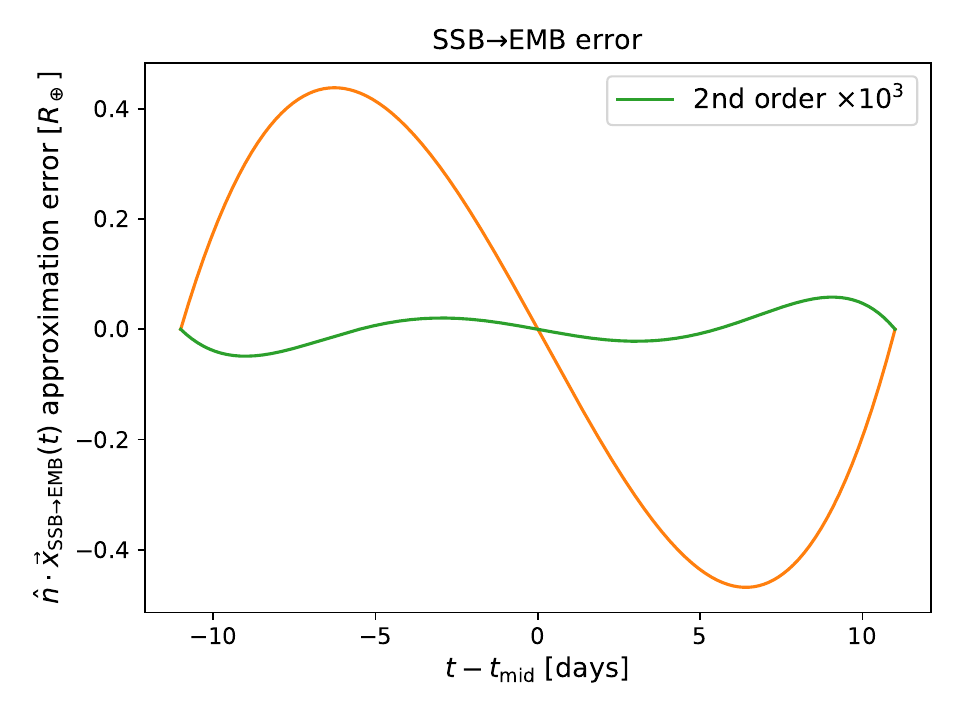}\\
\includegraphics[width=0.48\textwidth]{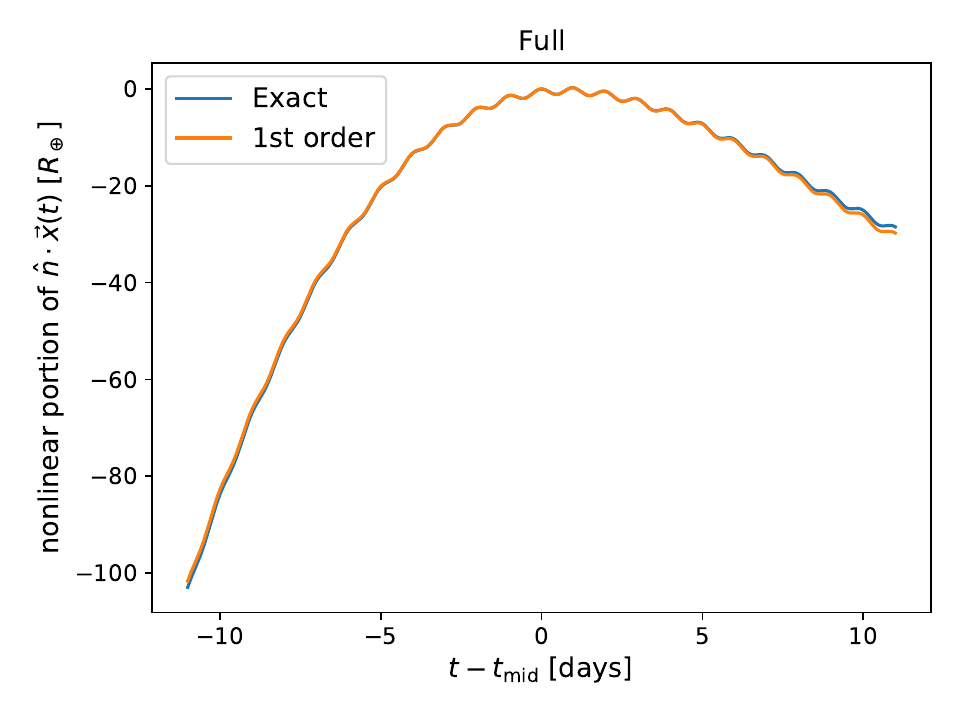}
\includegraphics[width=0.48\textwidth]{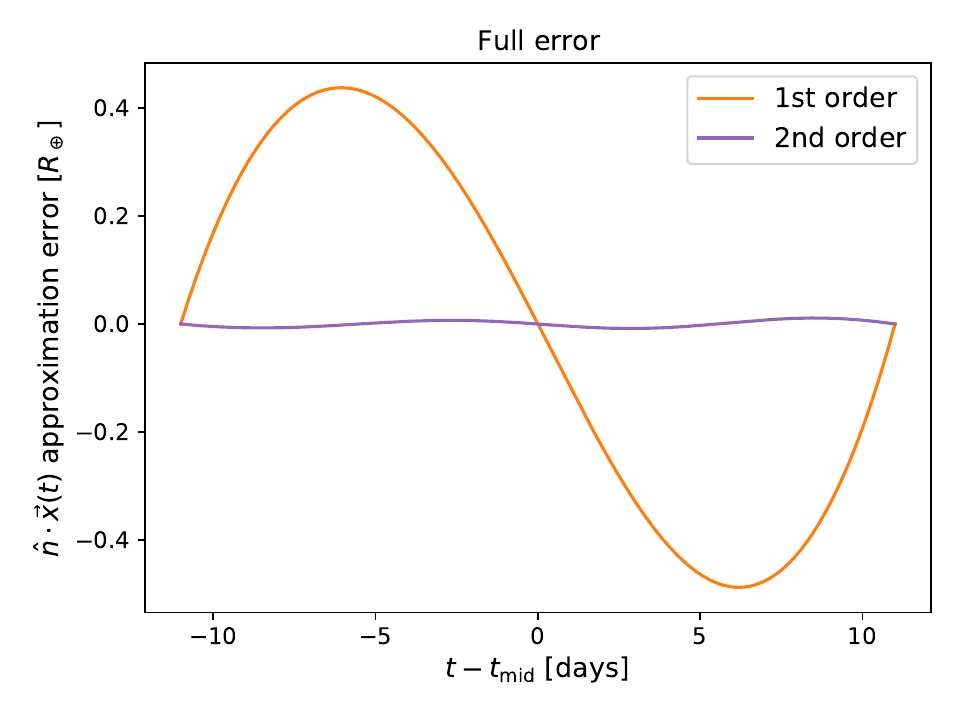}
\caption{Left-hand panels: The exact and first-order Fourier series approximations for the detector location with respect to the SSB projected along the source direction (bottom panel) and the contributions from the parts of the decomposition in equation~\eqref{eq:x_decomp} (upper panels). Right-hand panels: The errors of the first-order Fourier series approximation for these quantities, including the second-order approximation for the orbital quantities. These illustrations are for a source whose right ascension and declination are both zero, as observed from the LIGO Hanford site near a GPS time of $t_\text{mid} = 1291864218$~s (near Earth's perihelion). The exact values come from Astropy, while the first-order values use the approximation in equation~\eqref{eq:xdn_1st}, and the second-order expressions are given in \ref{app:Doppler_2nd_order}. The plots of the full and EMB orbital quantities in the left column just show the nonlinear part (computed as described in the text) to remove the very dominant linear contribution, though this is approximated accurately, so there is no need to remove it when computing the errors shown in the right column. The vertical axis scales of all the plots are different, in some cases significantly so. In particular, the upper right-hand plot shows the error in metres, while all the other plots are given in units of the Earth's equatorial radius $R_\oplus \simeq 6378$~km. Finally, we scale the EMB location error with the second-order approximation by $10^3$ so its shape is visible on the scale of the plot.}
\label{fig:Doppler_illustration}
\end{figure}

We illustrate the accuracy of this Fourier series approximation for the detector's location in figure~\ref{fig:Doppler_illustration} \footnote{We compute the nonlinear portion of the radial displacement shown in the figure by $\hat{n}\cdot[\vec{x}(t) - \vec{x}(t_\mathrm{mid}) - \dot{\vec{x}}(t_\mathrm{mid})(t - t_\mathrm{mid})]$, i.e.\ aligning at $t = t_\text{mid}$ and subtracting off the contribution from linear motion at the radial velocity at that time. We obtain the radial velocity using the Astropy function \texttt{get\_body\_barycentric\_posvel}.}. We also illustrate the improvement in accuracy obtained when using the second-order approximation to the orbital motion given in \ref{app:Doppler_2nd_order}. The $22$~day timespan shown in the figure is slightly longer than the length of an $0.2 + 0.2~M_\odot$ (detector frame mass) binary's dominant mode signal starting from $2$~Hz. The errors from the first-order Fourier series approximation are likely small enough for many applications, since the maximum error of $\sim 0.4R_\oplus$ is $< 20\%$ of the gravitational wavelength at $20$~Hz of $\sim 2R_\oplus$ (and the $0.2 + 0.2~M_\odot$ signal only lasts a bit more than an hour from an $m = 2$ frequency of $20$~Hz). However, we also include the second-order expressions to show how much improvement one gets with them, and because such higher-order expressions might be useful for analyses of continuous wave signals. With the second-order corrections the maximum error is $\sim 0.01R_\oplus$, which is $< 0.5\%$ of the gravitational wavelength at $20$~Hz.

We have checked that the maximum errors for a random sample of $1000$ central times and sky locations are not significantly larger than in the example given above: $<0.6R_\oplus$ and $<0.05R_\oplus$ for the first- and second-order expressions, respectively. We also checked that the errors are not larger when also varying $\Delta t$ values up to a maximum of $11$~days. If one reduces $\Delta t$ to $4$~days (giving an $8$ day timespan, a bit more than the length of an $0.2 + 0.2~M_\odot$ binary's dominant mode signal from $3$~Hz), then the maximum errors with the $1000$ central times and sky locations are $<0.03R_\oplus$ and $<3\times 10^{-4}R_\oplus \simeq 2$~km for the first- and second-order expressions. In all of these cases, the dominant error in the first-order case comes from the EMB motion, while the dominant error in the second-order case comes from the Earth's motion about the EMB. This is because (as illustrated in figure~\ref{fig:Doppler_illustration}) one only gets a factor of a few improvement in accuracy with the second-order expression for the Earth's motion about the EMB, while one gets several orders of magnitude improvement in accuracy with the second-order expression for the EMB motion. This difference is likely because the timespan considered is a significant fraction of the orbital period of the Earth around the EMB, and the Moon's mean motion is not the optimal period to use for this Fourier series representation. We leave studies of improved choices to future work.

We have not checked the accuracy for $\Delta t$ values larger than $11$~days. However, these could potentially be relevant in practice. For instance, for an $0.1 + 0.1~M_\odot$ binary, the lightest binary searched for in, e.g.\ \cite{Nitz:2022ltl}, the dominant mode signal lasts about $69$~days from $2$~Hz. Nevertheless, we leave investigating such cases to future work, not the least because analyzing such long signals (and possibly even signals of around the maximum length considered here) would likely require a hybrid of techniques currently employed for compact binary coalescences and continuous wave signals.

To obtain the Doppler corrections to the waveform, we just write the leading approximation (equivalent to the SPA from~\cite{Zhao_and_Wen}),
\begin{subequations}
\begin{align}
\label{eq:Doppler_full}
&\qquad\qquad\qquad\qquad\tilde{h}^\text{det}(f) \simeq e^{2\pi\ri f b(f)}\tilde{h}^\text{SSB}_\text{shift}(f),\\
\nonumber
b(f) &:= b_{1\text{c}}^\text{EMB}\{\cos[\Omega_\text{EMB}\bar{t}_\text{SPA}(f)] - \cos(\Omega_\text{EMB}t_\text{mid})\} + b_{1\text{s}}^\text{EMB}\{\sin[\Omega_\text{EMB}\bar{t}_\text{SPA}(f)] + \sin(\Omega_\text{EMB}t_\text{mid})\}\\
\nonumber
&\quad\; + b_{1\text{c}}^\text{EM}\{\cos[\Omega_\text{EM}\bar{t}_\text{SPA}(f)] - \cos(\Omega_\text{EM}t_\text{mid})\} + b_{1\text{s}}^\text{EM}\{\sin[\Omega_\text{EM}\bar{t}_\text{SPA}(f)] + \sin(\Omega_\text{EM}t_\text{mid})\}\\
&\quad\; + b_{1\text{c}}\{\cos[\Omega_\oplus\bar{t}_\text{SPA}(f)] - \cos(\Omega_\oplus t_\text{mid})\} + b_{1\text{s}}\{\sin\Omega_\oplus\bar{t}_\text{SPA}(f)] + \sin(\Omega_\oplus t_\text{mid})\},
\end{align}
\end{subequations}
where $\bar{t}_\text{SPA}(f) := \Psi'(f)/2\pi - t_\text{mid}$. To compute the full response of the detector to the waveform, one would first apply this expression to the polarizations and then apply one of the expressions from section~\ref{sec:rot} to include the effect of the Earth's rotation on the detector's response. (See~\cite{time_dep_response_Git} for our Python implementation of this calculation.)

\begin{figure}[tb]
\centering
\includegraphics[width=0.48\textwidth]{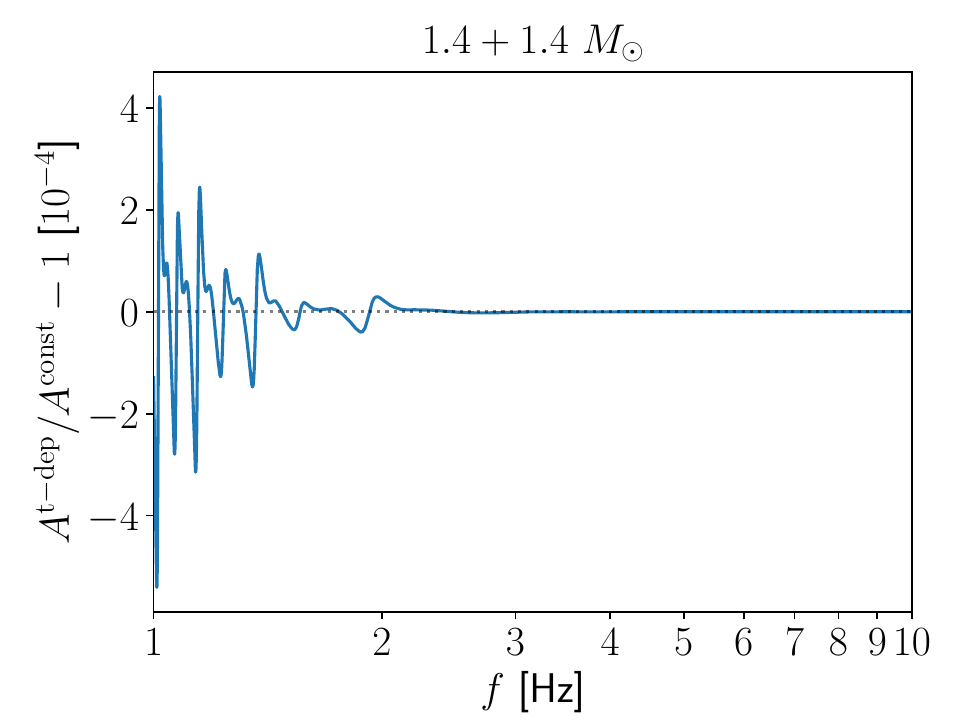}
\includegraphics[width=0.48\textwidth]{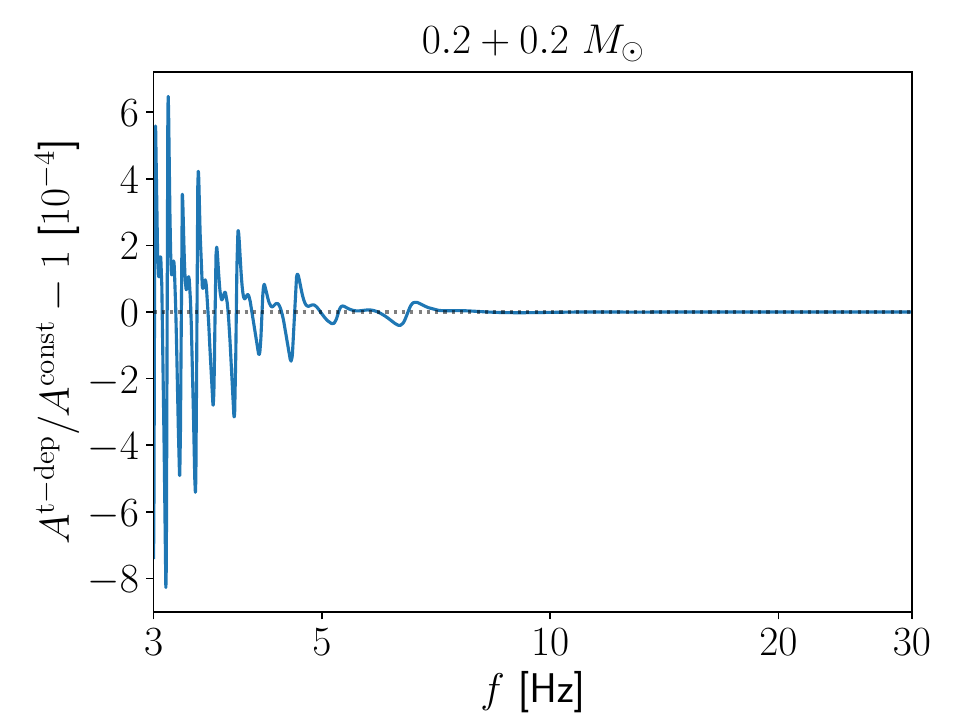}\\
\includegraphics[width=0.48\textwidth]{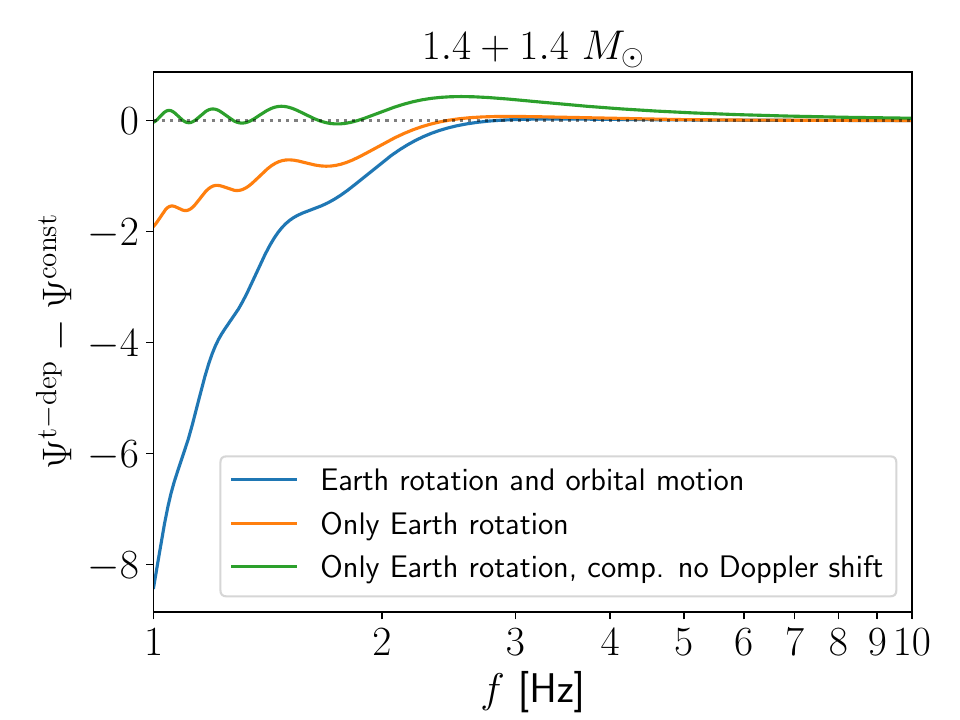}
\includegraphics[width=0.48\textwidth]{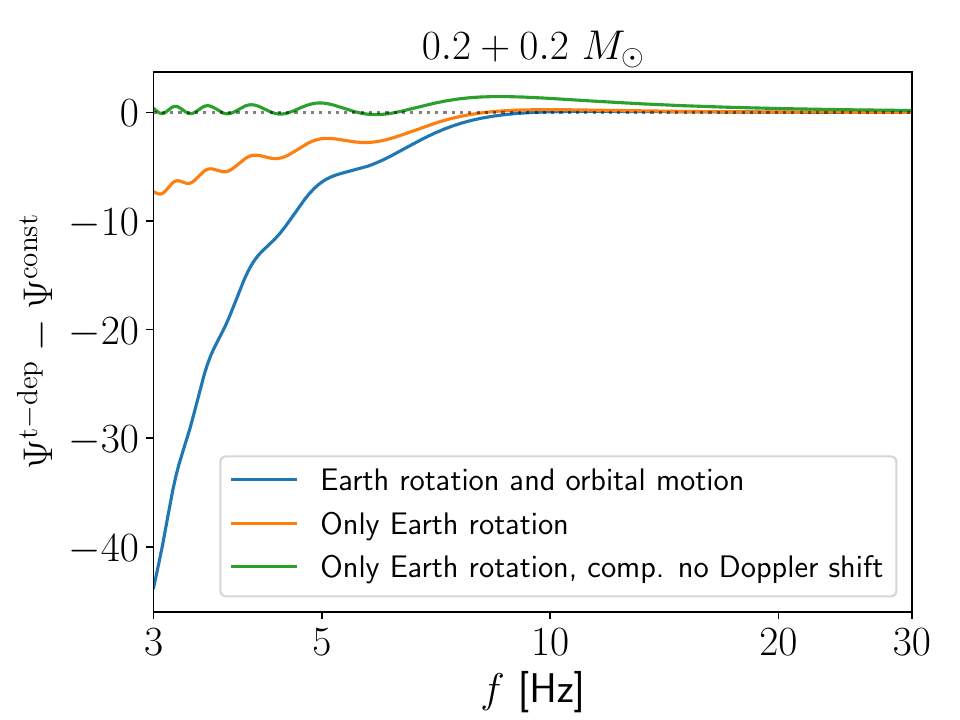}
\caption{The differences between time-dependent Doppler shift effects and constant Doppler shift effects from Earth's rotation and orbital motion in the amplitudes (top panel), and the phases (bottom panel), for the BNS (left) and the subsolar (right) systems considered in figure~\ref{fig:low_freq_SPA_corrections}. The bottom panel also shows the phase difference when considering time-dependent Doppler effects only due to Earth's rotation, comparing with the signal with the constant Doppler shift corresponding to the time of merger (orange) and without it (green). The sky location used is a right ascension of $5.13$~rad and a declination of $0$ and the detector is ET at the Virgo location. %\nkjm{I realized that the legend says ``no redshift'' when it should say ``no Doppler shift'' though I'm not sure there's room for this. If there's not, we can just comment on why we used ``no redshift'' in the caption, or use ``no Doppler'' or something similar. Apologies for using redshift in my suggestion.}\nkjm{In the legend, ``Only Earth rotation no redshift'' is probably unclear. Something like ``Only Earth rotation, comp.\ no redshift'' would be clearer (but is still not the best, so feel free to suggest something better).} %\nkjm{Ideally the only Earth's rotation curve would appear behind the curve that also includes the effect of its orbital motion, but this is really quite picky.}
}
\label{fig:doppler_shift}
\end{figure}

We now illustrate the effects of the time-dependent Doppler shift on the amplitude and phase of the signal in figure~\ref{fig:doppler_shift}. We thus compare with the signal with a constant Doppler shift given by the velocity at merger, i.e.\
\begin{subequations}
\begin{align}
\label{eq:const_dop}
&\qquad\qquad\qquad\qquad\qquad f \to \frac{f}{1 - v_\text{mrg}},\\
\nonumber
v_\text{mrg} &= \Omega_\text{EMB}[b_{1\text{c}}^\text{EMB}\sin(\Omega_\text{EMB}t_\text{mid}) + b_{1\text{s}}^\text{EMB}\cos(\Omega_\text{EMB}t_\text{mid})] + \Omega_\text{EM}[b_{1\text{c}}^\text{EM}\sin(\Omega_\text{EM}t_\text{mid})\\
&\quad + b_{1\text{s}}^\text{EM}\cos(\Omega_\text{EM}t_\text{mid})] + \Omega_\oplus[b_{1\text{c}}\sin(\Omega_\oplus t_\text{mid}) + b_{1\text{s}}\cos(\Omega_\oplus t_\text{mid})],
\end{align}
\end{subequations}
where we only make this substitution in the phase, since the SPA expression does not include amplitude corrections. We set $b_{1\text{c,s}}^\text{EMB} = b_{1\text{c,s}}^\text{EM} = 0$ to obtain the version with just the effects of Earth's rotation. We also include the comparison with the unredshifted waveform for only Earth's rotation effects. Figure~\ref{fig:doppler_shift} shows that the waveform with just the Doppler shift from Earth's rotation has the best agreement with the unredshifted waveform for low frequencies, though the waveform with the redshift at merger agrees better for higher frequencies (specifically above $\sim 2.5$~Hz for the BNS). Meanwhile, the Doppler shift due to Earth's orbital motion has the largest effect at low frequencies, as expected. 

The computation of the Doppler shift effects in the time-dependent response can be sped up by using the second order approximation in equation~(\ref{eq:htilde_response_approx}) instead of using the exact Fourier series, since we find that the accuracy of the derivative approximation to the time-dependent response is unchanged when including the Doppler shift. The time required for time-dependent Doppler shift effects using the full Fourier series method with direct finite differencing and the multiband frequency is $\sim 1.6$~s, for the same system used in table~\ref{tab:derivative_speed}. The time without the Doppler shift is $\sim 0.7$~s, so including the Doppler shift effect significantly increases the time needed to compute the time-dependent response. But when using the second order approximation, the computation time with time-dependent Doppler shift effects reduces to $\sim 0.7$ s, and it reduces further to $\sim 0.4$ s with the SPA-equivalent approximation, which has little increase from the time without Doppler shift effects in table~\ref{tab:derivative_speed}. However, most of the time spent in the Fourier series method is caused by the calling of Astropy functions used in computing the Fourier coefficients. By precomputing the Cartesian components of Fourier coefficients with three special sky coordinates---$(\alpha, \delta)$ being $(0,0)$, $(0,\pi/2)$ and $(\pi/2,0)$---and computing Fourier coefficients for any sky locations later with the equation under equation \eqref{eq:b_coef}, we can reduce the computation time significantly. The time for the SPA-equivalent approximation with Doppler shift effects reduces to $\sim 0.05$ s with precomputed Fourier coefficients over the multiband frequency mesh covering $5$--$2048$~Hz. For comparison, the implementation by Baral~\emph{et al}~\cite{Baral:2023xst} takes $\sim 0.11$ s, which is about twice that of our method.

%---------------
\subsection{Assessing the importance of the time-dependent Doppler shifts}
%---------------

For a quantitative comparison, we consider the overlap of the waveforms with and without the time-dependent Doppler shift effects. This overlap is defined using the noise-weighted inner product (see, e.g.\ \cite{Cutler:1994ys}, who use slightly different notation),
\<
\langle g, h \rangle := 4\Real \int_0^\infty \frac{\tilde{g}^*(f)\tilde{h}(f)}{S_n(f)}\mathrm{d}f,
\?
where $S_n(f)$ is the detector's noise power spectral density. The overlap of two signals $g$ and $h$ is then given by
\<
\cO := \frac{\langle g, h \rangle}{\sqrt{\langle g, g \rangle\langle h, h \rangle}},
\?
which is at most $1$, when the two waveforms are identical. We will consider overlaps quite close to $1$, so we will quote values in terms of $\bO := 1 - \cO$. The value of $\bO$ is particularly useful since it is only possible to distinguish two nearby signals of a given SNR if $\bO > D/(2\text{SNR}^2)$, if one is measuring $D$ parameters, assuming Gaussian noise (see, e.g.\ Appendix~G in~\cite{Chatziioannou:2017tdw}; this refers to the faithfulness instead of the overlap, though the faithfulness is usually defined with additional maximization over time and phase shifts, which is why we do not use that term here).

We first consider the overlap between the waveform with the time-dependent Doppler shift effects due to the Earth's rotation and orbital motion and the waveform with the constant Doppler shift at merger due to Earth's rotation and orbital motion. We consider the signal with that constant Doppler shift, since a constant Doppler shift is not observable if one does not know the binary's masses (or the relation between mass and tidal deformability; see, e.g.\ \cite{Messenger:2011gi}) beforehand, and this constant Doppler shift signal will be close to the signal with the time-dependent Doppler shift signal at higher frequencies. Thus, the overlaps we compute are expected to let us find a lower bound on the SNR required for the time-dependent Doppler shifts to be relevant for parameter estimation, though we shall see below that the actual story seems more complicated. Nevertheless, we still quote the overlaps to give a quantitative assessment of the significance of these effects for data analysis.

We compute this overlap for the BNS system in figure~\ref{fig:doppler_shift} with the $10$~km ET noise curve~\cite{ET_new} (used in~\cite{Branchesi:2023mws}) and a frequency range of $1$--$2048$~Hz, giving $\bO = 1.7\times 10^{-8}$. On the other hand, for CE with the $40$~km low-frequency noise curve~\cite{CE_new} \footnote{Here we take CE to have the same location and orientation as the LIGO Hanford detector and use the strain noise curve that assumes a source at $15^\circ$ from normal incidence when computing the detector size effects that are relevant at high frequencies.} with a frequency range of $5$--$2048$~Hz, $\bO = 6.6\times 10^{-7}$. It may seem surprising that $\bO$ is smaller for the ET case than the CE case even though ET's sensitivity extends to lower frequencies, where the effect of the time-dependent Doppler shifts is larger. However, the effect of the Doppler shift is different in the two cases, due to the different locations of the detectors on Earth, with the phase difference in the ET case at $5$~Hz slightly smaller than in the CE case. However, the larger effect is likely that the CE sensitivity is significantly better than the ET sensitivity at frequencies of $\sim 10$--$100$ Hz, as shown in figure~3.3 of~\cite{Evans:2021gyd}, making it more sensitive to the small time-dependent Doppler shift effects in the $\sim 10$--$20$~Hz range. Of course, both of these $\bO$ values are quite small, seemingly requiring very large SNRs for the time-dependent Doppler shift effects to be distinguishable ($\gtrsim 870$ and $5420$ for CE and ET, respectively, and likely at least a factor of a few more, since these bounds come from taking $D = 1$), but we will see below that these effects could be relevant in some realistic cases.

We can also assess how much of an effect the Earth's orbital motion has on the difference between these two signals by computing the same overlaps but just including the effect of the Earth's rotation. Here we obtain $\bO = 2.9\times10^{-8}$ for ET and $4.4\times10^{-7}$ for CE, while if we compute the overlap comparing with the signal with no Doppler shift (which gives better agreement for low frequencies, as illustrated in figure~\ref{fig:doppler_shift}), we obtain $\bO = 1.4\times10^{-5}$ for ET and $2.3\times10^{-7}$ for CE. It may seem strange that $\bO$ is smaller when including the Earth's orbital motion in addition to its rotation when comparing with the constant Doppler shift case for ET, given the significant differences between the phases in the two cases below $\sim 2.3$~Hz seen in figure~\ref{fig:doppler_shift}. However, these very low-frequency portions of the signal make a small contribution to the overlap in the case with just the Earth's rotation, where the $\bO$ value starting from $2.3$~Hz is only $\sim 4\%$ smaller than the value starting at $1$~Hz. The low-frequency portions of the signal make a larger difference in the overlap when also including the Earth's orbital motion, with the $\bO$ value starting from $2.3$~Hz being $\sim 30\%$ smaller than the value starting at $1$~Hz, but the smaller phase difference at frequencies above $\sim 2.7$~Hz leads to the smaller $\bO$ value. Similarly, the smaller $\bO$ value for CE with just the Earth's rotation when comparing with no Doppler shift is perhaps surprising, but in this case the phase difference starts positive and then becomes negative around $9$~Hz, while the phase difference for the comparison with the constant Doppler shift at merger is always negative. Thus, there is a smaller difference in the comparison of the full version with the version with no Doppler shift over the full frequency range considered. However, the $\bO$ value comparing with the constant Doppler shift at merger decreases rapidly as one increases the low-frequency cutoff, and is below the $\bO$ value comparing with no Doppler shift (which decreases much more slowly) for low-frequency cutoffs above $\sim 8$~Hz.

The full stochastic sampling results in Baral~\emph{et al}~\cite{Baral:2023xst} let us see how the small $\bO$ values we find can affect parameter estimation in a realistic observational scenario. Specifically, Baral~\emph{et al} consider the effect of the Earth's rotation (as well as detector finite size effects) on the ability to localize an SNR $1000$ BNS signal using a single CE detector. In their figures~6--8, they consider the different effects separately, including separating out the effect of the Earth's rotation on the antenna pattern functions (which they refer to as the Earth-rotation modulation) and the time-dependent Doppler shift (which they refer to as the Earth-rotation time delay). They find that omitting either effect can lead to completely incorrect inference of the binary's sky location and/or distance in some cases. We thus compute the overlap with and without the time delay due to Earth's rotation for the case that gives the best sky localization (shown in figure~6 of Baral~\emph{et al}), where the source is in the direction of one of the detector's dark spots (where it is less sensitive) at the time of merger.

The binary parameters considered by Baral~\emph{et al} in figure~6 are $m_1=1.44M_\odot$, $m_2=1.40M_\odot$, a luminosity distance of $\sim 22.2$~Mpc, inclination angle of $1.0$~rad, polarization angle of $\sim 1.57$, right ascension of $\sim 5.89$~rad, declination of $0$ and a GPS time at merger of $630696093$~s. They also use the same $5$--$2048$~Hz frequency range we consider. For these parameters, we obtain $\bO = 1.8\times 10^{-7}$, comparing the waveform with the time-dependent Doppler shift from Earth's rotation with one with the constant Doppler shift at merger. (One obtains $\bO = 4.9\times 10^{-4}$ when comparing instead with the waveform with no Doppler shift, so the waveform with the constant Doppler shift is the correct one to use to obtain a lower bound on the SNR for distinguishability.) Baral~\emph{et al} find a match of 95.4\% in this case when comparing the signal including the effects of Earth's motion and the finite size of the detector with the signal without those effects, including maximizing over time and phase shifts. However, they compare with the signal with no Doppler shift, which explains why they find a much smaller match than the overlap we compute, even though they include maximization over time and phase shifts. We have checked that the detector-size effects (which are only important at high frequencies) just make a $\sim10^{-9}$ difference in the overlap we computed (here using the displacement noise curve that is appropriate when accounting for detector-size effects explicitly).

Baral~\emph{et al} find that neglecting the time-dependent Doppler shift from the Earth's rotation leads to very significant biases in the sky location and distance (see their figure~6), even though the $\bO$ value we obtain is a factor of $\sim 3$ below the most optimistic threshold of $5.0\times10^{-7}$ for the SNR of 1000 they consider (and $D = 1$) \footnote{One might think that it is more appropriate to consider the overlap and distinguishability criterion for a lower upper frequency cutoff, since the  time-dependent Doppler shift is more important for lower frequencies. However, we checked that the distinguishability threshold increases more rapidly than $\bO$ when decreasing the upper frequency cutoff.}. Since the posteriors in this case are not very Gaussian, while the SNR criterion assumes Gaussianity, this might explain the apparent discrepancy. Since we obtain a larger value of $\bO = 4.0\times10^{-7}$ when including the Earth's orbital motion, the Baral~\emph{et al} results suggest that including this effect will also be important for parameter estimation, at least in the special case of a single CE detector. We will check this with direct stochastic sampling calculations in future work.

We also find that the first-order expressions for the orbital motion are indeed very accurate for these purposes by computing the overlap between the first- and second-order expressions, giving $\bO$ values of $4.7\times10^{-13}$ and $7.6\times10^{-11}$ for the BNS and the subsolar-mass cases in ET, respectively, which are likely well below any possibility of observability. (The subsolar-mass case considers a lower frequency of $3$~Hz, since the computational demands of going to lower frequencies are considerable.) For CE, one obtains $\bO$ values that are even smaller, as expected: $6.0 \times 10^{-16}$ for the BNS, and $5.9\times 10^{-11}$ for the subsolar-mass case (with both computations starting at $5$~Hz).

Finally, we check that the SPA expression for the effect of Earth's rotation on the antenna pattern functions is also very accurate, giving $\bO = 4.2\times10^{-14}$ when compared with the exact Fourier series expression for the BNS case with ET. It thus seems likely that the SPA calculation is sufficiently accurate for all BNS signals one can expect to observe in third-generation detectors, but this should be checked explicitly, given the failure of the standard SNR distinguishability criterion to account for the results in Baral~\emph{et al}.

%%%%%%%%%%%%%%%%%
\section{Sensitivity}
\label{sec:SNR}
%%%%%%%%%%%%%%%%%

\begin{figure}[tb]
\centering
\includegraphics[width=0.38\textwidth]{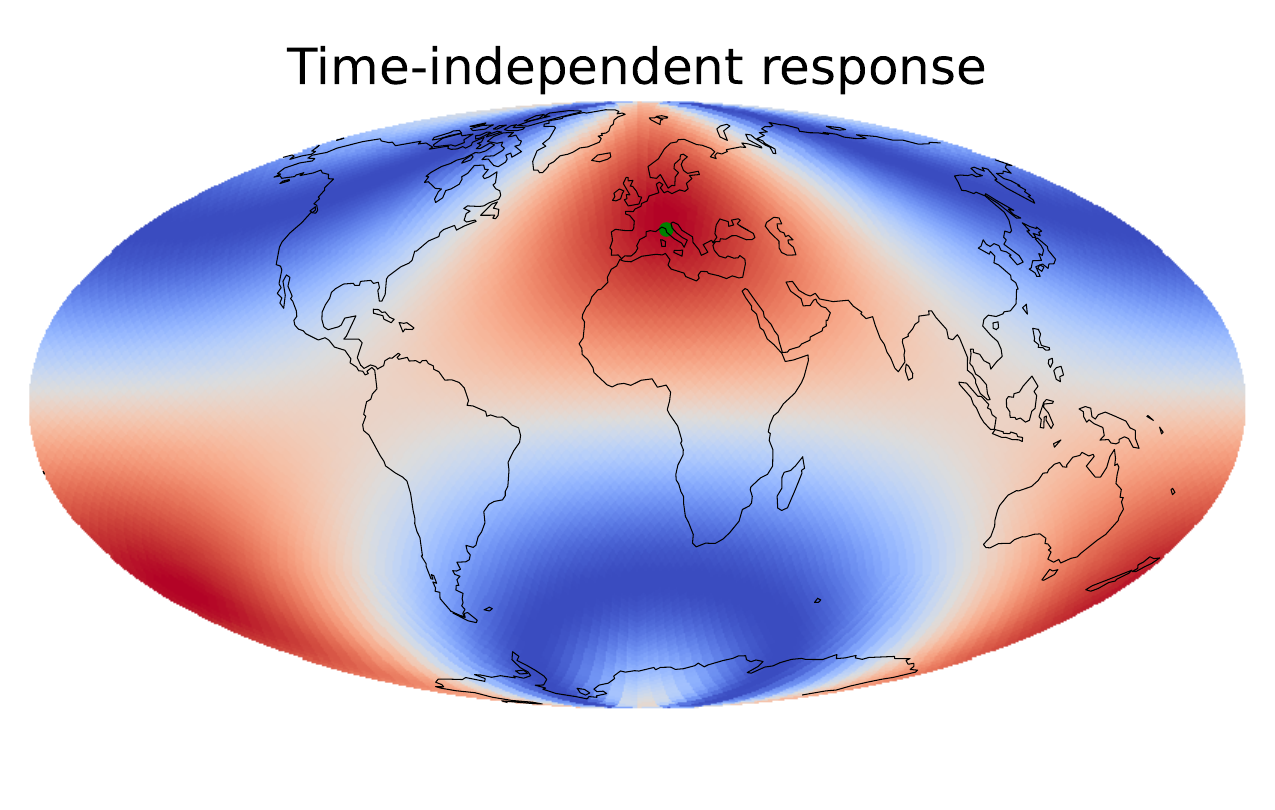}\\
\includegraphics[width=0.49\textwidth]{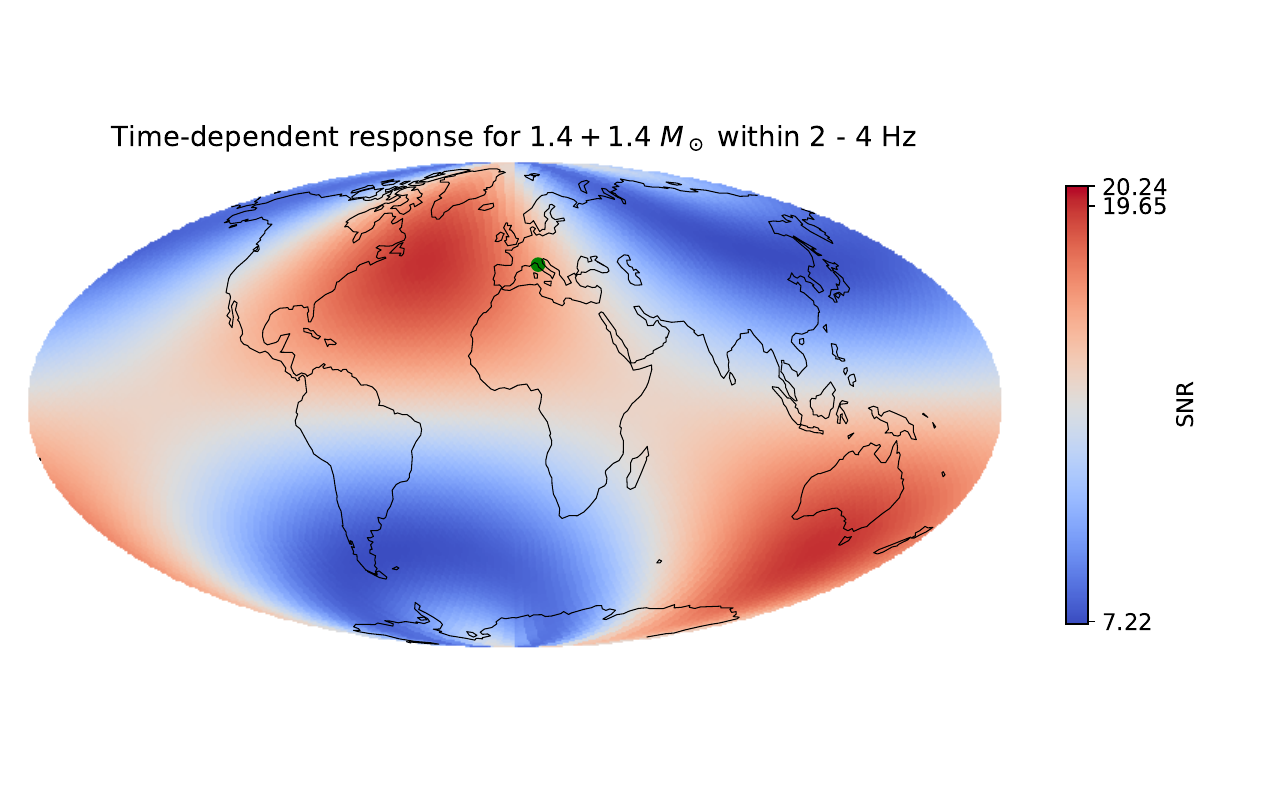}
\includegraphics[width=0.49\textwidth]{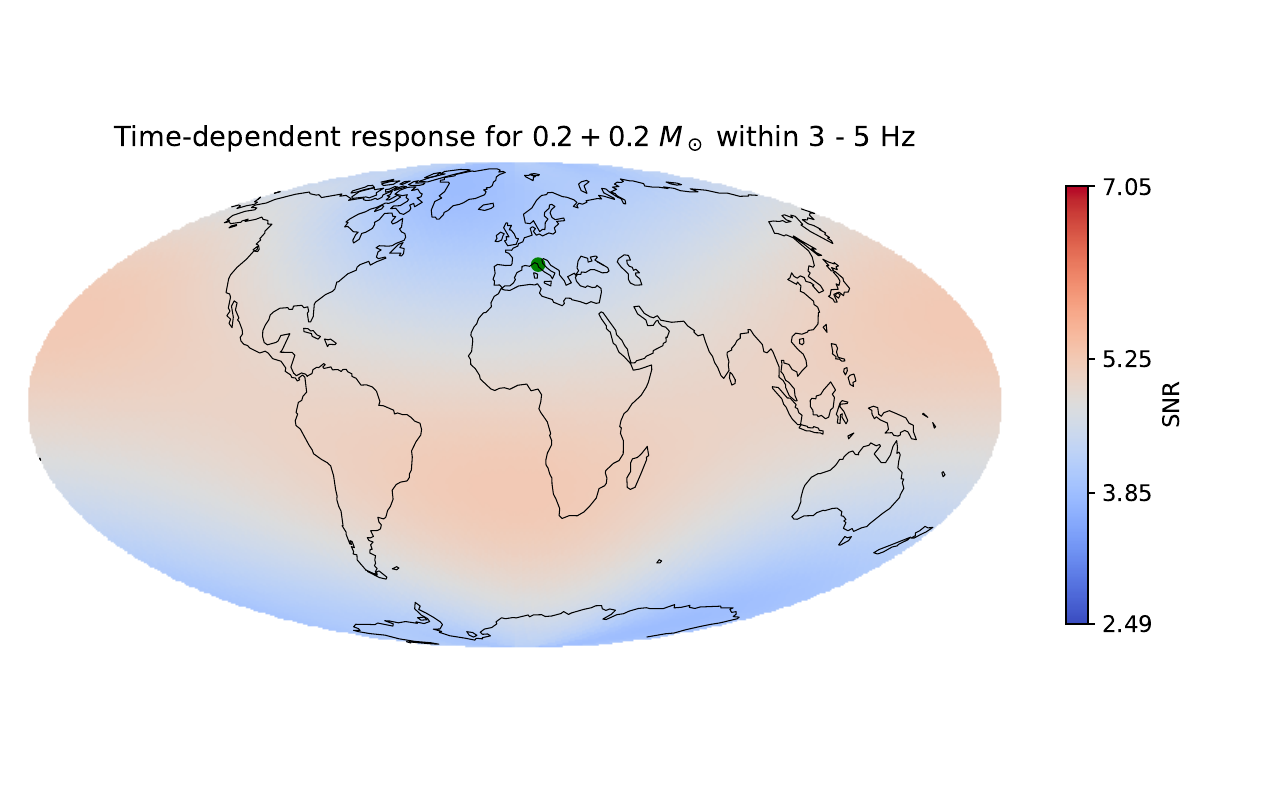}
\caption{The sky location dependence of the SNR for the time-independent and time-dependent responses for the $15$~km triangular ET detector, averaging over the polarization angle. The location of the detector is shown with a green dot and the orientation of the Earth is that at the coalescence time. The solar-mass and subsolar-mass waveforms used are the same as those in section~\ref{sec:rot}, except that the sky location and polarization angle are allowed to vary. The time-independent sky map is the same for both cases, except with different scales for the colour bar, given on the second row. The colour bars are scaled to the maximum and the minimum SNRs for the time-independent response. The inner sets of numbers on the colour bars show the maximum and minimum SNRs for the time-dependent cases. (The minimum is the same in both the time-independent and time-dependent cases for the solar-mass binary.)}
\label{fig:skymap}
\end{figure}

We now consider the effects of the time-dependent response on the sensitivity of the detector to sources at different sky locations. (This does not include the effect of the Doppler shift, since its effect on the amplitude of the frequency domain signal is negligible, as shown in figure~\ref{fig:doppler_shift}.) In figure~\ref{fig:skymap}, we show the sky location dependence of the total SNR for the three ET interferometers with time-dependent and time-independent responses in the band $2$--$4$~Hz for the solar-mass case, and $3$--$5$~Hz for the subsolar-mass case. We show the position of Earth at the coalescence time. We focus on low frequencies since the Earth's rotation makes the most difference in such cases. As in the previous section, we use the placeholder ET location and orientation for a triangular detector from~\cite{LALDetectors} (here considering all the detectors, i.e.\ `E1', `E2' and `E3'). Recall that this is the Virgo location with simple choices for the detectors' orientations. %\nkjm{The GPS time will not matter, since we're plotting things in the Earth-fixed frame.}
We use the $15$~km noise curve.
The sky map is produced using HEALPix~\footnote{\url{https://healpix.sourceforge.io}}.

The colours on the sky map indicate the SNRs averaged over polarization angle. The SNR with the time-independent response has the same sky location dependence for signals of different total masses, so the sky map only differs by the scaling of the colour bar. The time-dependent SNRs are always within the range of time-independent SNRs, as expected. The time-dependent SNR sky maps are also smoother than the time-independent ones, also as expected. The smoothing effect becomes more significant for lower-mass systems, as shown for a subsolar-mass case in the right bottom panel of figure~\ref{fig:skymap}, since lower-mass systems spend more time at higher frequencies where more SNR is accumulated. We illustrate this in table~\ref{tab:time_SNR} by giving the time the signals last in various subbands of frequency and the maximum and minimum SNRs accumulated in those bands (for different sky locations).

\begin{table}[t]
  \centering
  \caption{The time the GW signals in figure~\ref{fig:skymap} spend in various frequency bands (sub-bands of the band plotted in that figure) and the maximum and minimum SNRs accumulated over those bands for different sky locations and the $15$~km triangular ET detector. For the $(1.4+1.4)M_\odot$ BNS case, there is negligible SNR ($< 0.1$) accumulated below $2$~Hz.}
  \begin{tabular}{ccccc}        
    \hline 
   	Masses [$M_\odot$] & Frequency band [Hz] & Time [h] & Lowest SNR & Highest SNR \\ 
	\hline
	%\multirow{4}{*}{$1.4+1.4$} & $1.0-1.5$ & 74.6 & $0.0002$ & $0.0004$ \\ 
	% & $1.5-2.0$ & 20.5 & $0.03$ & $0.07$ \\ 
	 \multirow{4}{*}{$1.4+1.4$} & $2.0-2.5$ & 8.0 & $0.98$ & $2.74$ \\ 
	  & $2.5-3.0$ & 3.8 & $2.48$ & $6.92$ \\ 
	  & $3.0-3.5$ & 2.0 & $3.84$ & $10.80$ \\
	  & $3.5-4.0$ & 1.2 & $5.33$ & $15.05$ \\
	 \hline	
	 \multirow{4}{*}{$0.2+0.2$} & $3.0-3.5$ & 52.0 & $1.17$ & $1.58$ \\ 
	  & $3.5-4.0$ & 30.7 & $1.65$ & $2.20$ \\ 
	  & $4.0-4.5$ & 19.3 & $1.67$ & $3.05$ \\
	  & $4.5-5.0$ & 12.8 & $1.69$ & $3.98$ \\  
    \hline 
  \end{tabular}
  \label{tab:time_SNR}
\end{table}

%%%%%%%%%%%%%%%%%
\section{Parameter estimation}
\label{sec:information}
%%%%%%%%%%%%%%%%%

\begin{figure}[tb]
\centering
\includegraphics[width=0.4\textwidth]{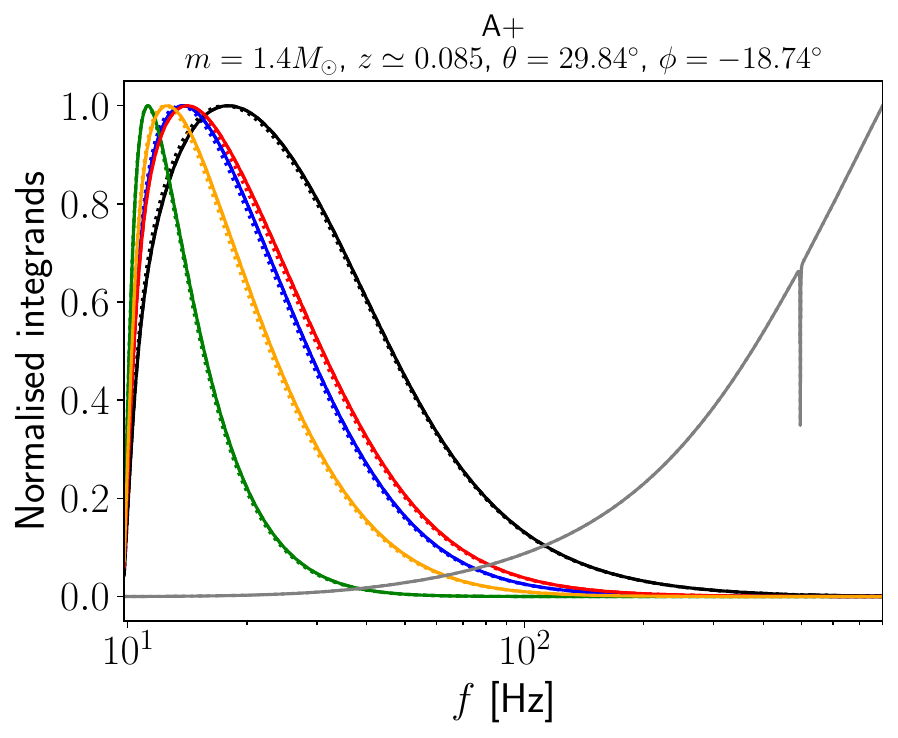}
\includegraphics[width=0.58\textwidth]{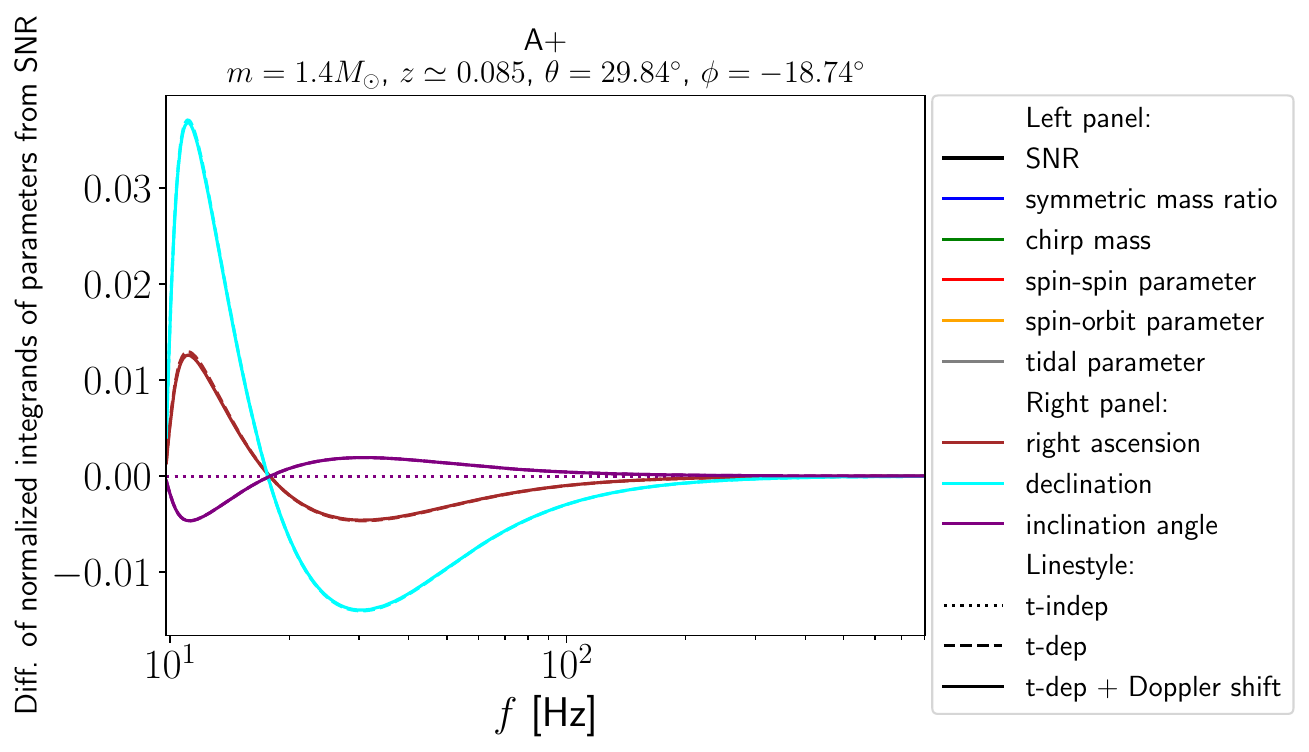}
\caption{Left panel: the Fisher matrix integrands for intrinsic parameters and the SNR for time-independent response (dotted lines), time-dependent response without Doppler effect (dashed lines), and time-dependent response with Doppler effect (solid lines) with the A+ design sensitivity noise curve for a randomly chosen sky location in the Earth fixed frame at merger of $\theta=29.84^{\circ}, \phi=-18.74^{\circ}$. Right panel: the differences between the normalised Fisher matrix integrands for extrinsic parameters and that for the SNR in the same scenario as the left panel.}
\label{fig:fisher_aLIGO}
\end{figure}

\begin{figure}[tb]
\includegraphics[width=0.4\textwidth]{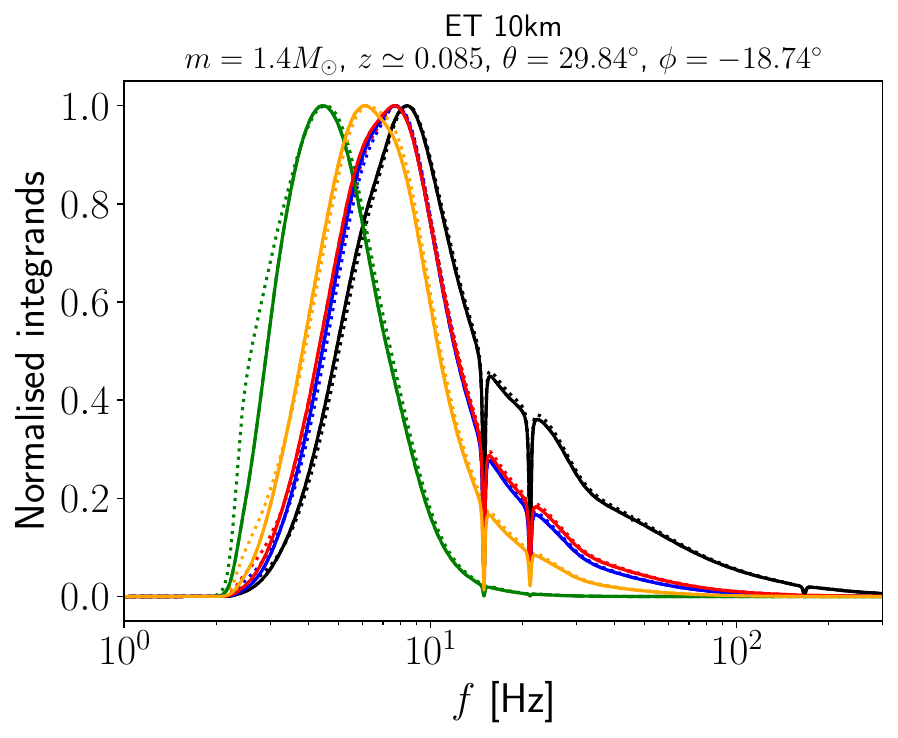}
\includegraphics[width=0.565\textwidth]{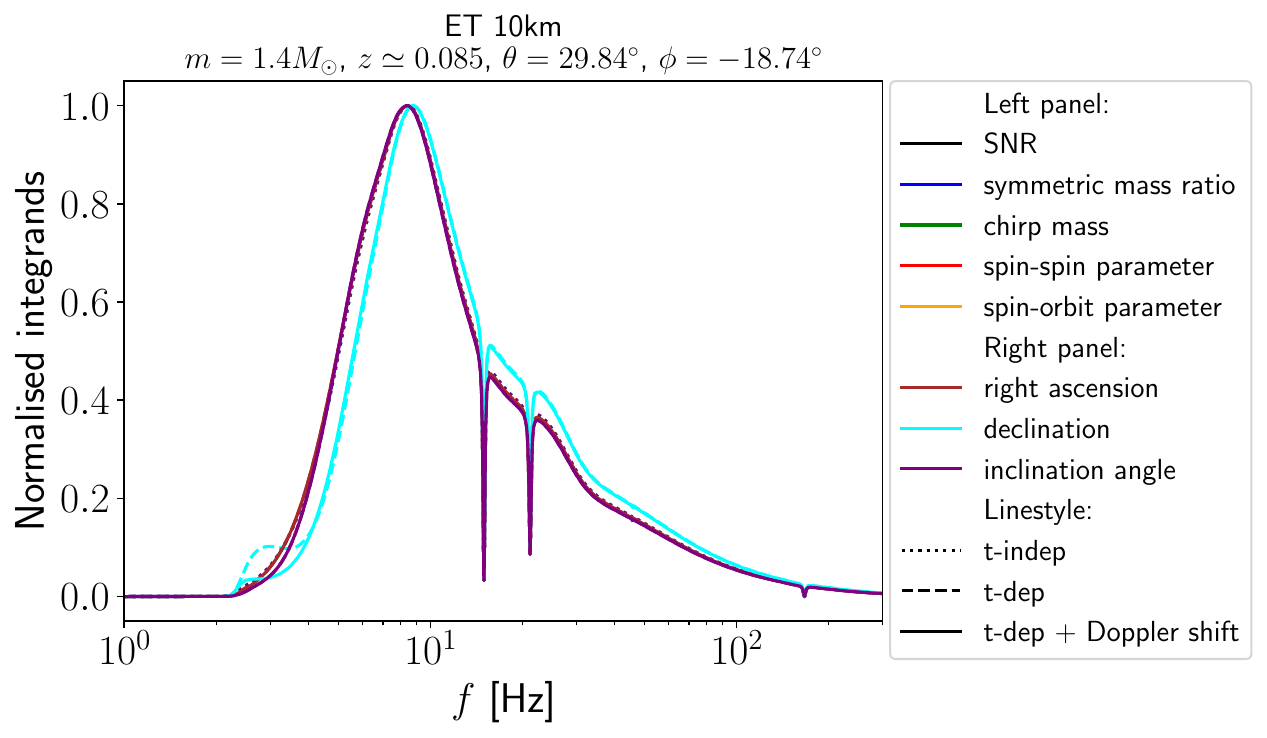}\\
\includegraphics[width=0.4\textwidth]{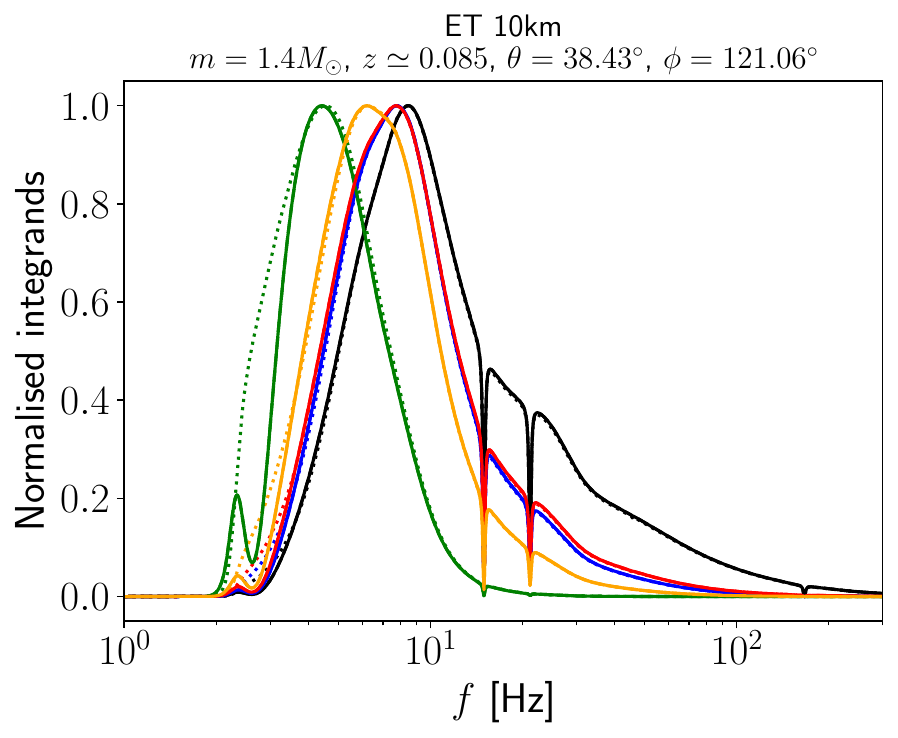}
\includegraphics[width=0.4\textwidth]{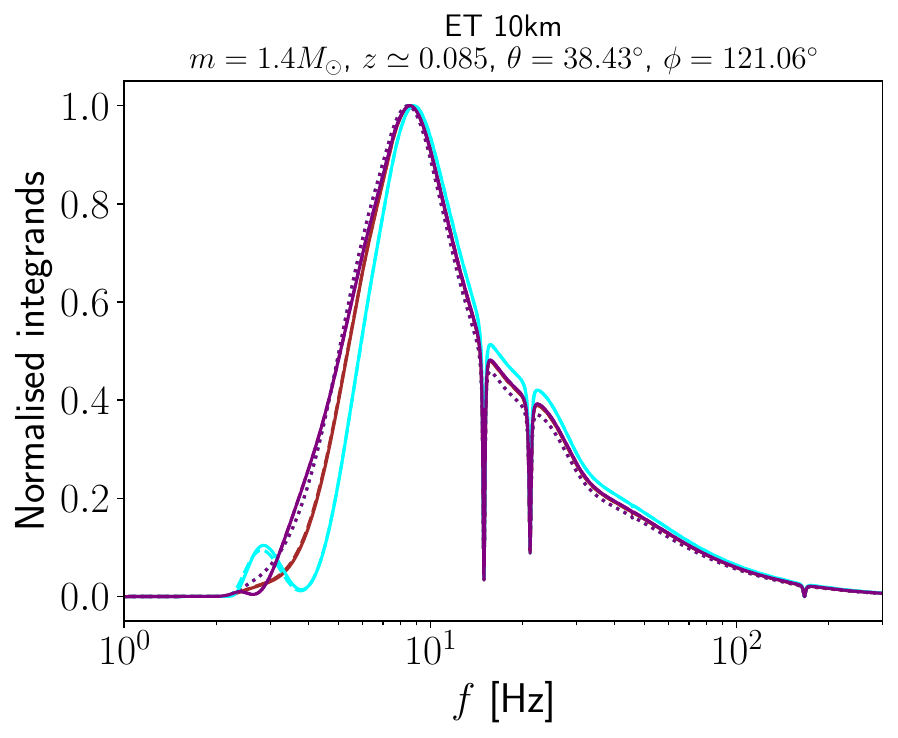}\\
\includegraphics[width=0.4\textwidth]{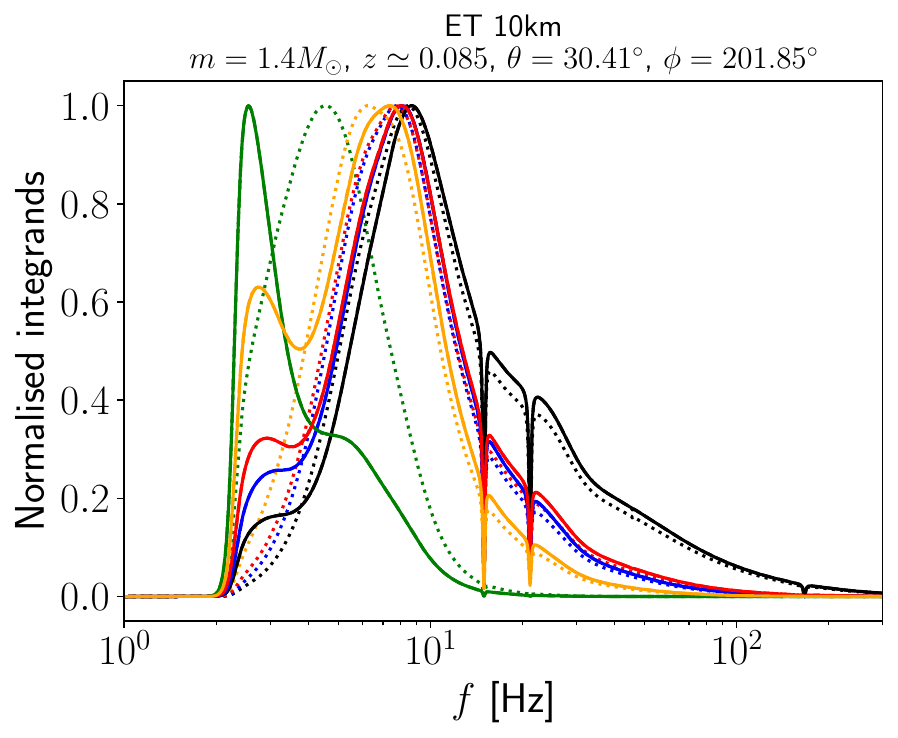}
\includegraphics[width=0.4\textwidth]{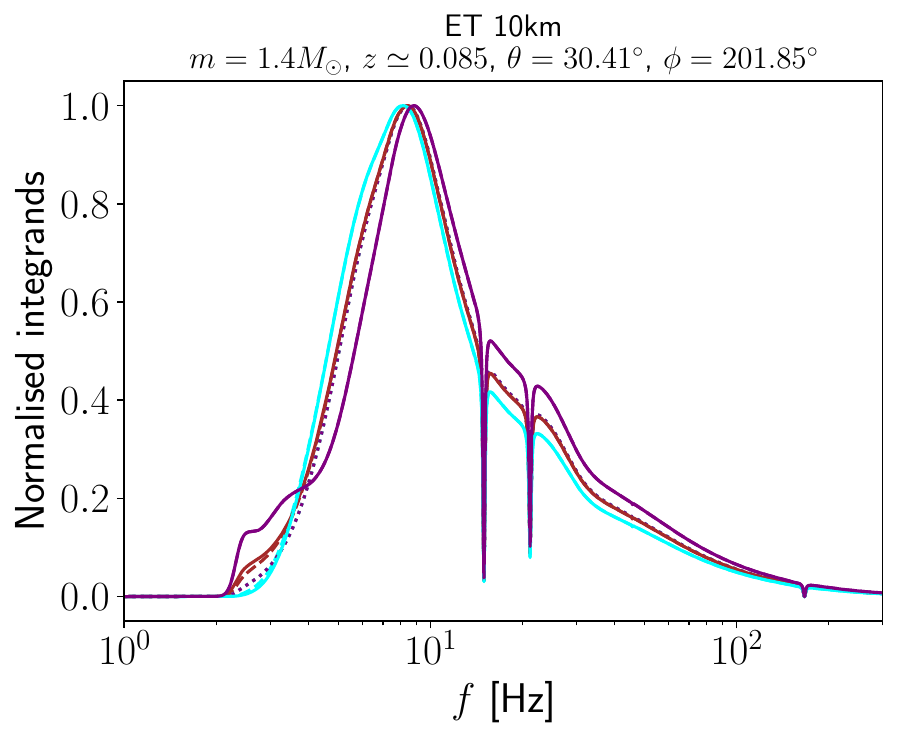}\\
\includegraphics[width=0.4\textwidth]{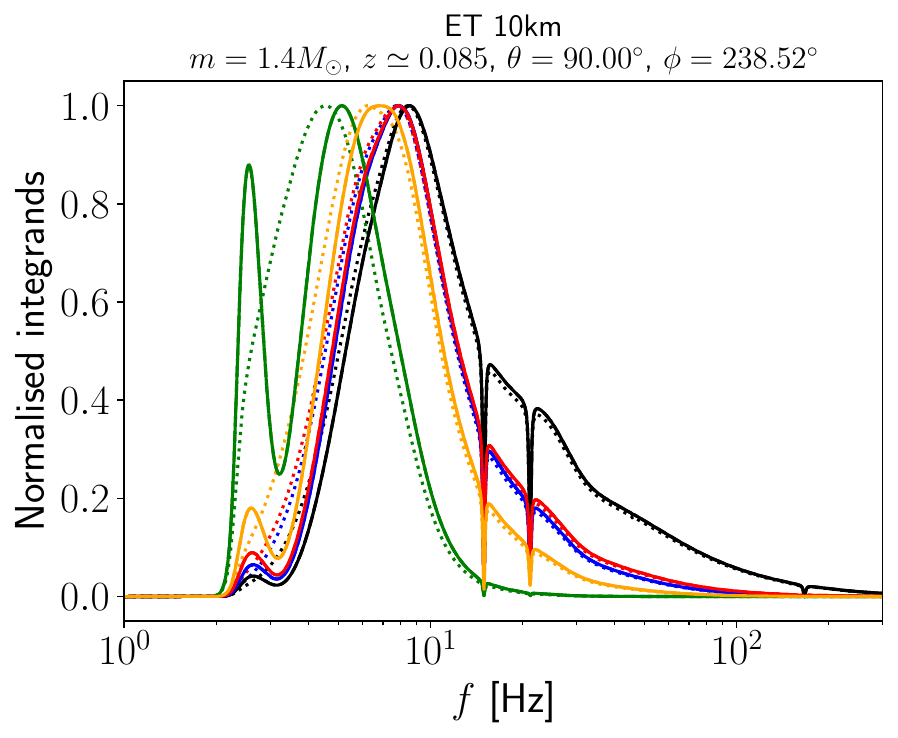}
\includegraphics[width=0.4\textwidth]{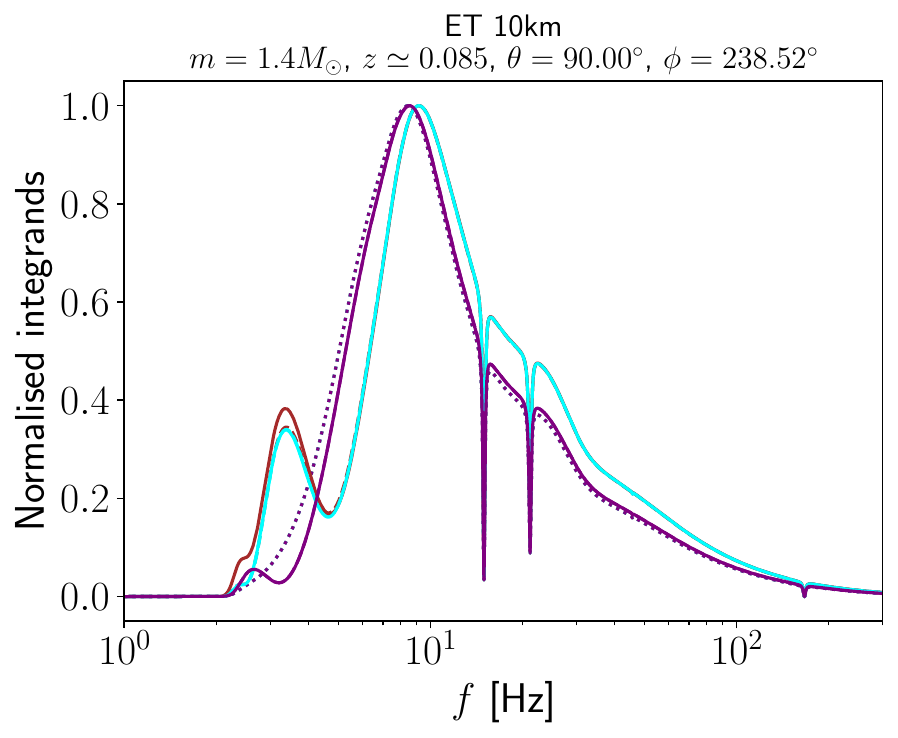}
\caption{The same as figure~\ref{fig:fisher_aLIGO} except for the ET sensitivity curve with 10 km arm length for a selection of sky locations (in the Earth fixed frame at merger). Additionally, the right panels show the normalized Fisher matrix integrands directly instead of the difference between these and the one for the SNR.}
\label{fig:fisher_ET10}
\end{figure}

\begin{figure}[tb]
\includegraphics[width=0.4\textwidth]{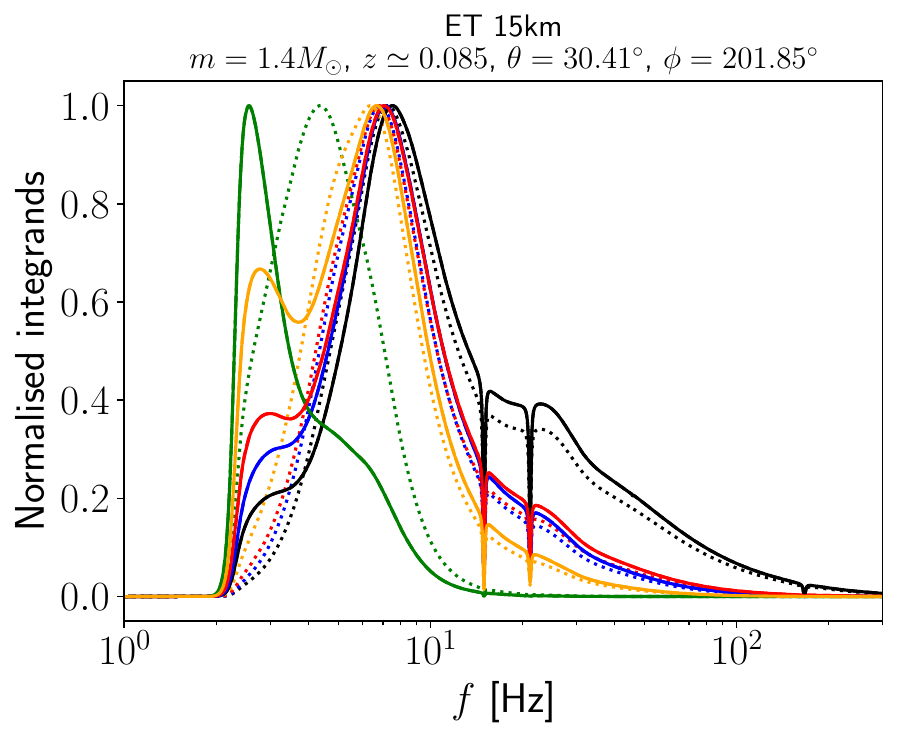}
\includegraphics[width=0.567\textwidth]{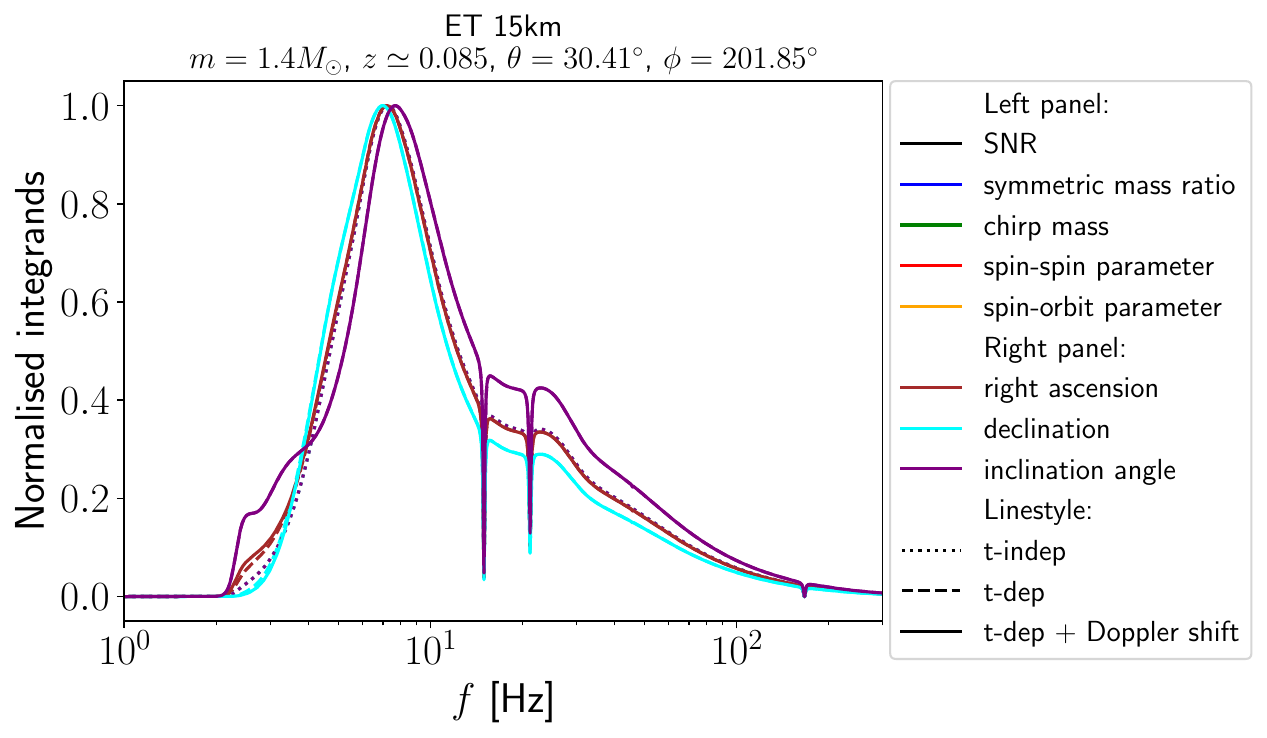}\\
\includegraphics[width=0.4\textwidth]{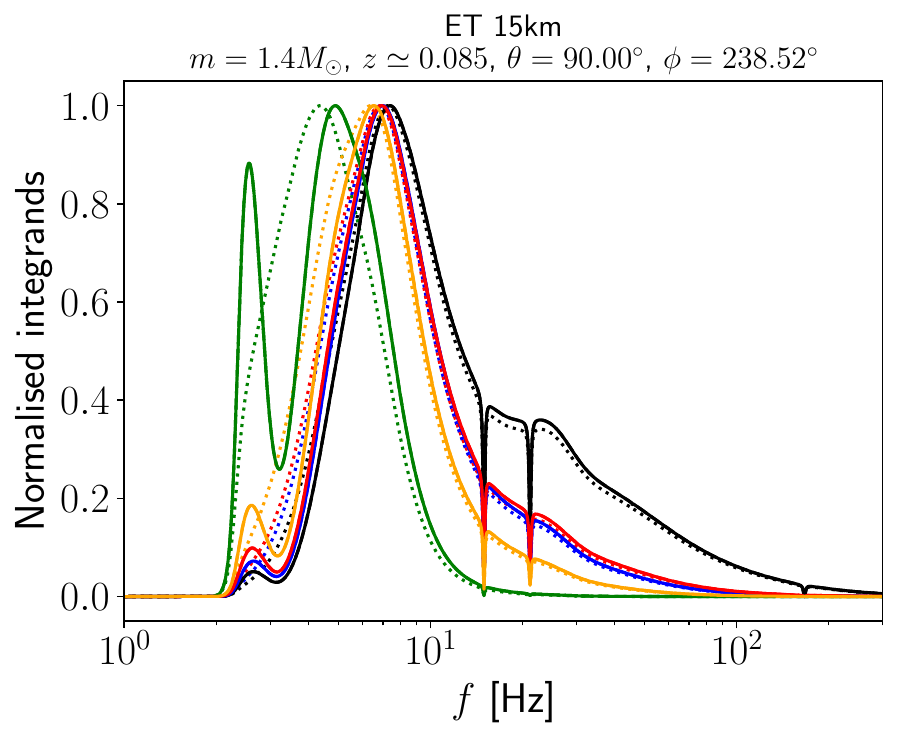}
\includegraphics[width=0.4\textwidth]{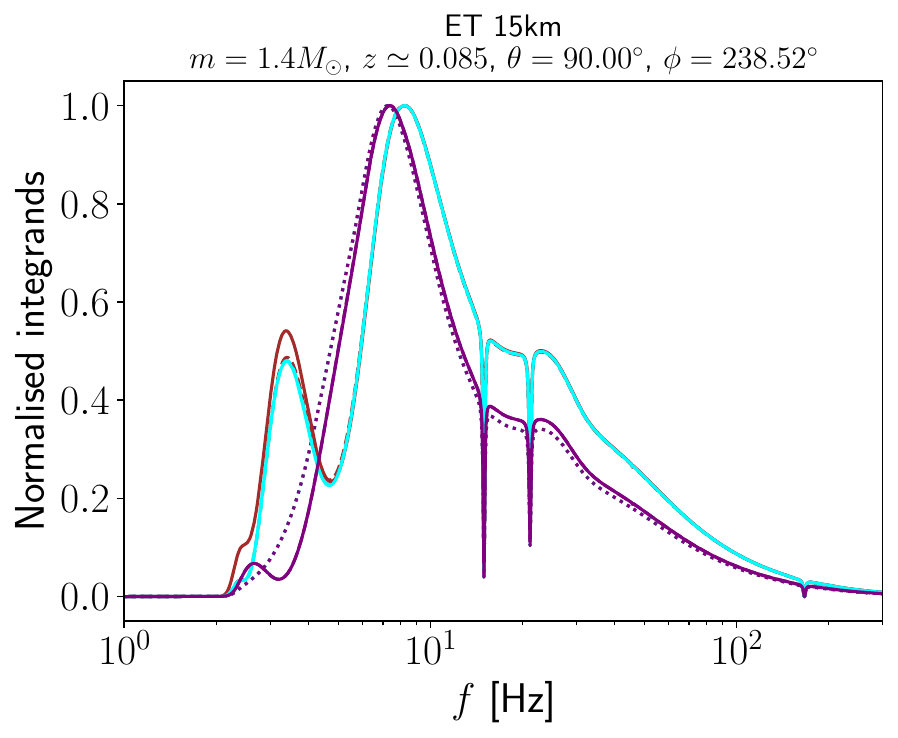}
\caption{The same as the lower two panels in figure~\ref{fig:fisher_ET10} except for the ET sensitivity curve with 15 km arm length.}
\label{fig:fisher_ET15}
\end{figure}

\begin{figure}[tb]
\includegraphics[width=0.4\textwidth]{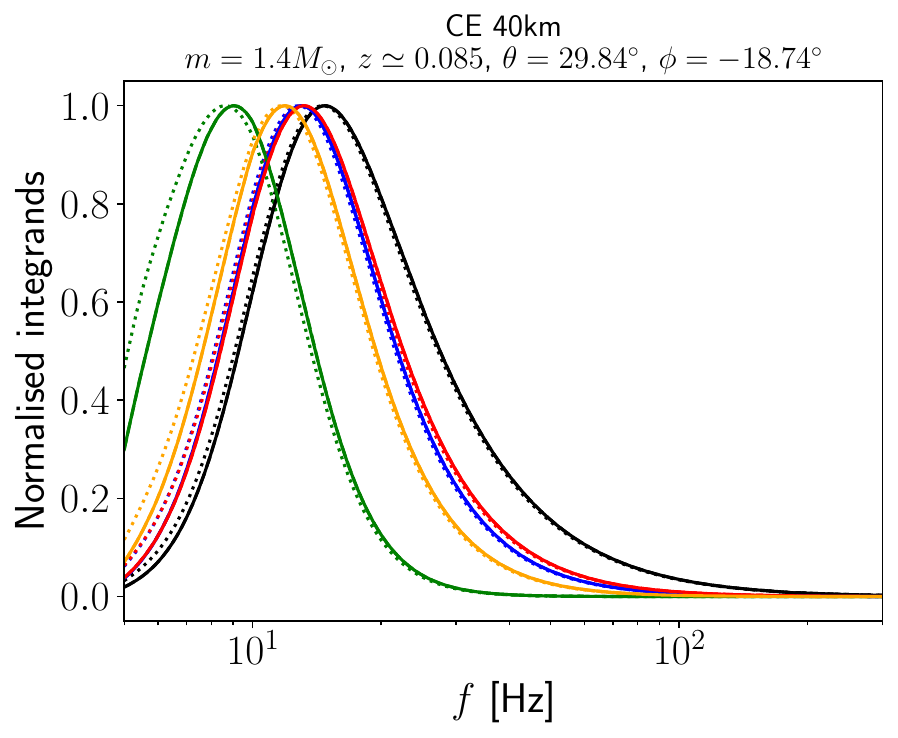}
\includegraphics[width=0.568\textwidth]{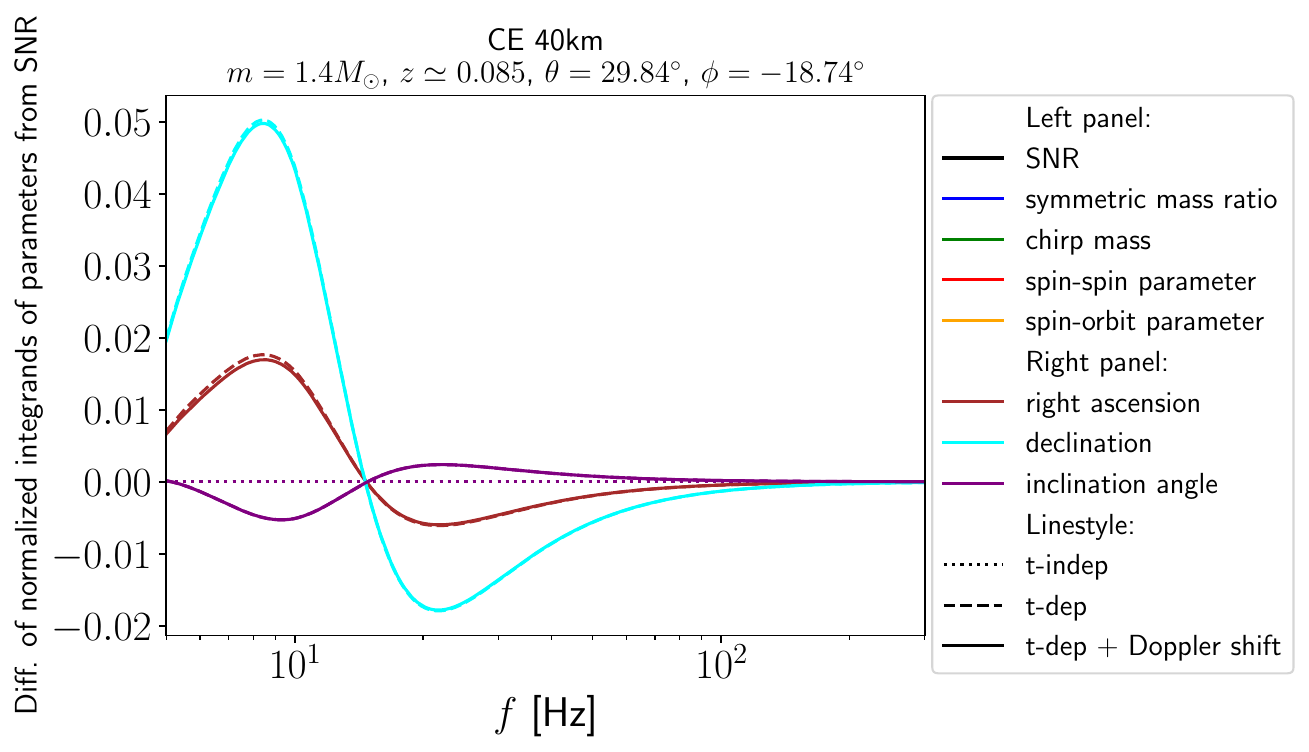}\\
\includegraphics[width=0.4\textwidth]{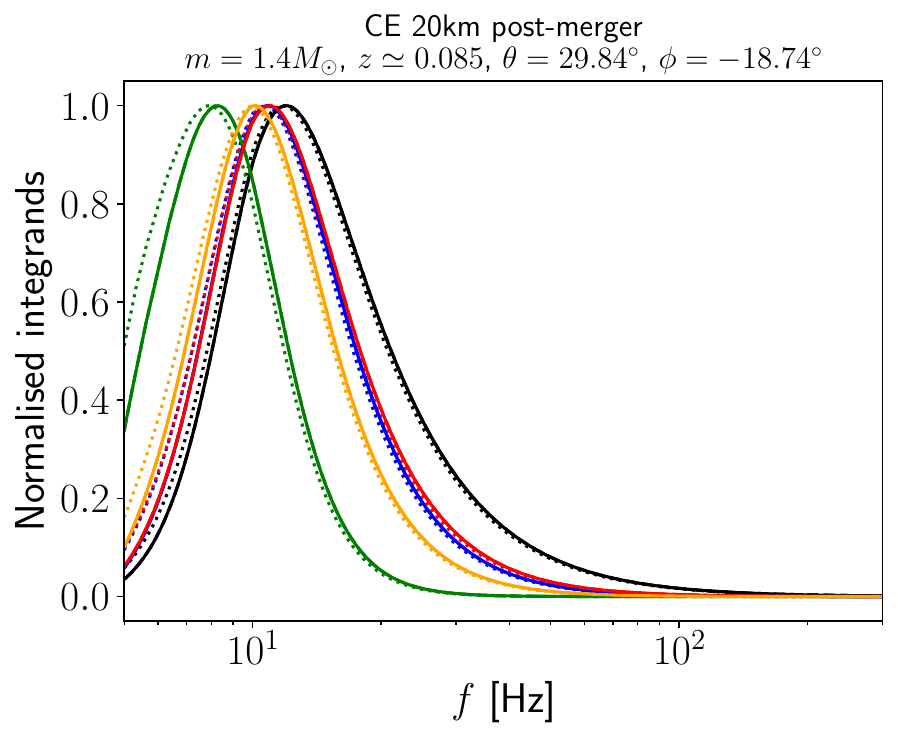}
\includegraphics[width=0.42\textwidth]{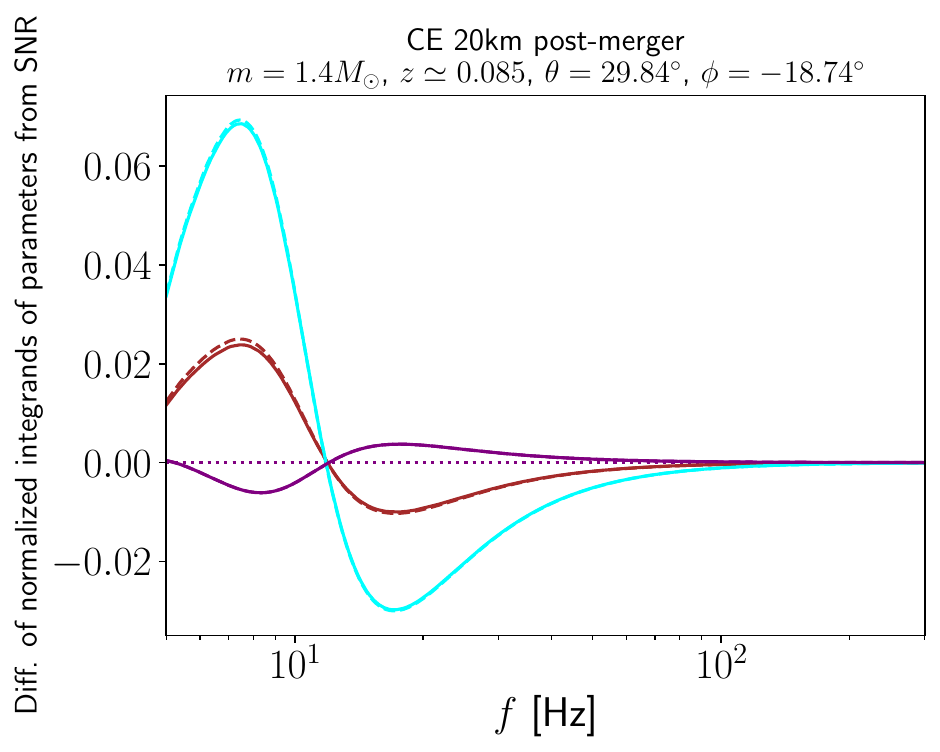}\caption{The same as figure~\ref{fig:fisher_aLIGO} except for a selection of CE sensitivity curves. }
\label{fig:fisher_CE}
\end{figure}

Here we illustrate the effect of the time-dependent response on the accumulation of information about the binary's parameters with frequency, using the normalised integrand of the Fisher matrix components, as in figure~2 in~\cite{Harry:2018hke}. See also figure~3 in~\cite{Damour:2012yf} for an earlier use of these sorts of plots and figures~15 and~18 in~\cite{Grimm:2020ivq} for illustrations of the effect of the time-dependent response on the accumulation of information about extrinsic parameters for BNSs observed by ET, though with an extra scaling with $f$ that we do not include. The Fisher matrix integrand is given by $|(\partial \tilde{h}/\partial \xi_i)|^2/S_n$, where $\xi_i$ is the parameter being considered, and $S_n$ is the detector's noise power spectral density. We use the same setup as in figure~2 in~\cite{Harry:2018hke} viz.\ the TaylorF2 waveform of a non-spinning equal-mass binary BNS system with component masses of $1.4~M_\odot$ in the source frame, tidal deformabilities of $1259.77$, and an inclination angle of $30^{\circ}$~\cite{Hinderer_PC}. The only small difference is that we consider the small cosmological redshift of $z \simeq 0.085$ corresponding to a distance of $400$~Mpc. We consider the same $3.5$PN point particle waveform~\cite{Buonanno:2009zt} as in section~\ref{sec:rot}, and now add the $6$PN tidal terms computed in~\cite{Vines:2011ud}. We also consider the accumulation of information about the leading spin-orbit and spin-spin terms (for aligned-spin systems) computed in~\cite{Kidder:1995zr}.

We first compare the time-dependent and time-independent responses with the A+ design sensitivity noise curve~\cite{Aasi:2013wya} in figure~\ref{fig:fisher_aLIGO}, considering the LIGO Livingston detector. The left panel of figure~\ref{fig:fisher_aLIGO} shows the expected very small differences in the frequency dependence of the accumulation of information between time-dependent and time-independent responses for the SNR and the intrinsic parameters we consider, which are the chirp mass, the symmetric mass ratio, the spin-spin parameter, the spin-orbit parameter and the tidal parameter (see~\cite{Harry:2018hke} for the definitions of these parameters). However, there is a significant difference between our figure and the one in~\cite{Harry:2018hke}, since that figure used the amplitude spectral density of the detector noise in the Fisher matrix integrand instead of the power spectral density by mistake~\cite{Hinderer_PC}. Additionally, while the caption in figure~2 in \cite{Harry:2018hke} mentions that the integrand contains a factor of $1/f$, the plot does not include this factor~\cite{Hinderer_PC}, so we also do not include it. We also included the extrinsic parameters (sky location, and inclination angle) that are not considered in~\cite{Harry:2018hke}, since their accumulation of information now has a different dependence on frequency than the SNR does, due to including the effect of Earth's rotation and orbital motion. (The distance still has the same frequency dependence as the SNR accumulation, so we do not show it.) When making this comparison, we multiply the component masses by the constant Doppler factor at merger [the denominator of equation \eqref{eq:const_dop}] when evaluating the derivatives of the waveform versus extrinsic parameters, since this constant Doppler shift is unobservable, as it is degenerate with a shift in the masses. We give the sky locations in spherical coordinates $(\theta, \phi)$ in the Earth fixed frame defined in~\cite{Anderson:2000yy} at the time of merger. Since the accumulation of information for the extrinsic parameters has a very similar frequency dependence to the SNR accumulation, we plot the differences with respect to the SNR accumulation in the right panel for better visualisation. These differences are zero with the time-independent response. %In fact, the information about the distance accumulates with the same frequency dependence as the SNR even including the time-dependent response, which is why we do not plot it. %This shows that the time-dependent effect also creates slight variation in the extrinsic Fisher matrix components.
We find that the time-dependent response leads to a $\sim 1\%$ difference in the frequency dependence of the accumulation of information about the declination, which is the largest effect for any of the parameters for the system we consider.

We next compare the Fisher matrix integrands for intrinsic and extrinsic parameters with the ET noise curves from~\cite{ET_new} (used in~\cite{Branchesi:2023mws}) for $10$~km and $15$~km arm length in figures~\ref{fig:fisher_ET10} and \ref{fig:fisher_ET15}, respectively. (The $20$~km armlength noise curve in that reference is for an L-shaped detector, rather than the triangular one we have been considering for ET.) We show the result for several different sky locations to illustrate the differences this makes in the accumulation of information. We again use the Virgo location and the orientation from~\cite{LALDetectors} for ET, and specifically consider the `E1' detector %at GPS time of 1152346754
in our analysis (i.e.\ we only consider the response of one pair of arms). The plots follow the same plan as figure~\ref{fig:fisher_aLIGO}. (For these cases, we do not include the tidal parameter, since we are concerned with the effects of the Earth's rotation at low frequencies, where there is negligible information accumulated about the tidal parameter.) However, since the Fisher matrix integrands for extrinsic parameters also show significant differences for time-dependent response, we thus plot them directly in the right panel. We see that the differences between the two responses becomes significant below $\sim 10$~Hz in several cases, and the sky location has a substantial impact on the effect of the time-dependent response. The differences in intrinsic parameters are more significant than the extrinsic parameters, with the largest difference in the chirp mass, which also shows the most significant dependence on the sky location. The Doppler shift does not have a noticeable effect on the accumulation of information about the intrinsic parameters, and only a small difference for the extrinsic parameters, with the largest difference occurring for the $\theta = 90.00^\circ$ case.

In addition, we investigate the Fisher matrix integrands with the baseline $40$~km and $20$~km post-merger CE noise curves from~\cite{CE_new} (the configurations are described in~\cite{Srivastava:2022slt}); we find that the baseline $20$~km and $40$~km low-frequency noise curves give results that are indistinguishable from the baseline $40$~km case. In all cases we use the strain noise curves for a source at $15^\circ$ from normal incidence. While the effects of detector size are small at the low frequencies we consider, there will be some small changes in the accumulation of information about the extrinsic parameters when including the dependence of the detector size effects on them. We leave investigating the size of such changes to future work. Figure~\ref{fig:fisher_CE} shows the Fisher matrix integrands for a sky location with significant differences between the time-dependent and time-independent response with these CE noise curves, with the same setup as figure~\ref{fig:fisher_aLIGO}. The differences are smaller than those with the ET noise curves, but larger than the aLIGO noise curve. Since the CE location is unknown, we adopt the location and orientation of LIGO Livingston for it, for simplicity (see, e.g.\ table~III in~\cite{Borhanian:2020ypi} for some possible locations for CE).

\begin{figure}[tb]
\includegraphics[width=0.4\textwidth]{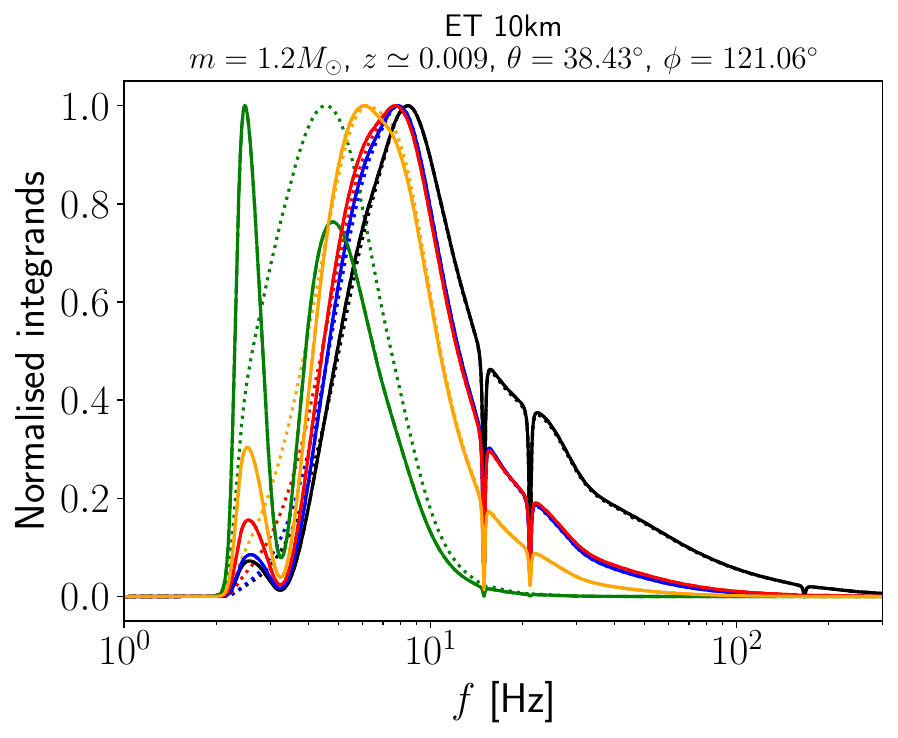}
\includegraphics[width=0.58\textwidth]{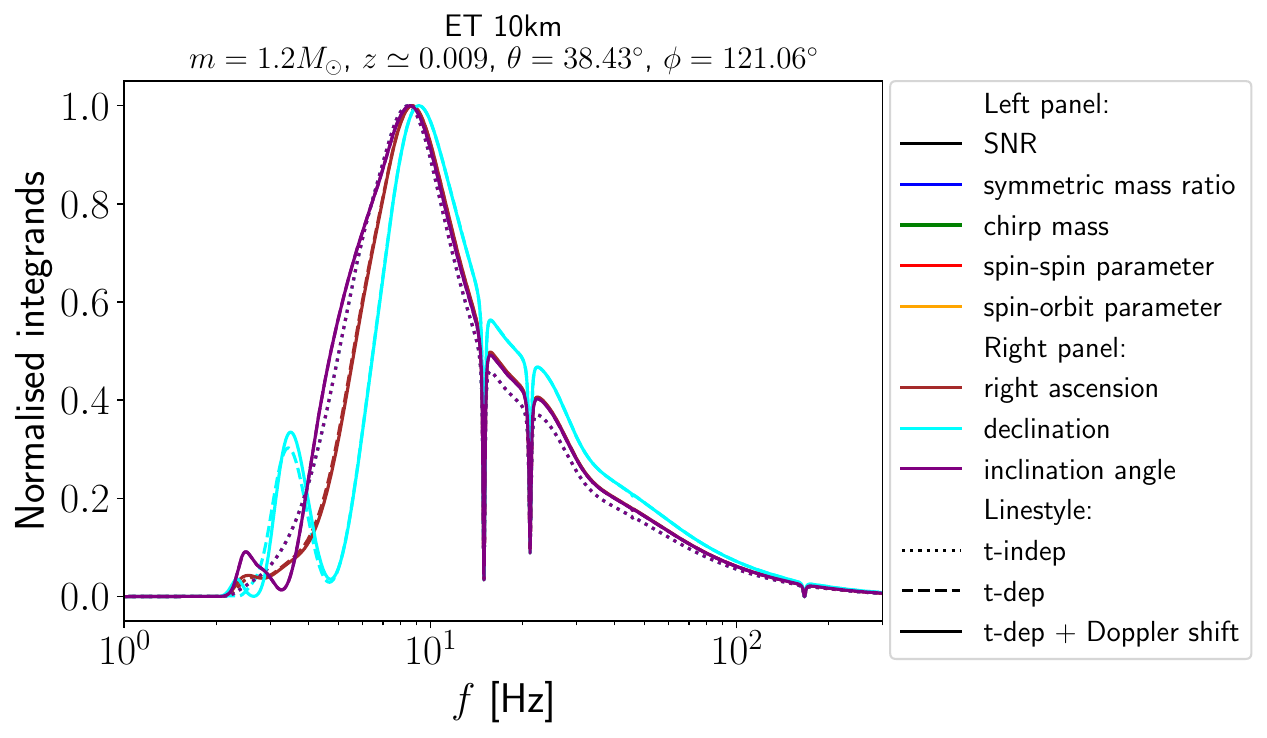}\\
\includegraphics[width=0.4\textwidth]{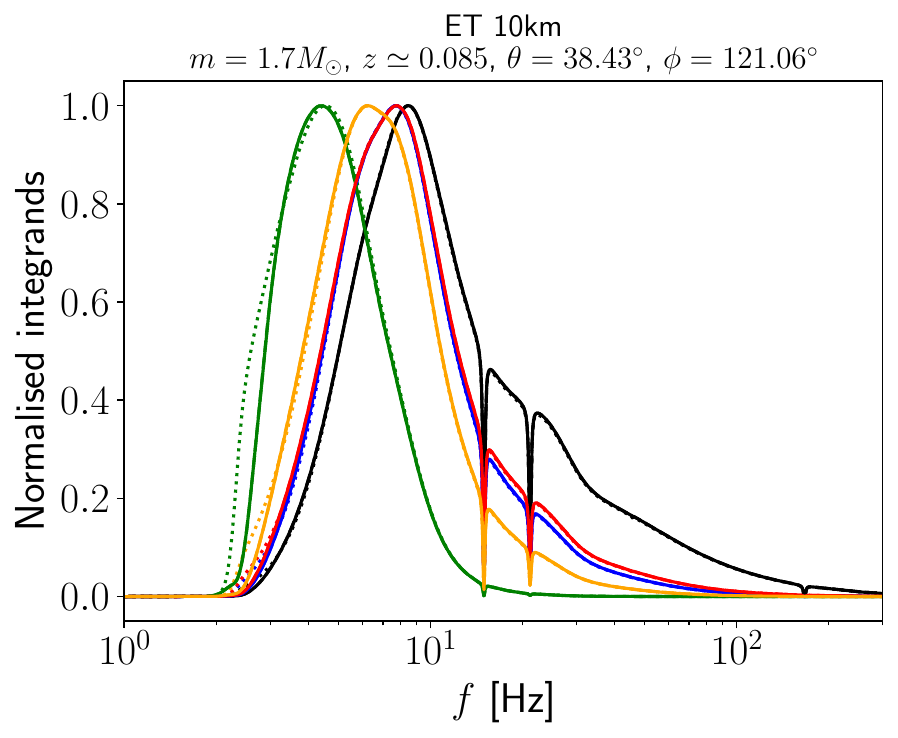}
\includegraphics[width=0.4\textwidth]{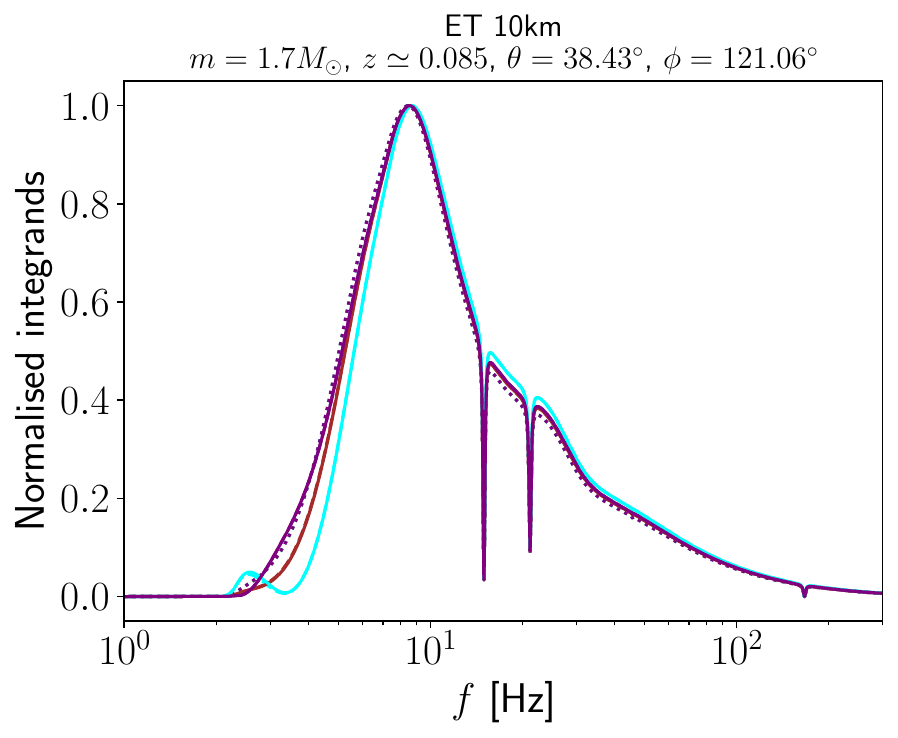}
\caption{The same as the second row in figure~\ref{fig:fisher_ET10} except for a binary with component masses of $1.2~M_\odot$ at $40$~Mpc and a binary with $1.7~M_\odot$ component masses (at $400$~Mpc).}
\label{fig:fisher_ET10_mass}
\end{figure}

Finally, we show the effect of different total masses and distances on the Fisher matrix components for ET in figure~\ref{fig:fisher_ET10_mass}. Specifically, we consider a low-mass BNS with the component mass of $1.2~M_\odot$ at a closer distance of $40$~Mpc ($z\simeq0.009$) and a high-mass BNS system at the same distance of $400$~Mpc considered previously. The sky location is the same as the second row in figure~\ref{fig:fisher_ET10_mass} and all other parameters are the same as in that figure. As expected, the effect of the time-dependent response is larger for the lower-mass system, but is still substantial for the higher-mass case. There are less significant differences in the aLIGO and CE cases for these systems compared to the $1.4~M_\odot$ system shown in figures~\ref{fig:fisher_aLIGO} and~\ref{fig:fisher_CE}, so we do not plot these.

%%%%%%%%%%%%%%%%%
\section{Conclusions and outlook}
\label{sec:concl}
%%%%%%%%%%%%%%%%%

We have explored a Fourier series method to compute the leading effect of the Earth's rotation on the response of ground-based gravitational-wave detectors in the frequency domain and extended it to compute the leading effects of the Doppler shift from the Earth's rotation and orbital motion. We found that the commonly used SPA calculation of these effects in the frequency domain is indeed quite accurate, likely sufficiently accurate for the anticipated applications to data analysis for third-generation detectors, though less accurate than a na{\"\i}ve estimate of the error. However, the SPA (at least with the obvious simple implementation) is an order of magnitude slower than our method. We also obtained an approximate Fourier series expression equivalent to the SPA for the dominant effect of the Earth's rotation as well as the next-order corrections to it, as well as a Fourier series expression for the Doppler shifts due to the Earth's rotation and orbital motion, where we found that the effects from the orbital motion are likely not negligible for parameter estimation of long compact binary signals in third-generation detectors, though they have so far only been included in studies using continuous-wave analysis techniques. Finally, we gave some applications of the method by illustrating the effect of the time-dependent response on the sky dependence of the SNR accumulated at low frequencies and the accumulation of information about various parameters with frequency.

The speed of this method should make including the time dependence of the response feasible as a standard setting in full stochastic sampling studies for future ground-based detectors, or even for performing parameter estimation on subsolar-mass triggers with existing detectors, using their full frequency band. Currently the inclusion of the time-dependent response in full stochastic sampling has only been carried out in~\cite{Smith:2021bqc,Nitz:2021pbr,Baral:2023xst}. We thus plan on implementing the method in the Bilby~\cite{Ashton:2018jfp} open source Bayesian inference package. (We have released a simple Python implementation of the method~\cite{time_dep_response_Git}.) This method can be applied immediately to signals with higher modes, precession, and/or eccentricity, though we leave an explicit application and study of the accuracy of the approximate versions to future work. Additionally, the generalization to scalar and vector polarizations to carry out studies of alternative theories of gravity (see, e.g.~\cite{Takeda:2019gwk}) would be almost immediate, given code to evaluate the detector response to such polarizations, such as exists in LALSuite~\cite{LALSuite} or PyCBC~\cite{PyCBC}. All that one would need to change in the expressions in this paper is the sky location dependence of the Fourier coefficients in \ref{app:resp_coeff_dep}. Finally, similar effects are present for proposed low-frequency gravitational wave detectors on the Moon~\cite{LSGA,Jani:2020gnz,Harms:2020kgl,Amaro-Seoane:2020ahu,Cozzumbo:2023gzs,Ajith:2024mie} and a simple generalization of these expressions would be able to treat such cases.

\paragraph{Acknowledgments} 

We thank Pratyusava Baral, Jolien Creighton, Evan Hall, Ian Harry, Stefan Hild, Tanja Hinderer, Andrew Miller and B~S~Sathyaprakash for useful discussions and comments. A~C acknowledges support from the China Scholarship Council No.~202008060014. Additionally, A~C acknowledges support from the Summer Undergraduate Research Exchange programme of the Department of Physics at the Chinese University of Hong Kong and thanks DAMTP for hospitality during the visit in which he started work on this project.
N~K~J-M acknowledges support from STFC Consolidator Grant No.~ST/L000636/1 and NSF grant AST-2205920. Also, this work has received funding from the European Union's Horizon 2020 research and innovation programme under the Marie Sk{\l}odowska-Curie Grant Agreement No.~690904. This is LIGO document P2400304.

This work made use of the following software: Astropy~\cite{astropy1,astropy2,astropy3}, gwent~\cite{Kaiser:2020tlg}, HEALPix~\cite{2005ApJ...622..759G}, healpy~\cite{Zonca2019}, LALSuite~\cite{LALSuite}, Matplotlib~\cite{Hunter:2007ouj}, NumPy~\cite{Harris:2020xlr}, PyCBC~\cite{PyCBC} and SciPy~\cite{Virtanen:2019joe}.

\appendix

%%%%%%%%%%%%%%%%
\section{Dependence of the Fourier coefficients of the detector response on polarization angle and sky location}
\label{app:resp_coeff_dep}
%%%%%%%%%%%%%%%%

We first consider the dependence of the Fourier coefficients of the detector response [defined in equation~\eqref{eq:Fourier_series}] on the polarization angle $\psi$. As seen in equations~(10,11) of~\cite{Jaranowski:1998qm}, these Fourier coefficients all have the form
\<
a^\polidx_{k\mathrm{c,s}} = a^{\polidx,\mathfrak{c}_\psi}_\gamma(\alpha,\delta)\cos 2\psi + a^{\polidx,\mathfrak{s}_\psi}_\gamma(\alpha,\delta)\sin 2\psi,
\?
$\polidx\in\{+,\times\}$, $\gamma\in\{0,1\mathrm{c},1\mathrm{s},2\mathrm{c},2\mathrm{s}\}$, where we have written the dependence on the right ascension and declination ($\alpha,\delta$) explicitly. Moreover, we have $a^{+,\mathfrak{c}_\psi}_\gamma(\alpha,\delta) = -a^{\times,\mathfrak{s}_\psi}_\gamma(\alpha,\delta)$ and $a^{+,\mathfrak{s}_\psi}_\gamma(\alpha,\delta) = a^{\times,\mathfrak{c}_\psi}_\gamma(\alpha,\delta)$, as expected from the definition of the polarization angle. Thus, one only needs to compute $\mathfrak{a}^+_\gamma(\alpha,\delta) := a^{+,\mathfrak{c}_\psi}_\gamma(\alpha,\delta)$ and $\mathfrak{a}^\times_\gamma(\alpha,\delta) := a^{\times,\mathfrak{c}_\psi}_\gamma(\alpha,\delta)$.

For the dependence on the sky location, one has [equations~(12,13) of~\cite{Jaranowski:1998qm}]
\begin{subequations}
\label{eq:sky_dependence}
\begin{align}
\mathfrak{a}^+_0(\alpha,\delta) &= \mathfrak{a}^{+,\delta}_0\cos^2\delta,\\
\mathfrak{a}^\times_0(\alpha,\delta) &= 0,\\
\mathfrak{a}^+_{1\mathrm{c,s}}(\alpha,\delta) &= (\mathfrak{a}^{+,\mathcal{C}_\alpha}_{1\mathrm{c,s}}\cos\alpha + \mathfrak{a}^{+,\mathcal{S}_\alpha}_{1\mathrm{c,s}}\sin\alpha)\sin 2\delta,\\
\mathfrak{a}^\times_{1\mathrm{c,s}}(\alpha,\delta) &= (\mathfrak{a}^{\times,\mathcal{C}_\alpha}_{1\mathrm{c,s}}\cos\alpha + \mathfrak{a}^{\times,\mathcal{S}_\alpha}_{1\mathrm{c,s}}\sin\alpha)\cos \delta,\\
\mathfrak{a}^+_{2\mathrm{c,s}}(\alpha,\delta) &= (\mathfrak{a}^{+,\mathcal{C}_\alpha}_{2\mathrm{c,s}}\cos 2\alpha + \mathfrak{a}^{+,\mathcal{S}_\alpha}_{2\mathrm{c,s}}\sin 2\alpha)(3 - \cos 2\delta),\\
\mathfrak{a}^\times_{2\mathrm{c,s}}(\alpha,\delta) &= (\mathfrak{a}^{\times,\mathcal{C}_\alpha}_{2\mathrm{c,s}}\cos 2\alpha + \mathfrak{a}^{\times,\mathcal{S}_\alpha}_{2\mathrm{c,s}}\sin 2\alpha)\sin \delta,
\end{align}
\end{subequations}
where the coefficients on the right-hand side only depend on the location, orientation, and angle of the arms of the interferometer. These coefficients can be computed by evaluating the Fourier coefficients at appropriate sky locations e.g.\ for $\mathfrak{a}^{+,(\mathcal{C}_\alpha,\mathcal{S}_\alpha)}_{1\mathrm{c,s}}$, $\alpha\in\{0,\pi/2\}$ and $\delta  = \pi/4$.

As one might expect, the dependence on the sky location for the $k$th order Fourier coefficients is given by the sum of the $\ell = 2$, $m = \pm k$ spin-$(-2)$-weighted spherical harmonics (see, e.g.\ the explicit list in table~B.1 of~\cite{CreightonAnderson}), with the real and imaginary parts giving the plus and cross polarizations, respectively. That is, if one sums the Fourier series over the $\mathfrak{a}_\bullet^{+,\times}$ coefficients, but still leaves the polarization angle dependence given explicitly, to obtain $\mathfrak{a}_{+,\times}$, with
\begin{subequations}
\begin{align}
R_+ &= \mathfrak{a}_+\cos 2\psi + \mathfrak{a}_\times\sin 2\psi,\\
R_\times &= \mathfrak{a}_\times\cos 2\psi  - \mathfrak{a}_+\sin 2\psi,
\end{align}
\end{subequations}
one has
\<
\mathfrak{a}_+ + \ri \mathfrak{a}_\times = \sum_{m = -2}^2\mathfrak{A}_m {}^{-2}Y_{2m}\left(\frac{\pi}{2} - \delta, \alpha - \Phi_\oplus\right).
\?
Here the complex coefficients $\mathfrak{A}_m$ depend only on the location, orientation, and angle of the arms of the interferometer, ${}^{-2}Y_{2 m}$ denotes the $\ell = 2$ spin-$(-2)$-weighted spherical harmonics and $\Phi_\oplus$ denotes the Earth's rotational phase.
Here it is necessary to recall that $\delta = \pi/2 - \theta$, where $\theta$ is the usual polar angle used to define the spin-weighted spherical harmonics. %There is thus a relation between the plus and cross coefficients in equations~\eqref{eq:sky_dependence} for a given $k$, but we do not write this down explicitly, since it is not clear how useful it is for applications.
The statement about the $k$th order Fourier coefficients above follows because the dependence of ${}^{-2}Y_{\ell m}(\theta,\phi)$ on $\phi$ is $e^{m\ri\phi}$.
However, the explicit expressions in equations~\eqref{eq:sky_dependence} are likely more useful for applications. See~\cite{1988MNRAS.234..663D} for other expressions relating the detector response to spin-weighted spherical harmonics.

%%%%%%%%%%%%%%%%
\section{Second-order expressions for the Fourier series approximation to Earth's orbital motion}
\label{app:Doppler_2nd_order}
%%%%%%%%%%%%%%%%

Here we extend the Fourier series describing the Earth's orbital motion to second order, to illustrate that it indeed gives an improvement in accuracy. Specifically, we take
\begin{subequations}
\label{eq:xdn_2nd}
\begin{align}
\nonumber
\hat{n}\cdot\vec{x}_{\text{SSB}\to\text{EMB}}(t) &\simeq \hat{n}\cdot\vec{x}_{\text{SSB}\to\text{EMB}}(t_\text{mid}) + b_{1\text{c}}^\text{EMB}\{\cos[\Omega_\text{EMB}(t - t_\text{mid})] - 1\} + b_{1\text{s}}^\text{EMB}\sin[\Omega_\text{EMB}(t - t_\text{mid})]\\
&\quad + b_{2\text{c}}^\text{EMB}\{\cos[2\Omega_\text{EMB}(t - t_\text{mid})] - 1\} + b_{2\text{s}}^\text{EMB}\sin[2\Omega_\text{EMB}(t - t_\text{mid})],\\
\nonumber
\hat{n}\cdot\vec{x}_{\text{EMB}\to\oplus}(t) &\simeq \hat{n}\cdot\vec{x}_{\text{EMB}\to\oplus}(t_\text{mid}) + b_{1\text{c}}^\text{EM}\{\cos[\Omega_\text{EM}(t - t_\text{mid})] - 1\} + b_{1\text{s}}^\text{EM}\sin[\Omega_\text{EM}(t - t_\text{mid})]\\
&\quad + b_{2\text{c}}^\text{EM}\{\cos[2\Omega_\text{EM}(t - t_\text{mid})] - 1\} + b_{2\text{s}}^\text{EM}\sin[2\Omega_\text{EM}(t - t_\text{mid})].
\end{align}
\end{subequations}
We obtain these coefficients using the same sort of collocation method as for the first order ones, now also evaluating at
\begin{subequations}
\begin{align}
d^{\text{EMB}, \pm/2} &:= \hat{n}\cdot\vec{x}_{\text{SSB}\to\text{EMB}}(t_\text{mid} \pm \Delta t/2) - \hat{n}\cdot\vec{x}_{\text{SSB}\to\text{EMB}}(t_\text{mid}),\\
d^{\text{EM}, \pm/2} &:= \hat{n}\cdot\vec{x}_{\text{EMB}\to\oplus}(t_\text{mid} \pm \Delta t/2) - \hat{n}\cdot\vec{x}_{\text{EMB}\to\oplus}(t_\text{mid}).
\end{align}
\end{subequations}
This gives (neglecting uncontrolled remainders)
\begin{subequations}
\begin{align}
b_{1\text{c}}^\bullet &= \frac{2(d^{\bullet, +/2} + d^{\bullet, -/2})[1 + \cos(\Omega_\bullet\Delta t)] - d^{\bullet, +} - d^{\bullet, -}}{16\sin^4(\Omega_\bullet\Delta t/4)[1 + 2\cos(\Omega_\bullet\Delta t/2)]},\\
b_{1\text{s}}^\bullet &= \frac{2(-d^{\bullet, +/2} + d^{\bullet, -/2})\cos(\Omega_\bullet\Delta t) + d^{\bullet, +} - d^{\bullet, -}}{8\sin^2(\Omega_\bullet\Delta t/4)\sin(\Omega_\bullet\Delta t/2)[1 + 2\cos(\Omega_\bullet\Delta t/2)]},\\
b_{2\text{c}}^\bullet &= \frac{-2(d^{\bullet, +/2} + d^{\bullet, -/2})[1 + \cos(\Omega_\bullet\Delta t/2)] + d^{\bullet, +} + d^{\bullet, -}}{16\sin^2(\Omega_\bullet\Delta t/4)\sin^2(\Omega_\bullet\Delta t/2)[1 + 2\cos(\Omega_\bullet\Delta t/2)]},\\
b_{2\text{s}}^\bullet &= \frac{2(d^{\bullet, +/2} - d^{\bullet, -/2})\cos(\Omega_\bullet\Delta t/2) - d^{\bullet, +} + d^{\bullet, -}}{8\sin^2(\Omega_\bullet\Delta t/4)\sin(\Omega_\bullet\Delta t)[1 + 2\cos(\Omega_\bullet\Delta t/2)]},
\end{align}
\end{subequations}
where $\bullet$ is either EMB or EM. Here we have to restrict $\Delta t$ to be at least a half a day in the EMB calculation to prevent numerical errors due to small denominators. In the $\oplus$ calculation, we restrict $\Delta t$ to be at least a fifth of a day, as in the first order version. The remaining first order expressions generalize immediately. (The versions of $b_{1\text{s}}^\bullet$ and $b_{2\text{s}}^\bullet$ coded in Python in~\cite{time_dep_response_Git} use equivalent forms obtained using trigonometric identities. The forms given here are chosen to show the similarities of the expressions.)

\bigskip

\bibliographystyle{iopart-num}
\bibliography{time_dependent_response}

\providecommand{\newblock}{}
\begin{thebibliography}{100}
\expandafter\ifx\csname url\endcsname\relax
  \def\url#1{{\tt #1}}\fi
\expandafter\ifx\csname urlprefix\endcsname\relax\def\urlprefix{URL }\fi
\providecommand{\eprint}[2][]{\url{#2}}

\bibitem{Hild:2010id}
Hild S {\em et~al.\/} 2011 {\em Class. Quantum Grav.\/} {\bf 28} 094013
  (\textit{Preprint} \eprint{1012.0908})

\bibitem{ET_design}
Abernathy M {\em et~al.\/} (ET Science Team) Einstein gravitational wave
  telescope conceptual design study
  \url{http://www.et-gw.eu/index.php/etdsdocument}

\bibitem{Reitze:2019iox}
Reitze D {\em et~al.\/} 2019 {\em Bull. Am. Astron. Soc.\/} {\bf 51} 035
  (\textit{Preprint} \eprint{1907.04833})

\bibitem{Evans:2021gyd}
Evans M {\em et~al.\/} 2021  (\textit{Preprint} \eprint{2109.09882})

\bibitem{Meacher:2015rex}
Meacher D, Cannon K, Hanna C, Regimbau T and Sathyaprakash B~S 2016 {\em Phys.
  Rev. D\/} {\bf 93} 024018 (\textit{Preprint} \eprint{1511.01592})

\bibitem{Samajdar:2021egv}
Samajdar A, Janquart J, Van Den~Broeck C and Dietrich T 2021 {\em Phys. Rev.
  D\/} {\bf 104} 044003 (\textit{Preprint} \eprint{2102.07544})

\bibitem{Pizzati:2021apa}
Pizzati E, Sachdev S, Gupta A and Sathyaprakash B~S 2022 {\em Phys. Rev. D\/}
  {\bf 105} 104016 (\textit{Preprint} \eprint{2102.07692})

\bibitem{Himemoto:2021ukb}
Himemoto Y, Nishizawa A and Taruya A 2021 {\em Phys. Rev. D\/} {\bf 104} 044010
  (\textit{Preprint} \eprint{2103.14816})

\bibitem{Relton:2021cax}
Relton P and Raymond V 2021 {\em Phys. Rev. D\/} {\bf 104} 084039
  (\textit{Preprint} \eprint{2103.16225})

\bibitem{Antonelli:2021vwg}
Antonelli A, Burke O and Gair J~R 2021 {\em Mon. Not. R. Astron. Soc.\/} {\bf
  507} 5069 (\textit{Preprint} \eprint{2104.01897})

\bibitem{Relton:2022whr}
Relton P, Virtuoso A, Bini S, Raymond V, Harry I, Drago M, Lazzaro C, Miani A
  and Tiwari S 2022 {\em Phys. Rev. D\/} {\bf 106} 104045 (\textit{Preprint}
  \eprint{2208.00261})

\bibitem{Hu:2022bji}
Hu Q and Veitch J 2023 {\em Astrophys. J.\/} {\bf 945} 103 (\textit{Preprint}
  \eprint{2210.04769})

\bibitem{Langendorff:2022fzq}
Langendorff J, Kolmus A, Janquart J and Van Den~Broeck C 2023 {\em Phys. Rev.
  Lett.\/} {\bf 130} 171402 (\textit{Preprint} \eprint{2211.15097})

\bibitem{Janquart:2023hew}
Janquart J, Baka T, Samajdar A, Dietrich T and Van Den~Broeck C 2023 {\em Mon.
  Not. R. Astron. Soc.\/} {\bf 523} 1699 (\textit{Preprint}
  \eprint{2211.01304})

\bibitem{Wang:2023ldq}
Wang Z, Liang D, Zhao J, Liu C and Shao L 2024 {\em Class. Quantum Grav.\/}
  {\bf 41} 055011 (\textit{Preprint} \eprint{2304.06734})

\bibitem{Alvey:2023naa}
Alvey J, Bhardwaj U, Nissanke S and Weniger C 2023  (\textit{Preprint}
  \eprint{2308.06318})

\bibitem{Dang:2023xkj}
Dang Y, Wang Z, Liang D and Shao L 2024 {\em Astrophys. J.\/} {\bf 964} 194
  (\textit{Preprint} \eprint{2311.16184})

\bibitem{Johnson:2024foj}
Johnson A~D, Chatziioannou K and Farr W~M 2024 {\em Phys. Rev. D\/} {\bf 109}
  084015 (\textit{Preprint} \eprint{2402.06836})

\bibitem{Cireddu:2023ssf}
Cireddu F, Wils M, Wong I~C~F, Pang P~T~H, Li T~G~F and Del~Pozzo W 2023
  (\textit{Preprint} \eprint{2312.14614})

\bibitem{Wong:2024hes}
Wong I~C~F, Pang P~T~H, Wils M, Cireddu F, Del~Pozzo W and Li T~G~F 2024
  (\textit{Preprint} \eprint{2407.08728})

\bibitem{Smith:2021bqc}
Smith R {\em et~al.\/} 2021 {\em Phys. Rev. Lett.\/} {\bf 127} 081102
  (\textit{Preprint} \eprint{2103.12274})

\bibitem{Essick:2017wyl}
Essick R, Vitale S and Evans M 2017 {\em Phys. Rev. D\/} {\bf 96} 084004
  (\textit{Preprint} \eprint{1708.06843})

\bibitem{Hall:2020dps}
Hall E~D {\em et~al.\/} 2021 {\em Phys. Rev. D\/} {\bf 103} 122004
  (\textit{Preprint} \eprint{2012.03608})

\bibitem{Harms:2022jth}
Harms J, Naticchioni L, Calloni E, De~Rosa R, Ricci F and D'Urso D 2022 {\em
  Eur. Phys. J. Plus\/} {\bf 137} 687 (\textit{Preprint} \eprint{2202.12841})

\bibitem{McClelland:2021wqy}
McClelland D {\em et~al.\/} 2021  (\textit{Preprint} \eprint{2111.06991})

\bibitem{Baral:2023xst}
Baral P, Morisaki S, Maga\~na Hernandez I and Creighton J~D~E 2023 {\em Phys.
  Rev. D\/} {\bf 108} 043010 (\textit{Preprint} \eprint{2304.09889})

\bibitem{Chan:2018csa}
Chan M~L, Messenger C, Heng I~S and Hendry M 2018 {\em Phys. Rev. D\/} {\bf 97}
  123014 (\textit{Preprint} \eprint{1803.09680})

\bibitem{Grimm:2020ivq}
Grimm S and Harms J 2020 {\em Phys. Rev. D\/} {\bf 102} 022007
  (\textit{Preprint} \eprint{2004.01434})

\bibitem{Singh:2021bwn}
Singh N and Bulik T 2022 {\em Phys. Rev. D\/} {\bf 106} 123014
  (\textit{Preprint} \eprint{2107.11198})

\bibitem{Li:2021mbo}
Li Y, Heng I~S, Chan M~L, Messenger C and Fan X 2022 {\em Phys. Rev. D\/} {\bf
  105} 043010 (\textit{Preprint} \eprint{2109.07389})

\bibitem{Singh:2021zah}
Singh N, Bulik T, Belczynski K and Askar A 2022 {\em Astron. Astrophys.\/} {\bf
  667} A2 (\textit{Preprint} \eprint{2112.04058})

\bibitem{Hu:2023hos}
Hu Q and Veitch J 2023 {\em Astrophys. J. Lett.\/} {\bf 958} L43
  (\textit{Preprint} \eprint{2309.00970})

\bibitem{Zhao_and_Wen}
{Zhao} W and {Wen} L 2018 {\em Phys. Rev. D\/} {\bf 97} 064031
  (\textit{Preprint} \eprint{1710.05325})

\bibitem{Nishizawa:2019rra}
Nishizawa A and Arai S 2019 {\em Phys. Rev. D\/} {\bf 99} 104038
  (\textit{Preprint} \eprint{1901.08249})

\bibitem{Takeda:2019gwk}
Takeda H, Nishizawa A, Nagano K, Michimura Y, Komori K, Ando M and Hayama K
  2019 {\em Phys. Rev. D\/} {\bf 100} 042001 (\textit{Preprint}
  \eprint{1904.09989})

\bibitem{Tsutsui:2020bem}
Tsutsui T, Nishizawa A and Morisaki S 2021 {\em Phys. Rev. D\/} {\bf 104}
  064013 (\textit{Preprint} \eprint{2011.06130})

\bibitem{Borhanian:2020ypi}
Borhanian S 2021 {\em Class. Quantum Grav.\/} {\bf 38} 175014
  (\textit{Preprint} \eprint{2010.15202})

\bibitem{Nitz:2021pbr}
Nitz A~H and Dal~Canton T 2021 {\em Astrophys. J. Lett.\/} {\bf 917} L27
  (\textit{Preprint} \eprint{2106.15259})

\bibitem{Tsutsui:2021izf}
Tsutsui T, Nishizawa A and Morisaki S 2022 {\em Mon. Not. R. Astron. Soc.\/}
  {\bf 512} 3878 (\textit{Preprint} \eprint{2107.12531})

\bibitem{Liu:2021dcr}
Liu C and Shao L 2022 {\em Astrophys. J.\/} {\bf 926} 158 (\textit{Preprint}
  \eprint{2108.08490})

\bibitem{Borhanian:2022czq}
Borhanian S and Sathyaprakash B~S 2022  (\textit{Preprint} \eprint{2202.11048})

\bibitem{Dupletsa:2022scg}
Dupletsa U, Harms J, Banerjee B, Branchesi M, Goncharov B, Maselli A, Oliveira
  A~C~S, Ronchini S and Tissino J 2023 {\em Astron. Comput.\/} {\bf 42} 100671
  (\textit{Preprint} \eprint{2205.02499})

\bibitem{Iacovelli:2022bbs}
Iacovelli F, Mancarella M, Foffa S and Maggiore M 2022 {\em Astrophys. J.\/}
  {\bf 941} 208 (\textit{Preprint} \eprint{2207.02771})

\bibitem{Zhou:2022nmt}
Zhou B, Reali L, Berti E, \c{C}al\i{}\c{s}kan M, Creque-Sarbinowski C,
  Kamionkowski M and Sathyaprakash B~S 2023 {\em Phys. Rev. D\/} {\bf 108}
  064040 (\textit{Preprint} \eprint{2209.01310})

\bibitem{Reali:2022aps}
Reali L, Antonelli A, Cotesta R, Borhanian S, \c{C}al\i{}\c{s}kan M, Berti E
  and Sathyaprakash B~S 2022  (\textit{Preprint} \eprint{2209.13452})

\bibitem{Singh:2023cxn}
Singh N, Bulik T, Belczynski K, Cieslar M and Calore F 2024 {\em Astron.
  Astrophys.\/} {\bf 681} A56 (\textit{Preprint} \eprint{2304.01341})

\bibitem{Gardner:2023znk}
Gardner J~W, Sun L, Borhanian S, Lasky P~D, Thrane E, McClelland D~E and
  Slagmolen B~J~J 2023 {\em Phys. Rev. D\/} {\bf 108} 123026 (\textit{Preprint}
  \eprint{2308.13103})

\bibitem{Yang:2023zxk}
Yang T, Cai R~G, Cao Z and Lee H~M 2024 {\em Phys. Rev. D\/} {\bf 109} 104041
  (\textit{Preprint} \eprint{2310.08160})

\bibitem{LIGOScientific:2014pky}
Aasi J {\em et~al.\/} (LIGO Scientific Collaboration) 2015 {\em Class. Quantum
  Grav.\/} {\bf 32} 074001 (\textit{Preprint} \eprint{1411.4547})

\bibitem{VIRGO:2014yos}
Acernese F {\em et~al.\/} (Virgo Collaboration) 2015 {\em Class. Quantum
  Grav.\/} {\bf 32} 024001 (\textit{Preprint} \eprint{1408.3978})

\bibitem{Abbott:2019qbw}
Abbott B~P {\em et~al.\/} (LIGO Scientific Collaboration and Virgo
  Collaboration) 2019 {\em Phys. Rev. Lett.\/} {\bf 123} 161102
  (\textit{Preprint} \eprint{1904.08976})

\bibitem{Nitz:2020bdb}
Nitz A~H and Wang Y~F 2021 {\em Phys. Rev. Lett.\/} {\bf 126} 021103
  (\textit{Preprint} \eprint{2007.03583})

\bibitem{Nitz:2021mzz}
Nitz A~H and Wang Y~F 2021 {\em Astrophys. J.\/} {\bf 915} 54
  (\textit{Preprint} \eprint{2102.00868})

\bibitem{Phukon:2021cus}
Phukon K~S, Baltus G, Caudill S, Clesse S, Depasse A, Fays M, Fong H, Kapadia
  S~J, Magee R and Tanasijczuk A~J 2021  (\textit{Preprint}
  \eprint{2105.11449})

\bibitem{Nitz:2021vqh}
Nitz A~H and Wang Y~F 2021 {\em Phys. Rev. Lett.\/} {\bf 127} 151101
  (\textit{Preprint} \eprint{2106.08979})

\bibitem{LIGOScientific:2021job}
Abbott R {\em et~al.\/} (LIGO Scientific Collaboration, Virgo Collaboration,
  and KAGRA Collaboration) 2022 {\em Phys. Rev. Lett.\/} {\bf 129} 061104
  (\textit{Preprint} \eprint{2109.12197})

\bibitem{Nitz:2022ltl}
Nitz A~H and Wang Y~F 2022 {\em Phys. Rev. D\/} {\bf 106} 023024
  (\textit{Preprint} \eprint{2202.11024})

\bibitem{LIGOScientific:2022hai}
Abbott R {\em et~al.\/} (LIGO Scientific Collaboration, Virgo Collaboration,
  and KAGRA Collaboration) 2023 {\em Mon. Not. R. Astron. Soc.\/} {\bf 524}
  5984 (\textit{Preprint} \eprint{2212.01477})

\bibitem{Prunier:2023cyv}
Prunier M, Morr\'as G, Siles J~F~N, Clesse S, Garc\'\i{}a-Bellido J and
  Ruiz~Morales E 2023  (\textit{Preprint} \eprint{2311.16085})

\bibitem{Magee:2018opb}
Magee R, Deutsch A~S, McClincy P, Hanna C, Horst C, Meacher D, Messick C,
  Shandera S and Wade M 2018 {\em Phys. Rev. D\/} {\bf 98} 103024
  (\textit{Preprint} \eprint{1808.04772})

\bibitem{Bandopadhyay:2022tbi}
Bandopadhyay A, Reed B, Padamata S, Leon E, Horowitz C~J, Brown D~A, Radice D,
  Fattoyev F~J and Piekarewicz J 2023 {\em Phys. Rev. D\/} {\bf 107} 103012
  (\textit{Preprint} \eprint{2212.03855})

\bibitem{Wolfe:2023yuu}
Wolfe N~E, Vitale S and Talbot C 2023 {\em J. Cosmol. Astropart. Phys.\/} {\bf
  11} 039 (\textit{Preprint} \eprint{2305.19907})

\bibitem{Morras:2023jvb}
Morr\'as G {\em et~al.\/} 2023 {\em Phys. Dark Univ.\/} {\bf 42} 101285
  (\textit{Preprint} \eprint{2301.11619})

\bibitem{Miller:2020kmv}
Miller A~L, Clesse S, De~Lillo F, Bruno G, Depasse A and Tanasijczuk A 2021
  {\em Phys. Dark Univ.\/} {\bf 32} 100836 (\textit{Preprint}
  \eprint{2012.12983})

\bibitem{Miller:2021knj}
Miller A~L, Aggarwal N, Clesse S and De~Lillo F 2022 {\em Phys. Rev. D\/} {\bf
  105} 062008 (\textit{Preprint} \eprint{2110.06188})

\bibitem{KAGRA:2022dwb}
Abbott R {\em et~al.\/} (LIGO Scientific Collaboration, Virgo Collaboration,
  and KAGRA Collaboration) 2022 {\em Phys. Rev. D\/} {\bf 106} 102008
  (\textit{Preprint} \eprint{2201.00697})

\bibitem{Miller:2023rnn}
Miller A~L, Singh N and Palomba C 2024 {\em Phys. Rev. D\/} {\bf 109} 043021
  (\textit{Preprint} \eprint{2309.15808})

\bibitem{Miller:2024fpo}
Miller A~L, Aggarwal N, Clesse S, De~Lillo F, Sachdev S, Astone P, Palomba C,
  Piccinni O~J and Pierini L 2024  (\textit{Preprint} \eprint{2402.19468})

\bibitem{Zhu:2017ujz}
Zhu S~J, Papa M~A and Walsh S 2017 {\em Phys. Rev. D\/} {\bf 96} 124007
  (\textit{Preprint} \eprint{1707.05268})

\bibitem{DAntonio:2021cfv}
D'Antonio S, Palomba C, Frasca S, Astone P, La~Rosa I, Leaci P, Mastrogiovanni
  S, Piccinni O~J, Pierini L and Rei L 2021 {\em Phys. Rev. D\/} {\bf 103}
  063030

\bibitem{Valluri:2020cqe}
Valluri S~R, Dergachev V, Zhang X and Chishtie F~A 2021 {\em Phys. Rev. D\/}
  {\bf 104} 024065 (\textit{Preprint} \eprint{2010.13647})

\bibitem{Marsat:2018oam}
Marsat S and Baker J~G 2018  (\textit{Preprint} \eprint{1806.10734})

\bibitem{Chen:2020fzm}
Chen A, Johnson-McDaniel N~K, Dietrich T and Dudi R 2020 {\em Phys. Rev. D\/}
  {\bf 101} 103008 (\textit{Preprint} \eprint{2001.11470})

\bibitem{Harry:2018hke}
Harry I and Hinderer T 2018 {\em Class. Quantum Grav.\/} {\bf 35} 145010
  (\textit{Preprint} \eprint{1801.09972})

\bibitem{time_dep_response_Git}
{\url{https://git.ligo.org/anson.chen/time_dependent_response}}

\bibitem{Ashton:2018jfp}
Ashton G {\em et~al.\/} 2019 {\em Astrophys. J. Suppl.\/} {\bf 241} 27
  (\textit{Preprint} \eprint{1811.02042})

\bibitem{Kaplan:2006nv}
{Kaplan} G~H 2005 {\em U.S. Naval Observatory Circulars\/} {\bf 179}
  (\textit{Preprint} \eprint{astro-ph/0602086})

\bibitem{Astone:2010zz}
Astone P, D'Antonio S, Frasca S and Palomba C 2010 {\em Class. Quantum Grav.\/}
  {\bf 27} 194016

\bibitem{Jaranowski:1998qm}
Jaranowski P, Kr{\'o}lak A and Schutz B~F 1998 {\em Phys. Rev. D\/} {\bf 58}
  063001 (\textit{Preprint} \eprint{gr-qc/9804014})

\bibitem{LALSuite}
{LIGO--Virgo--KAGRA Algorithm Library Suite (LALSuite),
  \url{https://doi.org/10.7935/GT1W-FZ16}}

\bibitem{PyCBC}
Nitz A~H {\em et~al.\/} {PyCBC software}
  \urlprefix\url{https://doi.org/10.5281/zenodo.596388}

\bibitem{Buonanno:2009zt}
Buonanno A, Iyer B~R, Ochsner E, Pan Y and Sathyaprakash B~S 2009 {\em Phys.
  Rev. D\/} {\bf 80} 084043 (\textit{Preprint} \eprint{0907.0700})

\bibitem{Dietrich:2017aum}
Dietrich T, Bernuzzi S and Tichy W 2017 {\em Phys. Rev. D\/} {\bf 96} 121501(R)
  (\textit{Preprint} \eprint{1706.02969})

\bibitem{Dietrich:2018uni}
Dietrich T {\em et~al.\/} 2019 {\em Phys. Rev. D\/} {\bf 99} 024029
  (\textit{Preprint} \eprint{1804.02235})

\bibitem{Khan:2015jqa}
Khan S, Husa S, Hannam M, Ohme F, P\"urrer M, Jim\'enez~Forteza X and Boh\'e A
  2016 {\em Phys. Rev. D\/} {\bf 93} 044007 (\textit{Preprint}
  \eprint{1508.07253})

\bibitem{Kaiser:2020tlg}
Kaiser A~R and McWilliams S~T 2021 {\em Class. Quantum Grav.\/} {\bf 38} 055009
  (\textit{Preprint} \eprint{2010.02135})

\bibitem{Anderson:2000yy}
Anderson W~G, Brady P~R, Creighton J~D and Flanagan {\'E}~{\'E} 2001 {\em Phys.
  Rev. D\/} {\bf 63} 042003 (\textit{Preprint} \eprint{gr-qc/0008066})

\bibitem{Aghanim:2018eyx}
Aghanim N {\em et~al.\/} (Planck Collaboration) 2020 {\em Astron. Astrophys.\/}
  {\bf 641} A6 (\textit{Preprint} \eprint{1807.06209})

\bibitem{LALDetectors}
\url{https://git.ligo.org/lscsoft/lalsuite/blob/master/lal/lib/tools/LALDetectors.h}

\bibitem{Vinciguerra:2017ngf}
Vinciguerra S, Veitch J and Mandel I 2017 {\em Class. Quantum Grav.\/} {\bf 34}
  115006 (\textit{Preprint} \eprint{1703.02062})

\bibitem{Garcia-Quiros:2020qlt}
Garc\'\i{}a-Quir\'os C, Husa S, Mateu-Lucena M and Borchers A 2021 {\em Class.
  Quantum Grav.\/} {\bf 38} 015006 (\textit{Preprint} \eprint{2001.10897})

\bibitem{Morisaki:2021ngj}
Morisaki S 2021 {\em Phys. Rev. D\/} {\bf 104} 044062 (\textit{Preprint}
  \eprint{2104.07813})

\bibitem{1994A&A...282..663S}
{Simon} J~L, {Bretagnon} P, {Chapront} J, {Chapront-Touz{\'e}} M, {Francou} G
  and {Laskar} J 1994 {\em Astron. Astrophys.\/} {\bf 282} 663

\bibitem{2002A&A...387..700C}
{Chapront} J, {Chapront-Touz{\'e}} M and {Francou} G 2002 {\em Astron.
  Astrophys.\/} {\bf 387} 700 see also
  \url{https://hpiers.obspm.fr/eop-pc/models/constants.html}

\bibitem{astropy1}
Robitaille T~P {\em et~al.\/} (Astropy Collaboration) 2013 {\em Astron.
  Astrophys.\/} {\bf 558} A33 (\textit{Preprint} \eprint{1307.6212})

\bibitem{astropy2}
Price-Whelan A~M {\em et~al.\/} (Astropy Collaboration) 2018 {\em Astron. J.\/}
  {\bf 156} 123 (\textit{Preprint} \eprint{1801.02634})

\bibitem{astropy3}
Price-Whelan A~M {\em et~al.\/} (Astropy Collaboration) 2022 {\em Astrophys.
  J.\/} {\bf 935} 167 (\textit{Preprint} \eprint{2206.14220})

\bibitem{Cutler:1994ys}
Cutler C and Flanagan {\'E}~{\'E} 1994 {\em Phys. Rev. D\/} {\bf 49} 2658
  (\textit{Preprint} \eprint{gr-qc/9402014})

\bibitem{Chatziioannou:2017tdw}
Chatziioannou K, Klein A, Yunes N and Cornish N 2017 {\em Phys. Rev. D\/} {\bf
  95} 104004 (\textit{Preprint} \eprint{1703.03967})

\bibitem{Messenger:2011gi}
Messenger C and Read J 2012 {\em Phys. Rev. Lett.\/} {\bf 108} 091101
  (\textit{Preprint} \eprint{1107.5725})

\bibitem{ET_new}
Danilishin S and Zhang T 2023 {ET sensitivity curves used for CoBA Science
  Study} \urlprefix\url{https://apps.et-gw.eu/tds/ql/?c=16492}

\bibitem{Branchesi:2023mws}
Branchesi M {\em et~al.\/} 2023 {\em J. Cosmol. Astropart. Phys.\/} {\bf 07}
  068 (\textit{Preprint} \eprint{2303.15923})

\bibitem{CE_new}
Kuns K {\em et~al.\/} 2023 {Cosmic Explorer Strain Sensitivity}
  \urlprefix\url{https://dcc.cosmicexplorer.org/CE-T2000017/public}

\bibitem{Damour:2012yf}
Damour T, Nagar A and Villain L 2012 {\em Phys. Rev. D\/} {\bf 85} 123007
  (\textit{Preprint} \eprint{1203.4352})

\bibitem{Hinderer_PC}
Hinderer T (private communication)

\bibitem{Vines:2011ud}
Vines J, Flanagan {\'E}~{\'E} and Hinderer T 2011 {\em Phys. Rev. D\/} {\bf 83}
  084051 (\textit{Preprint} \eprint{1101.1673})

\bibitem{Kidder:1995zr}
Kidder L~E 1995 {\em Phys. Rev. D\/} {\bf 52} 821 (\textit{Preprint}
  \eprint{gr-qc/9506022})

\bibitem{Aasi:2013wya}
Abbott B~P {\em et~al.\/} (KAGRA Collaboration, LIGO Scientific Collaboration,
  and Virgo Collaboration) 2020 {\em Living Rev. Relativity\/} {\bf 23} 3 noise
  curves available from \url{https://dcc.ligo.org/T2000012/public}
  (\textit{Preprint} \eprint{1304.0670})

\bibitem{Srivastava:2022slt}
Srivastava V, Davis D, Kuns K, Landry P, Ballmer S, Evans M, Hall E~D, Read J
  and Sathyaprakash B~S 2022 {\em Astrophys. J.\/} {\bf 931} 22
  (\textit{Preprint} \eprint{2201.10668})

\bibitem{LSGA}
Katsanevas S {\em et~al.\/} 2020 {Lunar Seismic and Gravitational Antenna
  (LSGA), Idea I-2020-03581 in \emph{Ideas for exploring the Moon with a large
  European lander (ESA)}},
  \url{https://ideas.esa.int/servlet/hype/IMT?documentTableId=45087648405769217&userAction=Browse&templateName=&documentId=a315450fae481074411ef65e4c5b7746}

\bibitem{Jani:2020gnz}
Jani K and Loeb A 2021 {\em J. Cosmol. Astropart. Phys.\/} {\bf 06} 044
  (\textit{Preprint} \eprint{2007.08550})

\bibitem{Harms:2020kgl}
Harms J {\em et~al.\/} (LGWA Collaboration) 2021 {\em Astrophys. J.\/} {\bf
  910} 1 (\textit{Preprint} \eprint{2010.13726})

\bibitem{Amaro-Seoane:2020ahu}
Amaro-Seoane P, Bischof L, Carter J~J, Hartig M~S and Wilken D 2021 {\em Class.
  Quantum Grav.\/} {\bf 38} 125008 (\textit{Preprint} \eprint{2012.10443})

\bibitem{Cozzumbo:2023gzs}
Cozzumbo A, Mestichelli B, Mirabile M, Paiella L, Tissino J and Harms J 2023
  (\textit{Preprint} \eprint{2309.15160})

\bibitem{Ajith:2024mie}
Ajith P {\em et~al.\/} 2024  (\textit{Preprint} \eprint{2404.09181})

\bibitem{2005ApJ...622..759G}
{G{\'o}rski} K~M, {Hivon} E, {Banday} A~J, {Wandelt} B~D, {Hansen} F~K,
  {Reinecke} M and {Bartelmann} M 2005 {\em Astrophys. J.\/} {\bf 622} 759
  (\textit{Preprint} \eprint{astro-ph/0409513})

\bibitem{Zonca2019}
Zonca A, Singer L~P, Lenz D, Reinecke M, Rosset C, Hivon E and G{\'o}rski K~M
  2019 {\em J. Open Source Softw.\/} {\bf 4} 1298

\bibitem{Hunter:2007ouj}
Hunter J~D 2007 {\em Comput. Sci. Eng.\/} {\bf 9} 90

\bibitem{Harris:2020xlr}
Harris C~R {\em et~al.\/} 2020 {\em Nature (London)\/} {\bf 585} 357
  (\textit{Preprint} \eprint{2006.10256})

\bibitem{Virtanen:2019joe}
Virtanen P {\em et~al.\/} 2020 {\em Nat. Methods\/} {\bf 17} 261
  (\textit{Preprint} \eprint{1907.10121})

\bibitem{CreightonAnderson}
Creighton J~D~E and Anderson W~G 2011 {\em Gravitational-Wave Physics and
  Astronomy\/} (Weinheim: John Wiley \& Sons, Ltd)

\bibitem{1988MNRAS.234..663D}
{Dhurandhar} S~V and {Tinto} M 1988 {\em Mon. Not. R. Astron. Soc.\/} {\bf 234}
  663

\end{thebibliography}

\end{document}